\documentclass[sts]{imsart_arxiv}

\RequirePackage{amsthm,amsmath,amsfonts,amssymb}
\RequirePackage[numbers]{natbib}

\startlocaldefs
\usepackage{commands}
\newcommand{\new}[1]{#1}

\theoremstyle{plain}

\newtheorem{theorem}{Theorem}[section]
\newtheorem{lemma}[theorem]{Lemma}
\newtheorem{assumption}{Assumption}
\newtheorem{prop}{Proposition}[section]
\newtheorem{mydef}{Definition}
\newtheorem*{example}{Example}

\newcommand{\indep}{\mathrel{\perp\!\!\!\perp}} 
\usepackage{hyperref}
\endlocaldefs

\begin{document}

\begin{frontmatter}
\title{Replicable Bandits for Digital Health Interventions}
%\title{A Sample Article Title with Some Additional Note\thanksref{t1}}
\runtitle{Replicable Bandits for Digital Health Interventions}
%\thankstext{T1}{A sample of additional note to the title.}

% transmittal number for your manuscript is STS1017

\begin{aug}
%%%%%%%%%%%%%%%%%%%%%%%%%%%%%%%%%%%%%%%%%%%%%%%
%% Additional information such as            %%
%% identifying the corresponding author must %%
%% be included in in the Acknowledgments     %%
%% section if necessary.                     %%
\author[A]{\fnms{Kelly W.}~\snm{Zhang}\ead[label=e1]{kelly.zhang@imperial.ac.uk}},
\author[B]{\fnms{Nowell}~\snm{Closser}\ead[label=e2]{nowellclosser@g.harvard.edu},} %\orcid{0000-0000-0000-0000}}
\author[C]{\fnms{Anna L.}~\snm{Trella}\ead[label=e3]{annatrella@g.harvard.edu},} %\orcid{0000-0000-0000-0000}}
\and
\author[D]{\fnms{Susan A.}~\snm{Murphy}\ead[label=e4]{samurphy@g.harvard.edu}} %\ead[label=u1,url]{www.foo.com}}
%\thanksref{t1}
%\thankstext{t1}{Some comment}

\address[A]{Kelly W. Zhang is an Assistant Professor at Imperial College London \printead[presep={\ }]{e1}.}

\address[B]{Nowell Closser is a PhD student at Harvard University \printead[presep={\ }]{e2}.}

\address[C]{Anna L. Trella is a PhD student at Harvard University \printead[presep={\ }]{e3}.}

\address[D]{Susan Murphy is Professor at Harvard University\printead[presep={\ }]{e4}.}
\end{aug}

\begin{abstract}
%\kwz{Rebuttal is here \url{https://www.overleaf.com/1562899898xtttgkzrcbhn#c09279}}
Adaptive treatment assignment algorithms, such as bandit  algorithms, are increasingly used in digital health intervention clinical trials. Frequently the data collected %resulting 
from these trials is used to conduct causal inference and related data analyses to decide how to refine the intervention, and  whether to roll-out the intervention more broadly. \new{This work studies inference for estimands that depend on the adaptive algorithm itself; a simple example is the mean reward under %resulting from the use of 
the adaptive algorithm. Specifically, we investigate the replicability of statistical analyses concerning such estimands %based on 
when using data from trials deploying adaptive treatment assignment algorithms.} We demonstrate that many standard statistical estimators can be inconsistent and fail to be replicable across repetitions of the clinical trial, even as the sample size grows large. We show that this non-replicability is intimately related to properties of the adaptive algorithm itself. We introduce a formal definition of a "replicable bandit algorithm" and prove that under such algorithms, a wide variety of common  statistical estimators are guaranteed to be consistent and asymptotically normal. We present both theoretical results and simulation studies based on a mobile health oral health self-care intervention. Our findings underscore the importance of designing adaptive algorithms with replicability in mind, especially for settings like digital health where deployment decisions rely heavily on replicated evidence. We conclude by discussing open questions on the connections between algorithm design, statistical inference, and experimental replicability.
\end{abstract}


\end{frontmatter}
%%%%%%%%%%%%%%%%%%%%%%%%%%%%%%%%%%%%%%%%%%%%%%
%%%% Main text entry area:

%%%%%%%%%%%%%%%%%%%%%%%%%%%%%%%%%%%%%%%%%%%%%%%%%%%%%%%
%%%%%%%%%%%%%%%%%%%%%%%%%%%%%%%%%%%%%%%%%%%%%%%%%%%%%%%

\section{Introduction}

Adaptive treatment assignment algorithms, e.g., bandit \cite{lattimore2020bandit}  and more general reinforcement learning (RL) algorithms \cite{sutton2018reinforcement}, are increasingly used in clinical trials, social science experiments, and by tech companies to personalize treatment decisions to individuals \citep{figueroa2021adaptive,shaikh2019balancing,trella2022reward,trellaPCS,liao2020personalized,offer2019adaptive,athey2023can,wu2022interpretable,maystre2023optimizing}. 
These algorithms are adaptive because they learn during the course of the trial, i.e., they are updated periodically using accruing data collected during the trial.
After an adaptive algorithm is deployed, various statistical analyses are needed to answer questions such as: What was the average reward under the deployed algorithm? What was the effect of treatment on other outcomes of interest? What about in a particular state? Further, statistical analyses can be used to inform the design of the adaptive algorithm for the next implementation, e.g., design choices regarding the state space, action space, and/or choice of reward.

In this work, we investigate the replicability of statistical analyses for data from trials deploying adaptive treatment assignment algorithms. 
We focus on digital health intervention clinical trials; replicability of clinical trials is particularly critical as these trials are intended to provide evidence regarding future implementation on the broader population of such individuals. We use the term replicability in the spirit of \href{https://www.ncbi.nlm.nih.gov/books/NBK547524/}{Reproducibility and Replicability in Science} \cite{national2019reproducibility}, where they say that replicability means ``obtaining \textit{consistent} results across studies aimed at answering the same scientific question, each of which has obtained its own data.'' The definition of replicability in clinical trials refers to replications of the same study design yielding similar results with high probability. This typically involves:
\begin{itemize}[leftmargin=15pt]
    \item \bo{Using the same methodology:} Identical procedures, inclusion/exclusion criteria, outcome measures, and data curation and analysis techniques are applied. This also means the same bandit 
    or reinforcement learning algorithm (with the same hyperparameters) in adaptive experiments.
    \item \bo{New, independent sample from the same population:} Repetitions of the trial conducted on different samples of individuals from the same population.
    \item \bo{Same or different researchers:} Replication can be done by the original team or independent researchers.
\end{itemize}
Replicability ensures that other scientists will reach similar conclusions, i.e., have similar post-trial data analysis results, if they implement the same adaptive algorithm with a different group of individuals sampled from the same population. Replicability is closely related to the concept of ``Stability'' in Vertical Data Science (Predictability, Computability, and Stability) from \citet{yu2020veridical}. 

There are many notions of replicability discussed in the literature. In the reinforcement learning literature there are a variety of definitions of what it means for the \textit{algorithm} or the policies learned by the algorithm to be replicable \citep{karbasi2024replicability,eaton2024replicable,komiyama2024replicability,esfandiari2022replicable}.  At a high level, algorithmic replicability involves repeated use of the algorithm on inputs,  sampled from the same underlying distribution, leading to the same or similar outputs. 
In contrast, we aim for ``inferential'' replicability.  This is replicability of the statistical inferential methods across trial replications \citep{goodman2016does}. This work will begin to shed light on the ways that these two types of replicability are related. 

The main focus of this work is on how to ensure the replicability of statistical analyses for digital health intervention experiments that utilize adaptive algorithms for treatment assignment. While several works have shown that valid statistical inference  is possible, they generally assume idealized settings, e.g., each individual's responsiveness to treatment does not change over time and/or the
post-trial inference uses a correctly specified  
outcome model.
In this work, we discuss how basic notions of replicability of statistical analyses can fail when the true environment begins to deviate from these idealized settings.  In health, the true environment is highly likely to differ from these idealized setting as the environment concerns complex human behaviors. 

In particular, we show that on adaptively collected data, the replicability of many standard statistical estimators  
is intimately related to the replicability of the policies formed by the adaptive algorithm used to collect the data (see Section \ref{sec:replicableBandits} for the formal definition of replicable adaptive algorithms). 
This is true specifically for estimators of quantities, like the expected value of a health outcome on average across individuals in the trial, which depends on the treatment assignments made by the adaptive algorithm.
We show that standard statistical estimators for such quantities can be inconsistent and fail to be replicable across repetitions of an experiment, if the policies formed by the adaptive algorithm are ``non-replicable''. For example, we show in simple settings that common treatment effect estimators will be positive (strictly bounded above zero) for a constant fraction of trial replications and negative (strictly bounded below zero) for another constant fraction of replications; moreover, this is the case even as the number of individuals in each replicated trial goes to infinity.
Furthermore, we show that many standard bandit algorithms, like $\epsilon$-greedy and Thompson sampling \cite{russo2018tutorial}, can fail to satisfy such replicability requirements in these very simple settings.

This work  underscores the importance of ensuring inferential replicability when using adaptive algorithms.  Inferential replicability is critical for the wider adoption and application of these algorithms in settings like digital health and experimental design. Furthermore, we argue that researchers in the area of adaptive experimentation must prioritize the development of algorithms and methodologies that ensure inferential replicability. 

%%%%%%%%%%%%%%%%%%%%%%%%%%%%%%%%%%%%%%%%%%%%%%%%%%%%%%%
\subsection{Relationship to existing literature}
\label{sec:relatedWork}

There is a long history of works on adaptive experimentation in the biostatistics literature \citep{chow2008adaptive,hu2006theory,jennison1999group,chow2005statistical,greenhill2020bayesian,berry2012adaptive,park2018critical,chow2017adaptive,schafer2006adaptive,zhang2007asymptotic}. This literature primarily focuses on decision-making settings that involve making a single decision per individual (some exceptions that consider longitudinal data are \citep{jeffries2015longitudinal,yin2005adaptive} which focus on adaptive stopping rules and \citep{atkinson2017optimal} which considers Gaussian outcomes). These works primarily develop post-trial inference methods that are specific to the adaptive treatment assignment algorithm (e.g., algorithms like randomized play-the-winner and drop-the-loser are commonly considered).

In contrast, a recent line of work develops inference methods that make minimal assumptions on the adaptive algorithm, and are compatible with common bandit algorithms like Thompson sampling and $\epsilon$-greedy. This line of work includes methods that focus
%Recent work focuses 
on off-policy evaluation (i.e., evaluating the value of a policy) \citep{waudby2024anytime,bibaut2021post,zhan2021off} or inference on the mean reward under a particular treatment in a multi-arm bandit \citep{hadad2021confidence,bibaut2021post,howard2021time,hirano2023asymptotic,zhang2020inference,nair2023randomization}.  Another line of work aims to infer the parameters in a correctly specified outcome model \citep{deshpande,zhang2021mestimator,ying2024adaptive,khamaru2021near}. See \citet{bibaut2024demistifying} for a nice overview.
The finding in many of these previous works is that standard estimators (e.g., means, least squares, M/Z-estimators) on adaptively data are  \textit{consistent} (i.e., converging to the correct value). However, the distribution of these standard estimators can be asymptotically non-normal and converge non-uniformly in distribution, depending on properties of the bandit algorithm itself \citep{zhang2020inference,hadad2021confidence,deshpande}. For this reason, various works consider altered versions of standard estimators that incorporate ``adaptive'' weights that depend on the treatment propensities formed by the bandit algorithm \citep{hadad2021confidence,bibaut2021post,zhang2021mestimator,zhang2020inference,khamaru2021near,zhan2021off,deshpande,bibaut2024demistifying}; these weights can help ensure that standard estimators are still asymptotically normal on adaptively collected data.

Generally the above works %consider bandit algorithms and 
assume that the  environment is stationary and that outcome models are correctly specified.
In this paper, we investigate inference on adaptively collected data in the digital health setting, where these assumptions may fail to hold, specifically under nonstationary environments and misspecified outcome models.
See the next section for a discussion of why misspecified outcome models are common in  digital health.  Classically, estimators formed by minimizing a loss function based on a misspecified model will be consistent none-the-less for the minimizers of the loss function (e.g., negative log likelihood, $L_2$ loss). 
In this work, we show that on adaptively collected data, %we will show that 
under these types of deviations in the environment (non-stationarity and outcome model misspecification), standard estimators can be  \textit{inconsistent}  for the parameters that minimize the associated loss function. This is a  very severe consequence of using adaptive algorithms, which has significant consequences for inferential replicability.

\new{Another line of work uses randomization-based methods for inference on adaptively collected data \citep{nair2023randomization,freidling2024selective}. \citet{freidling2024selective} only consider inference for finite populations, which is not the focus of this work. \citet{nair2023randomization} provide a variety of methods (e.g. conditional independence testing in Markov decision processes, prediction intervals) and the one that is closest to our work considers inference for a fixed treatment effect in a contextual bandit setting. This differs from our work in that (i) they do not accommodate longitudinal data, (ii) their estimand cannot depend on the data collection algorithm (e.g. cannot be used to infer the average reward under the algorithm), and (iii) their method only applies to location and scale family outcomes (e.g. does not work for binary outcomes).}

Finally, there are works that formalize the replicability of a decision-making rule or algorithm \citep{karbasi2024replicability,eaton2024replicable,komiyama2024replicability,esfandiari2022replicable,parmigiani2023defining}. Many of these works build on the framework introduced in \citet{impagliazzo2022reproducibility}, which formalizes what it means for a randomized algorithm to be replicable (these works are discussed further in Section \ref{sec:relatedReplicable}). 
In this work, we focus on the replicability of statistical analyses, rather than on ensuring replicability of the adaptive algorithm; however, we show that the replicability of post-trial statistical analyses is often closely related to the replicability of the adaptive data collection algorithm (see Sections \ref{sec:nonrepExamples} and \ref{sec:replicableBandits}).

\section{Problem Setting} %%%%%%%%%%%%%%%%%%%%%%%%%%%%%%%%%%%%
\label{sec:problemSetting}

%%%%%%%%%%%%%%%%%%%%%%%%%%%%%%%%%%%%%%%%%%%%%%%%%%%%%%%
\subsection{Common characteristics of digital health intervention trials}
\label{sec:characteristics}

The motivation for this work is the increasing use of adaptive algorithms in digital intervention experiments. These interventions are designed to encourage individuals to engage in healthy behaviors. These interventions focus on a variety of health areas---including physical activity \citep{liao2020personalized,yom2017encouraging,figueroa2021adaptive,mintz2020nonstationary,zhou2018personalizing,khan2022kalman,tragos2023keeping,zhou2018evaluating,mintz2023behavioral}, medication adherence \citep{lauffenburger2021reinforcement}, smoking cessation \citep{battalio2021sense2stop,faro2023effect,albers2022addressing}, mental health \citep{beltzer2022building}, reducing alcohol consumption \citep{liu2023thompson}, chronic pain \citep{piette2022artificial}, dieting \citep{forman2019randomized,goldstein2020refining,rabbi2015automated}, and assisting elderly individuals \citep{hoey2010automated,mihailidis2008coach}. 
In digital health interventions, the adaptive algorithm itself is often considered a component of the entire intervention package. That is, if the digital intervention is found to be cost-effective, it would be rolled out with the adaptive algorithm component.  Thus the effectiveness of the intervention package will depend on the quality of the algorithm and the decisions it makes.

\begin{example}[Oralytics]
    In this work, we use the series of Oralytics clinical trials as a running example \citep{nahum2024optimizinh,trellaPCS,trella2022reward,trella2024oralytics}. Oralytics is a dental health intervention designed to augment standard dental care by providing additional support  over a 10-week period. In the Oralytics trial, individuals are given a Bluetooth-enabled toothbrush and asked to download the Oralytics mobile app.  During the the trial, individuals may receive engagement messages encouraging toothbrushing via the app before their usual morning and evening brushing windows (i.e., up to twice daily). An adaptive decision-making algorithm (e.g., a contextual bandit) is used to optimize (personalize) whether or not to deliver an engagement message.  Individual's toothbrushing behaviors are automatically detected with the Bluetooth-enabled toothbrush. 

    Oralytics involves a series of clinical trials, 
    where each trial uses a contextual bandit algorithm.  Between trials, the contextual bandit algorithm is redeveloped \citep{trellaPCS,trella2022reward,nahum2024optimizinh, trella2024oralytics}. 
    At this time, the Oralytics algorithm is  being revised for deployment in a two-arm trial for late spring 2025. In this two-arm trial, individuals will be randomized between the standard of care versus standard of care augmented with the digital intervention delivered using the revised contextual bandit algorithm. The primary outcome in this trial will be the difference in measures of overall dental health between the two groups at the end of the trial period. Further Oralytics digital intervention trials are in the planning stages. 
    \label{ex:oralytics}
\end{example}

We now highlight some common characteristics of digital health intervention trials that use adaptive algorithms (including Oralytics).
\begin{enumerate}[leftmargin=15pt]
    \item \bo{Multiple decision-times over time per individual:} Digital health interventions involve making multiple decisions on the same individual over time. For example, treatment decisions could be whether or not to deliver engagement messages to encourage healthy behavior, as is the case in Oralytics. Typically the number of decision times per individual are fixed and particular to the domain, e.g., twice per day over a 70-day period to improve brushing behaviors. 
    %%%%%%%%%%%%%%%%%%%%%%%%%%%%%%%%%%%%%%%%
    \item \bo{Data-sharing bandit algorithms:} Digital intervention settings often have low signal relative to the amount of noise in outcomes. This is because treatments are delivered to individuals as they go about their everyday lives, as opposed to in a clinic. Moreover, clinical trials generally only involve a limited number of individuals. 
    Thus, in the digital health setting developers of adaptive algorithms must carefully trade-off bias and variance due to the high variance in outcomes typical in digital health. As a result, contextual bandit algorithms (simple RL algorithms) are most often used \citep{liao2020personalized,piette2022artificial,figueroa2021adaptive,mintz2020nonstationary,beltzer2022building,liu2023thompson,yom2017encouraging,lauffenburger2021reinforcement,nahum2024optimizinh}. 
    
    \quad \quad To further trade-off bias and variance, it is also common for bandit algorithms to combine or ``pool'' data across heterogeneous individuals in the trial to improve learning, i.e., the algorithm may adapt the treatment decisions for one individual based on previous treatment outcomes from others \citep{yom2017encouraging,figueroa2021adaptive,piette2022artificial,nahum2024optimizinh}. These data-sharing or pooling algorithms reduce noise and potentially learn faster.
    \item \bo{The primary data analysis is prespecified in clinical trials:} All National Institute of Health (NIH) funded clinical trials, including the post-trial statistical analyses, are required to be preregistered. This means that the analyses that will be performed cannot be changed without jeopardizing the validity of the trial.
\end{enumerate}
In this work, we consider digital health settings in which adaptive treatment assignment algorithms pool data over individuals to learn. % make treatment decisions.  
Further, we make the simplifying assumption that individuals are all recruited into the trial all at once (which is  common in trials that recruit on the internet). In general, digital health intervention trials typically recruit individuals into the trial incrementally or in batches over time.  Though we do not explicitly handle incremental recruitment in this paper, we conjecture that the key ideas below will generalize to that setting. % without difficulty.

%%%%%%%%%%%%%%%%%%%%%%%%%%%%%%%%%%%%%%%%%%%%%%%%%%%%%%%
\subsection{Observed data and bandit algorithm}

We now formalize the problem setting. 
Throughout this work, we consider contextual bandit algorithms which are commonly used in real-life digital health intervention trials \citep{liao2020personalized,yom2017encouraging,figueroa2021adaptive,mintz2020nonstationary,lauffenburger2021reinforcement,beltzer2022building,liu2023thompson,piette2022artificial,rabbi2015automated}.  Further, note that by designing the reward carefully, contextual bandit algorithms can be thought of as crudely approximating a full reinforcement learning algorithm \citep{trella2022reward,liao2020personalized}.\footnote{
\new{Consider a Markov Decision Process setting with binary action space. The main idea is that  the optimal policy selects the action $A_t = 1$ if $\E \big[ R_t  \mid S_t = s, A_t=1 \big]-\E \big[ R_t \mid S_t = s, A_t=0\big] + \gamma\E \big[ V^{\pi^*}(S_{t+1}) \mid S_t = s, A_t = 1 \big] - \gamma\E \big[ V^{\pi^*}(S_{t+1}) \mid S_t = s, A_t = 0 \big] > 0$,
where $\gamma$ is the discount rate and $V^{\pi^*}(s)=E_{\pi^*}\left[\sum_{j=0}^\infty \gamma^{j-1} R_{t+j}\mid S_{t+1}=s\right] $ is the value of the optimal policy. The papers \citet{liao2020personalized} and \citet{trella2022reward} attempt to improve upon contextual bandit algorithms by adding to the  reward, a function that approximates $\gamma\E \big[ V^{\pi^*}(S_{t+1}) \mid S_t = s, A_t = 1 \big] - \gamma\E \big[ V^{\pi^*}(S_{t+1}) \mid S_t = s, A_t = 0 \big]$.  In the above two papers the function is constructed using  data from prior trials.
}
}

The trial consists of $n$ individuals drawn independently from a population of interest. The algorithm makes a total of $T$ treatment decisions per individual. \new{Note that $T$ is a fixed duration specified by the intervention duration.\footnote{\new{Recall that the number of decision points per individual, $T$, is determined by the intervention as well as the particular health disorder.}}} %; for expositional simplicity for much of the paper we will focus on a $T = 2$ setting. 
For each individual $i \in [1 \colon n]$ and each decision time $t \in [1 \colon T]$, \new{the adaptive algorithm observes a vector of time-varying covariate features $X_{i,t} \in \MC{X}$, which is often called the ``context'' or ``state''. The algorithm uses $X_{i,t}$ to inform its treatment decision ``action'' $A_{i,t} \in \MC{A}$, where $\MC{A}$ is the action space.\footnote{In general the action space may be constrained by the covariates $X_{i,t}$, e.g., $X_{i,t}$ may indicate whether the individual is ``available'' for treatment.}} %For simplicity, we only consider binary treatments, $|\MC{A}|=2$.  
Then, a multivariate post-treatment outcome vector $Y_{i,t}$ is observed. The algorithm's reward is a known, real-valued function, $R$, of treatment outcomes $Y_{i,t}$: $R(Y_{i,t}) \in \real$.

A contextual bandit algorithm forms a working model of the conditional mean of the reward:
    \begin{align*}
        \E \big[ R(Y_{i,t}) \mid X_{i,t}, A_{i,t} \big] = f_b(X_{i,t}, A_{i,t}),
    \end{align*}
where $b$ is model parameter vector. 
%$b \in \real^{d_b}$. 
As discussed in the previous section, in order to trade off bias and variance, the working model is likely misspecified.
The algorithm uses the data accrued during the trial to iteratively update a statistic $\hat{\beta}_t^{(n)} \in \real^{d_\beta}$; this statistic is often comprised of an estimator of the model parameters $b$, as well as estimates of the variance of that estimator.

Bandit algorithms are usually designed to select actions to maximize expected rewards, possibly subject to some constraints (e.g., the algorithm may be constrained to ensure a certain level of power of a post-trial statistical test \citep{yao2020power,simchi2023multi}). To do this, the contextual bandit algorithm uses the working reward model to form the treatment selection policies 
at each decision time $t$.
Prior to the start of the trial, the bandit algorithm forms an initial policy $\pi_1$ that is used to select actions $\{ A_{i,1} \}_{i=1}^n$, i.e., the first treatment decision for all individuals. $\pi_1 : \MC{X} \by \MC{A} \mapsto [0, 1]$, where
\begin{align*}
    \pi_1(x, a) = \PP \left( A_{i,1} = a \mid X_{i,1} = x \right).
\end{align*}
We use $\HH_{i,1} = \{ X_{i,1}, A_{i,1}, Y_{i,1} \}$ to denote individual $i$'s history and $\HH_{1:n,1} = \{ \HH_{i,1} \}_{i=1}^n$
%\triangleq \{ X_{i,1}, A_{i,1}, Y_{i,1} \}_{i=1}^n$ 
to denote the data collected from the first decision time from all $n$ individuals.

At each subsequent decision time $t \in [2 \colon T]$, the bandit algorithm uses 
the previously-collected data $\HH_{1:n,t-1} = \{ \HH_{i,t-1} \}_{i=1}^n = \{ X_{i,t'}, A_{i,t'}, Y_{i,t'} 
 : t' \in [1 \colon t-1] \}_{i=1}^n$
from \textit{all $n$ individuals} to form the statistic $\hat{\beta}_{t-1}^{(n)} \in \real^{d_\beta}$.
The statistic is then used to form a policy %$\hat\pi_t^{(n)}$ 
for selecting action $A_{i,t}$ for each individual $i \in [1 \colon n]$; see Figure \ref{fig:pooling} for an illustration. Specifically, there is a policy class $\{ \pi(\cdotspace; \beta) \}_{\beta \in \real^{d_\beta}}$ and the policy $\hat{\pi}_t^{(n)}(\cdotspace): \MC{X} \by \MC{A} \mapsto [0,1]$ is such that
%= \pi_t \big( \cdotspace; \hat{\beta}_{t-1}^{(n)} \big)
\begin{multline}
    \label{eqn:pihat}
    \hat{\pi}_t^{(n)}(x,a) = 
    \pi\big( x, a; \hat\beta_{t-1}^{(n)} \big) \\
    = \PP \big( A_{i,t} = a \mid X_{i,t} = x, \hat\beta_{t-1}^{(n)} \big).
\end{multline}
See Section \ref{sec:nonrepBanditAlgs} for examples of common bandit algorithms.

%%%%%%%%%%%%%%%%%%%%%%%%%%%%%%%%%%%%%%%%%%%%%%%%%%%%%%%
\subsection{Post-trial analyses}

\subsubsection{Potential outcomes.}
%Notice that 
To discuss the post-trial statistical analyses (including causal analyses),  we use potential outcome notation. Note that the potential outcomes model we consider for the post-trial analyses does not necessarily match the reward model used by the adaptive treatment assignment algorithm.
The post-trial statistical analyses needs to be interpretable, replicable, and valid even if the the adaptive algorithm makes simplifying modeling assumptions and its reward model is misspecified. Recall from Section \ref{sec:characteristics} that simplified contextual bandit algorithms are commonly used in order to reduce variance, so it is not surprising that the algorithm's model may be misspecified.

Let $a_{1:t} \in \MC{A}^t$ to denote a sequence of actions; here $\MC{A}^t$ denotes the $t$-fold cartesian product of $\MC{A}$.   We assume no interference \citep{hudgens2008toward}
as in the experiments we consider, there is no contact between individuals; that is, the potential outcomes for one individual are unaffected by the treatment actions received by other individuals.  Thus, for the $i^{\TN{th}}$ individual, we use $Y_{i,t}(a_{1:t})$ to denote the $t^{\TN{th}}$ decision-time potential outcome and $X_{i,t+1}(a_{1:t})$ to denote the $(t+1)^{\TN{st}}$ decision-time potential covariate observed after taking the action sequence $a_{1:t}$.  We use $O_i$ to denote all potential outcomes for an individual $i$:
\begin{multline}
    \label{eqn:poDist}
    O_i \triangleq \big\{ X_{i,1}, Y_{i,1}(a_1), X_{i,2}(a_1), Y_{i,2}(a_{1:2}), \dots, \\
    %X_3(a_{1:2}), Y_3(a_{1:3}), 
    X_{i,T}(a_{1:T-1}), Y_{i,T}(a_{1:T}) : a_{1:T} \in \{0, 1 \}^T \big\}.
\end{multline}
We assume that the individuals' potential outcomes are drawn i.i.d. from a joint distribution $P \in \MC{P}$:
\begin{align}
    \label{eqn:OdistP}
    O_1, O_2, \dots, O_n \sim P.
    %\big( X_1, Y_1(0), Y_1(1) \big), \big( X_2, Y_2(0), Y_2(1) \big), \dots, \big( X_T, Y_T(0), Y_T(1) \big) \iidsim P.
\end{align}
\new{The above assumption practically means that individuals are drawn i.i.d. from a population of interest, $P$.}

\begin{figure}[t]
    \centering
    \includegraphics[width=\linewidth]{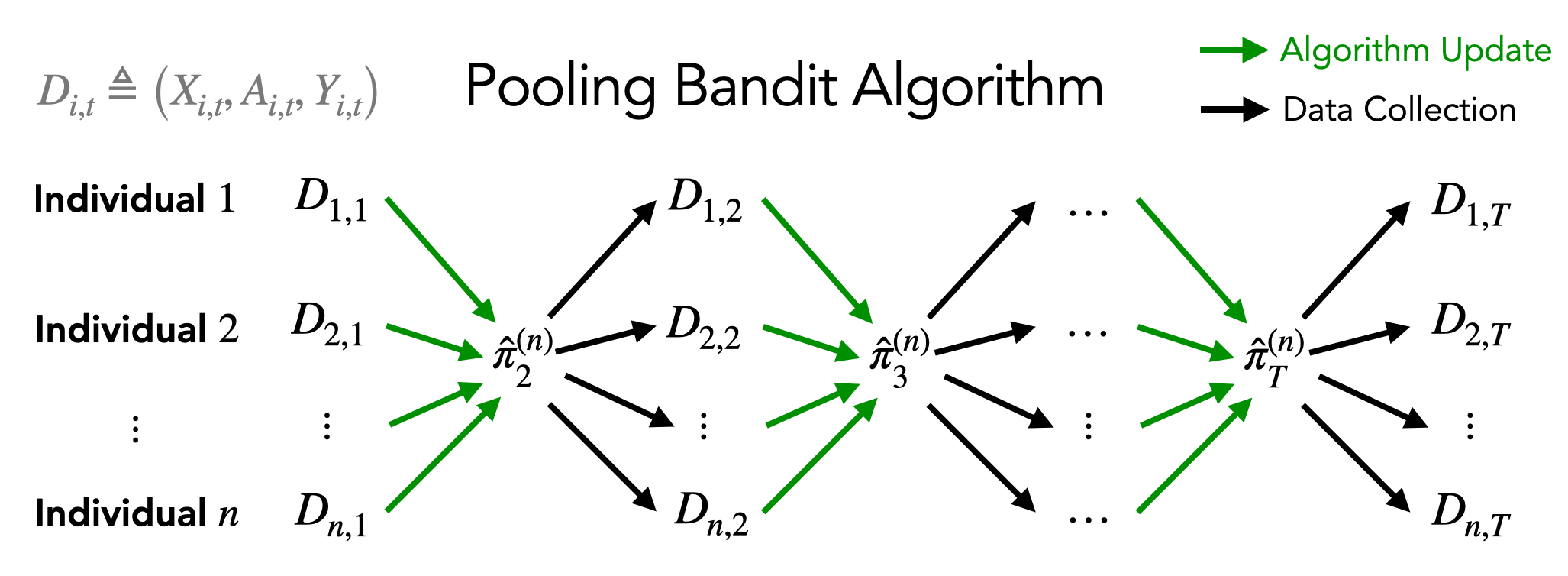}
    \vspace{-4mm}
    \caption{\bo{Pooling bandit algorithm.} Depicting how the bandit algorithm combines or ``pools'' data across individuals to form and update the policies over time. \vspace{-3mm}}
    \label{fig:pooling}
\end{figure}

Finally, we make a stable unit value treatment assumption \citep{rubin1980randomization,imbens2015causal} in that the observed outcome is $Y_{i,t} \triangleq Y_{i,t}(A_{i,1:t})$ and the observed state $X_{i,t} \triangleq X_{i,t}(A_{i,1:t-1})$. Note that even though the $n$ potential outcomes are independent, the observed outcomes $\{Y_{i,t}, X_{i,t} \}_{i=1}^n$ are dependent due to pooling by the bandit algorithm to select actions; see  Figure \ref{fig:pooling} for a pictorial representation. 

%%%%%%%%%%%%%%%%%%%%%%%%%%%%%%%%%%%%%%%%%%%%%%%%%%%%%%%
\subsubsection{Estimands under population-deployed policies.}
\label{sec:popDeployed}

When bandit algorithms pool or combine data across individuals to learn, the policies $\{ \pi_t( \cdotspace; \hat\beta_{t-1}^{(n)}) \}_{t = 1}^T$ that the bandit algorithm learns depend on $n$, the number of individuals in the trial (as seen in display \eqref{eqn:pihat} and Figure \ref{fig:pooling}). This means that quantities like the expected average reward per individual (averaged over time) 
under a bandit policy can also depend on $n$:
% , of individuals in the trial:
\begin{align*}
    \E^{(n)} \left[ \frac{1}{T} \sum_{t=1}^T R(Y_{i,t}) \right].
\end{align*}
Above, we use $\E^{(n)}$ to denote expectations under the deployed bandit algorithm; the expectation averages over the distribution of the potential outcomes of the $n$ individuals, the stochasticity in the bandit algorithm itself, and the treatment decisions the algorithm makes. Formally, the expectation $\E^{(n)}$ is indexed by $n$, the number of individuals in the trial because \textit{the observed outcomes $Y_{i,t}$ depend on the distribution of the treatment selection policies $\{ \hat{\pi}_t^{(n)} \}_{t = 1}^T$ 
%\HH_{1:n,t-1}
%\hat\beta_{t-1}^{(n)}) \}_{t = 1}^T$ 
formed by the bandit algorithm, which can change with the number of individuals in the trial}.

The dependence of quantities like the expected reward on the number of individuals in the trial requires careful thought about estimands. In this work, we consider estimands under  \textit{population-deployed policies}: estimands consistent with deployment of the intervention package in the population.  Indeed, the goal of many digital health trials using adaptive algorithms is to evaluate  the performance of the intervention package (of which the contextual bandit algorithm is one component) if deployed on the wider population. For this reason, we are particularly interested in statistical inference under the policy that \new{would be learned under the limit as the number of individuals in the trial grows  ($n\to\infty$).} For example, we might consider the following estimand, which is the expected reward:
\begin{align}
    \label{eqn:diffAveReward}
    \theta^* = \lim_{n \to \infty} \E^{(n)} \left[ \sum_{t=1}^T R(Y_{i,t}) \right].
\end{align}
The above quantity can be interpreted as the expected reward if the bandit algorithm was deployed on the entire population of interest. This type of estimand is relevant for the two-arm trial phase of Oralytics (see Example in Section \ref{ex:oralytics}). We provide formal conditions under which the limit in display \eqref{eqn:diffAveReward} exists, later in Proposition \ref{prop:thetastarWellDefined}.

\new{
\subsubsection{Replicable post-trial analyses.}
After the trial concludes, we are interested in performing statistical analyses using the dataset $\HH_{1:n,T}$ collected by the adaptive algorithm. We consider two properties that help ensure replicable statistical analyses; the first directly concerns the replicability of an estimator's value across trial replications, while the second, though distinct from replicability itself, concerns being able to characterize the variability of that estimator across hypothetical replications.

After the trial concludes, we will form an estimator $\hat{\theta}^{(n)} \triangleq f(\HH_{1:n,T})$ using some real-valued function $f$. For example, to estimate \eqref{eqn:diffAveReward}, we could use $f(\HH_{1:n,T}) = \frac{1}{n} \sum_{i=1}^n \frac{1}{T} \sum_{t=1}^T R(Y_{i,t})$. We say this estimator is replicable if its value is ``close'' with high probability across independent replications of the clinical trial. Moreover, as the number of individuals in the trial $n$ grows, the estimates across replications must get closer. We formalize this in Definition \ref{def:replicableStat} below.
\begin{mydef}[Replicable estimator] 
    \label{def:replicableStat}
    Let $\HH_{1:n,T}$ and $\tilde{\HH}_{1:n,T}$ be histories associated with independent replications of a trial. Let $\hat{\theta}^{(n)} \triangleq f(\HH_{1:n,T})$ and $\tilde{\theta}^{(n)} \triangleq f(\tilde{\HH}_{1:n,T})$ for a real-valued function $f$. We say $f$ is a replicable estimator if for each $\epsilon > 0$, as $n \to \infty$,
    \begin{align*}
        % \label{eqn:replicableDefStat}
        \sup_{P \in \MC{P}} \PP_P^{(n)} \big( \big| \hat{\theta}^{(n)} -  \tilde{\theta}^{(n)} 
        \big| > \epsilon \big) 
        \to 0.
    \end{align*}
\end{mydef}

In our definition above, we take the supremum over the class of outcome distributions $P \in \MC{P}$.  The use of the supremum in the definition means that, replicability as defined here, holds \textit{uniformly} across a range of possible data-generating distributions, including the worst cases.

The second property that we would like for our post-trial statistical analyses is that we are able to use $\HH_{1:n,T}$ to form a consistent estimate for the variance of our estimator $\hat{\theta}^{(n)}$. Having an accurate estimate of the variance ensures that using data from our \textit{single} trial run, $
\HH_{1:n,T}$, we are able to characterize the variability of $\hat{\theta}^{(n)}$ across many hypothetical replications of the trial. Accurate variance estimation is typically required for using asymptotic approximations (e.g. asymptotic normal approximations) to form confidence intervals for estimands of interest. In Definition \ref{def:replicableVar} below, we formally define what we mean by consistently estimating the variance of $\hat{\theta}^{(n)}$.
\begin{mydef}[Consistent variance estimator] 
    \label{def:replicableVar}
    Let $\HH_{1:n,T}$ be the data collected in an adaptive trial and let $\hat{\theta}^{(n)} \triangleq f(\HH_{1:n,T})$ be an estimator defined by a real-valued function $f$. We say that $v(\HH_{1:n,T})$ is a consistent estimator for the variance of $\hat{\theta}^{(n)}$, i.e., $\Var(\hat{\theta}^{(n)})$, if for any $\epsilon > 0$, as $n \to \infty$,
    \begin{align*}
        % \label{eqn:replicableDefStat}
        \sup_{P \in \MC{P}} \PP_P^{(n)} \bigg( \bigg| \frac{ \Var(\hat{\theta}^{(n)}) }{v(\HH_{1:n,T})} - 1 \bigg| > \epsilon \bigg) 
        \to 0.
    \end{align*}
\end{mydef}
In our definition above, we ensure that the ratio of the variance estimate and the true variance is close to $1$.
Additionally, just as in Definition \ref{def:replicableStat}, we take a supremum over the class of outcome distributions $P \in \MC{P}$, which ensures our variance estimator is consistent even for worst-case possible distributions.
}

%%%%%%%%%%%%%%%%%%%%%%%%%%%%%%%%%%%%%%%%%%%%%%%%%%%%%%%
%%%%%%%%%%%%%%%%%%%%%%%%%%%%%%%%%%%%%%%%%%%%%%%%%%%%%%%
\section{Two Examples of Non-Replicability}
\label{sec:nonrepExamples}

This section aims to demonstrate the need to think carefully about replicability when using bandit algorithms in digital health. We consider two examples in which we attempt to replicate statistical analyses when using common bandit algorithms in digital health trials. In these examples, the estimator \textit{averaged across replications} converges to the desired estimand. However, in any particular replication, the estimator is \textit{inconsistent} and does not converge to the desired estimand as the number of individuals $n \to \infty$.  These examples are disturbing since each digital health scientist only has access to their trial and not a population of hypothetical replications.

The first example considers estimation of the mean reward based on data collected by a bandit using a misspecified reward model. In particular, the true conditional mean of the reward is changing with time (non-stationary), but the bandit's reward model is stationary. 
%in a (misspecified) non-stationary bandit environment. 
In the second example, both the reward model of the bandit algorithm used to make adaptive treatment assignments during the trial, as well as the outcome model used in the statistical analysis after the trial, are misspecified.
For simplicity and clarity, the examples consider a $T=2$ setting. These simplistic examples capture the main problems that one may encounter in real digital health intervention trials (we examine settings with $T>2$ in simulations in Section \ref{sec:simulations}).

\subsection{Mean reward in a misspecified non-stationary environment} %%%%%%%%%%%%%%%%%%%%%%%%%%%%%%%%%%%%
\label{sec:nonstationary}

In this example, we are interested in estimating the average reward under the bandit algorithm when deployed on the population (under the \textit{population-deployed policy} from Section \ref{sec:popDeployed})
as described in \eqref{eqn:diffAveReward}: 
%For simplicity, here we just consider inferring the first term in this difference, the average reward under the population-deployed policy:
\begin{align}
    \label{eqn:misspecifiedThetastar}
    \theta^* = \lim_{n \to \infty} \E^{(n)} \left[ \frac{1}{T} \sum_{t=1}^T R(Y_{i,t}) \right]. %- \E_{\pi^{\TN{baseline}}} \left[ \frac{1}{T} \sum_{t=1}^T R(Y_{i,t}) \right]
\end{align}
It is assumed for now that the above limit is well-defined; we prove this for the setting we consider below (see Proposition \ref{prop:nonrepNonstat}).

A  natural estimator  is the empirical average reward, i.e.,
\begin{align}
    \label{eqn:aveReward}
    \hat{\theta}^{(n)} \triangleq \frac{1}{n} \sum_{i=1}^n \frac{1}{T} \sum_{t=1}^T R(Y_{i,t}). %- \frac{1}{n} \sum_{i=n+1}^{2n} \frac{1}{T} \sum_{t=1}^T R(Y_{i,t}) 
\end{align}
Next we describe a simple environment, with $T=2$ and  with a non-stationary mean reward  in which  $\hat{\theta}^{(n)}$ does not converge in probability to $\thetastar$ and fails to be replicable in the sense of Definition 
\ref{def:replicableStat}. \new{We also show empirically that the sample variance $\hat{s}^2 = 
\frac{1}{n} \sum_{i=1}^n 
\big( \frac{1}{T} \sum_{i=1}^T R(Y_{i,t}) - \hat{\theta}^{(n)} \big)^2$ fails to be an accurate estimate for variance of $\hat{\theta}^{(n)}$ across trial replications.}

\subsubsection{Nonstationary data generating environment.} %%%%%%%%%%%%%%%%%%%%%%
\label{sec:nonstationaryDGP}
It is well known that in health settings the impact of treatment may change over time \citep{mintz2020nonstationary,simchi2023non}. For example, individuals may initially be more responsive earlier in the trial, but become increasingly disengaged with the intervention over time \citep{amagai2022challenges,weiEngagement2020,park2023understanding}. 
In this example, the underlying environment is non-stationary, but the bandit algorithm's model of the environment is stationary, and thus misspecified.

Consider a trial with just $T=2$ decision times. Actions are binary, so  $\MC{A} = \{ 0, 1 \}$. Suppose the rewards are generated as follows:
\begin{align*}
    R\big( Y_{i,t} \big) = \mu_0 + \Delta_t A_{i,t} + \epsilon_{i,t},
\end{align*}
where $\epsilon_{i,t} \sim \N(0, 1)$ i.i.d. across $i$ and $t$.

In the following, $\mu_0 = 0$ and to mimic disengagement with the intervention over time, there is a decreasing trend in the treatment effect $\Delta_t$: $\Delta_1 = 0$ and $\Delta_2 = -0.25$.

\subsubsection{Sample mean is non-replicability under common bandit algorithms.} %%%%%%%%%%%%%%%%%%%%%%
\label{sec:nonrepBanditAlgs}
We consider two simple, stationary multi-armed
bandit algorithms (there are no context features): (i) $\epsilon$-greedy, and (ii) Thompson sampling with a conjugate Gaussian-Gaussian model. We choose these algorithms not because the conclusions are limited to these two algorithms, but rather because these algorithms are often used in practice \citep{nahum2024optimizinh,figueroa2021adaptive,liao2020personalized,liu2023thompson}. 
In both the (i) $\epsilon$-greedy  and (ii) Thompson sampling  examples, the   $\{A_{i,1}\}_{i=1}^n$ at time $t=1$ are i.i.d. Bernoulli ($0.5$).

\begin{example}[$\epsilon$-greedy multi-armed bandit]
Given $\epsilon \in (0, 0.5)$, the $\epsilon$-greedy bandit algorithm selects treatment at time $t=2$ with probability:
\begin{align*}
    \PP \big( A_{i,t} = 1 \mid \hat{\beta}_{t-1}^{(n)} \big) = \begin{cases}
        1-\epsilon & \TN{if}~ \hat{\beta}_{t-1}^{(n)} > 0 \\
        \epsilon & \TN{otherwise}
    \end{cases},
\end{align*}
for $\betahat{t-1} = \frac{\sum_{i=1}^n \sum_{t'=1}^{t-1} A_{i,t'} R(Y_{i,t'}) }{ \sum_{i=1}^n \sum_{t'=1}^{t-1} A_{i,t'} } - \frac{\sum_{i=1}^n \sum_{t'=1}^{t-1} (1-A_{i,t'}) R(Y_{i,t'}) }{ \sum_{i=1}^n \sum_{t'=1}^{t-1} (1-A_{i,t'}) }$.
\end{example}

\begin{example}[Thompson sampling multi-armed bandit]
The Thompson sampling bandit algorithm \citep{russo2018tutorial} forms a Bayesian model of the rewards:
\begin{align*}
     R(Y_{i,t}) = \mu_a + \epsilon_{i,t},
\end{align*}
with prior $\mu_a \sim \N(0, 1)$ independently for each $a\in\{0,1\}$ and  error, $\epsilon_{i,t} \sim \N(0, 1)$ i.i.d. across $i$ and $t$. The algorithm selects treatment using:
\begin{align*}
    \PP \big( A_{i,t} = 1 \mid \HH_{1:n,t-1}
    \big) 
    = \PP \big( \mu_1 > \mu_0 \mid \HH_{1:n,t-1} \big).
\end{align*}
The right hand side probability above averages over the posterior draw of $\mu_1, \mu_0$ given the collective user history $\HH_{1:n,t-1}$.
Recall that $\HH_{1:n,t-1} \triangleq \{ \HH_{i,t-1} \}_{i=1}^n$ for individual user histories $\HH_{i,t-1} \triangleq \{ X_{i,t'}, A_{i,t'}, Y_{i,t'} \}_{t'=1}^{t-1}$.
\end{example}

Figure \ref{fig:nonstationary} below shows empirically that under these common bandit algorithms in the non-stationary environment described above, $\hat{\theta}^{(n)}$ does not converge in probability to a constant, 
as $n \to \infty$ (this is formalized in Proposition \ref{prop:nonrepNonstat} below). 
Since at $t=1$, both the $\epsilon$-greedy and Thompson sampling algorithms select $A_{i,t=1} = 1$ with probability $0.5$ and furthermore there is no treatment effect at time $t=1$ ($\Delta_1=0$), one might expect that these two algorithms would perform much like a baseline policy 
$\pi^{\TN{baseline}}$ that selects  $A_{i,t} = 1$ with probability $0.5$ for all $i,t$. The expected average (over $T$) reward under this baseline policy is $\E_{\pi^{\TN{baseline}}} \left[ \frac{1}{T} \sum_{t=1}^T R(Y_{i,t}) \right] = -0.0625$.
If one uses the either of the two bandit algorithms in the trial, and estimates the expected average (over $T$) reward using the estimator \eqref{eqn:aveReward}, a constant fraction of replication trials will have $\hat{\theta}^{(n)} > 0.0625 + \delta$ and another constant fraction of replications will have $\hat{\theta}^{(n)} < -0.0625 - \delta$ for some $\delta > 0$; this is true even as $n \to \infty$.

\begin{figure}[h]
    \centering
    \includegraphics[width=0.49\linewidth]{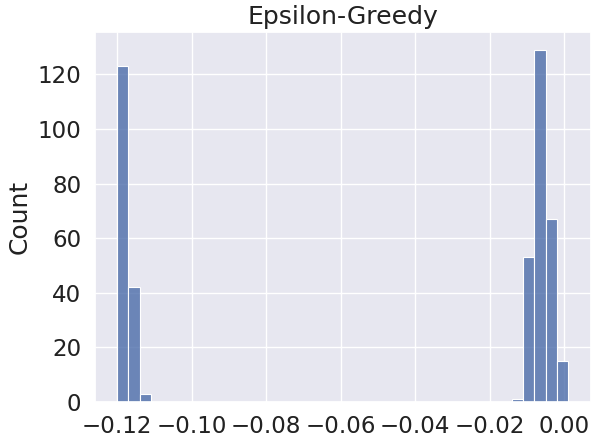}
    ~
    \includegraphics[width=0.48\linewidth]{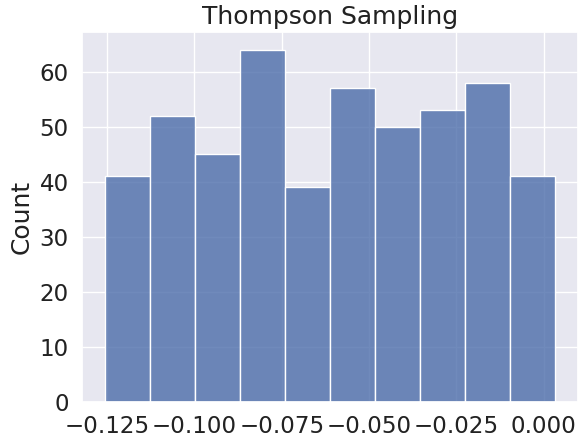}
    \caption{\bo{Histograms of Average Reward Estimators \eqref{eqn:aveReward} under Non-Stationarity.} Experiments with $n=100,000$, $T=2$, and $500$ Monte Carlo repetitions. Even for very large $n$, $\thetahat$ does not converge to a constant across trial replications. 
    }
    \label{fig:nonstationary}
\end{figure}

\begin{prop}[Non-replicability under non-stationarity]
    \label{prop:nonrepNonstat}
    $\epsilon$-greedy (for any $\epsilon < 0.5$) or Gaussian-Gaussian Thompson sampling (with standard normal priors) is deployed in the non-stationary environment described above.

    In this setting, $\thetastar$ from \eqref{eqn:misspecifiedThetastar} is well-defined. Moreover, $\hat{\theta}^{(n)}$ does not converge in probability to $\theta^*$ as $n \to \infty$.
    Additionally, there exists a $\delta>0$ for which $\lim_{n\to\infty} \PP^{(n)} \big( \hat{\theta}^{(n)} > -0.0625 + \delta\big)>0$ and \\
    $\lim_{n\to\infty} \PP^{(n)}\big(\hat{\theta}^{(n)} < -0.0625 + \delta\big)>0$. We characterize the limiting distribution of $\hat{\theta}^{(n)}$ in Appendix \print{\ref{app:nonstationary}}{A.1}.
\end{prop}

We now describe the intuition behind this inferential non-replicability phenomenon.
Since at the first decision time $(t=1)$ the true treatment effect $\Delta_1 = 0$, the bandit algorithm should not learn to favor selecting action $1$ or action $0$ on average across trial replications. Indeed, under common bandit algorithms, including $\epsilon$-greedy and Thompson sampling the action selection probabilities $\PP(A_{i,2} = 1 \mid \HH_{1:n,1})$ on average across trial replications equal $0.5$, i.e.,
\begin{align*}
    \E^{(n)} \big[ \PP(A_{i,2} = 1 \mid \HH_{1:n,1}) \big] = 0.5
\end{align*}
for each $n$. However, the treatment probability $\PP(A_{i,2} = 1 \mid \HH_{1:n,1})$ in any \textit{individual} trial does not concentrate, that is, $\PP(A_{i,2} = 1 \mid \HH_{1:n,1})$ \textit{does not} converge in probability to $0.5$ as $n \to \infty$. This occurs because when the treatment effect is zero, the algorithm has no reason to favor one treatment over the other (formalized in Appendix \print{\ref{app:nonstationary}}{A.1}). In other words, the bandit algorithm's treatment propensities are not replicable across trials, which leads to the variability in the average reward estimator $\hat\theta^{(n)}$, from display \eqref{eqn:aveReward}, in this non-stationary environment.

\begin{figure*}[h]
    \vspace{-3mm}
    \centering
    \includegraphics[width=0.32\linewidth]{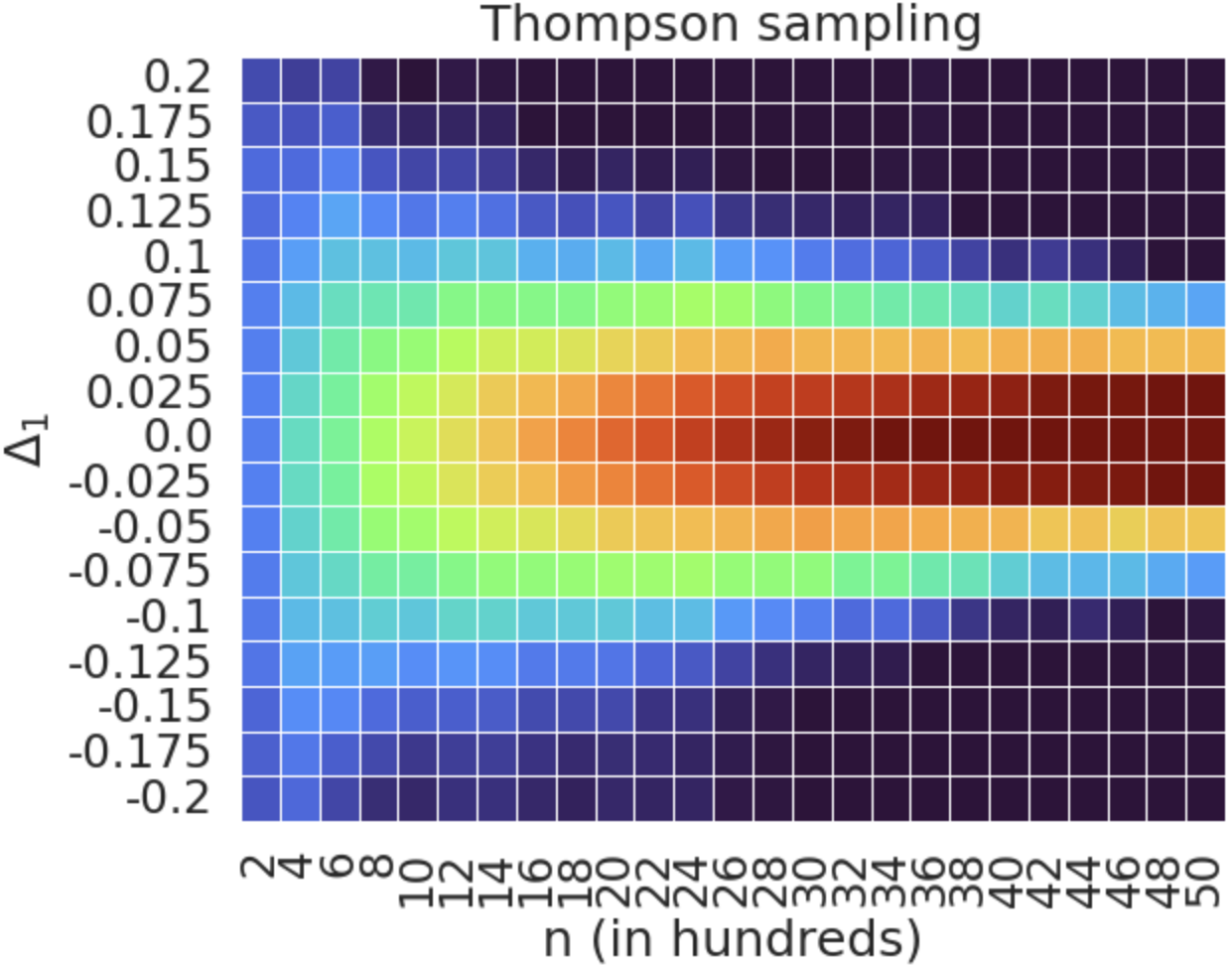}
    ~~
    \includegraphics[width=0.262\linewidth]{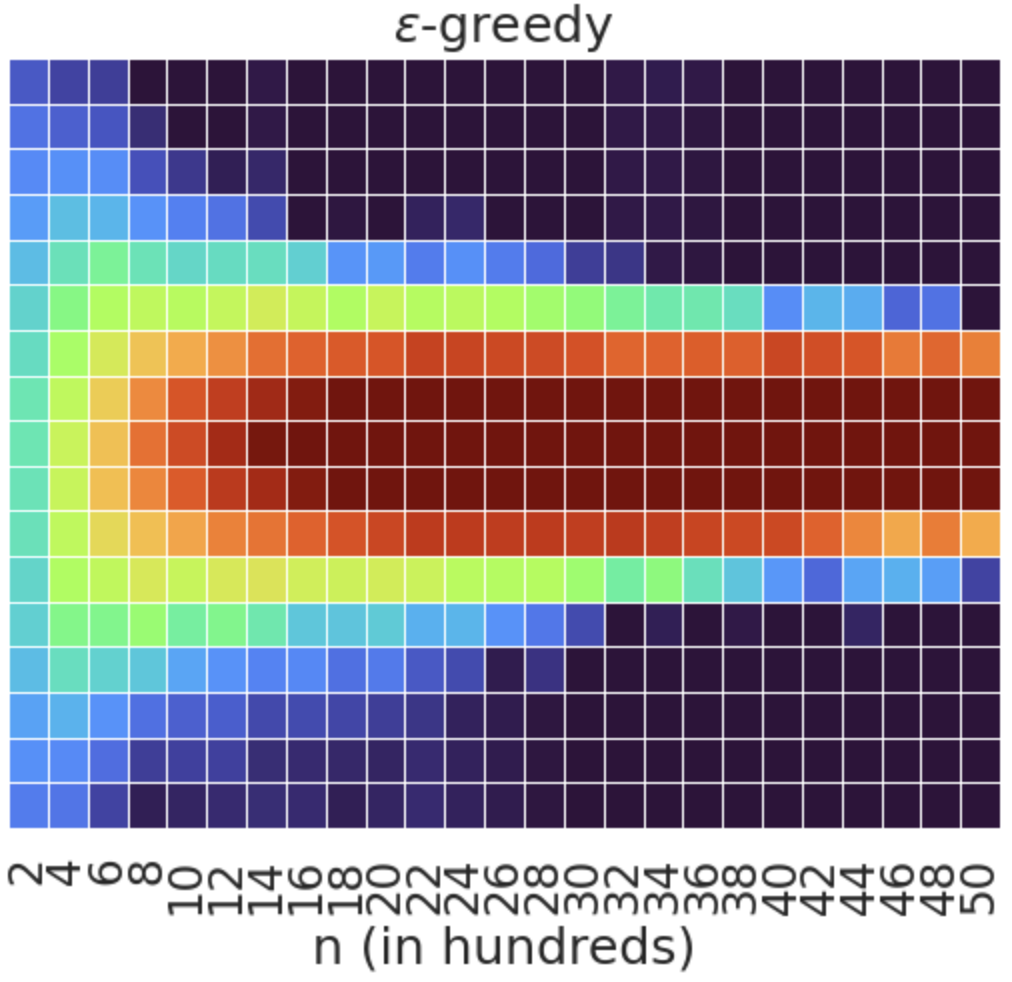}
    ~~
    \includegraphics[width=0.322\linewidth]{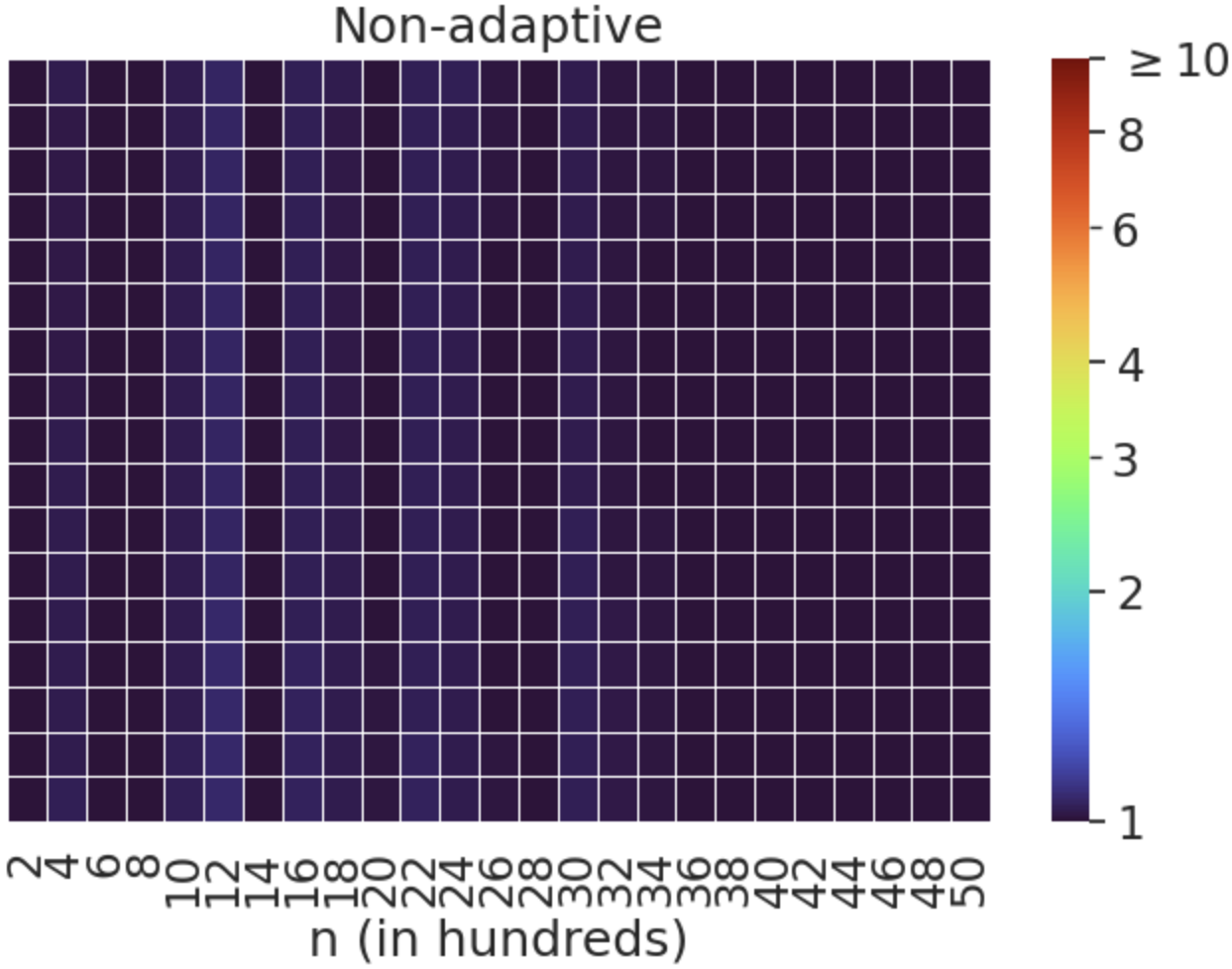}
    \vspace{-2mm}
    \caption{\new{\bo{Misspecified nonstationary setting (Section \ref{sec:nonstationary}): Sample variance underestimates the true variance under common bandit algorithms.} We plot the ratio of the estimated variance and the true empirical variance of $\hat{\theta}^{(n)}$, specifically $\frac{\widehat\Var(\hat{\theta}^{(n)})}{\hat{s}^2/n}$ under different algorithms across $1000$ replications for different values of $\Delta_1$ and $n$. Our heatmap uses a log-scale and rounds down all values greater than $10$. %Under common bandit algorithms, $\hat{s}^2/n$ often underestimates the true variance by a factor of over 10! 
    \vspace{-5mm}}}
    \label{fig:reward-heatmap}
\end{figure*}

\new{
\subsubsection{Sample variance underestimates the true variance under common bandit algorithms.} %%%%%%%%%%%%%%%%%%%%%%
\label{sec:heatmap-nonstationarity}

Building on the experimental setup from Section \ref{sec:nonstationaryDGP}, we now show the issues that
%that poor replicability 
can occur regarding variance estimation, even when $\Delta_1$ does not exactly equal $0$. We use the same algorithms (Thompson sampling and $\epsilon$-greedy), and keep most hyperparameters the same ($\mu_0 = 0$, $\Delta_2 = -0.25$, $T=2$). We only vary $\Delta_1$ and the number of individuals $n$ in each trial. 

In each simulated trial we compute the mean estimator $\hat{\theta}^{(n)} = \frac{1}{n} \sum_{i=1}^n \frac{1}{T} \sum_{t=1}^T R(Y_{i,t})$. We also compute a variance estimator for $\hat{\theta}^{(n)}$ using the sample variance:
\begin{align*}
    \hat{s}^2 = \frac{1}{n} \sum_{i=1}^n \bigg( \frac{1}{T} \sum_{t=1}^T R(Y_{i,t}) - \hat{\theta}^{(n)} \bigg)^2.
\end{align*}
Typically, we expect that $
\frac{\Var(\hat{\theta}^{(n)})}{\hat{s}^2 / n} \approx 1$.
Note that across all our simulations ($1000$ replications for each value of $\Delta_1$ and $n$), the variance estimate, $\hat{s}^2$, took values in the range $[0.45, 0.55]$. In other words, $\hat{s}^2$ was very stable across a variety of different environments with different values of $\Delta_1$ and $n$; we used values $\Delta_1 \in [-0.2, 0.2]$ and $n \in [200, 5000]$.

We compare $\hat{s}^2/n$ to the true variance $\Var(\hat{\theta}^{(n)})$. We form our approximate estimate $\widehat\Var(\hat{\theta}^{(n)})$ by taking the sample variance of $\hat{\theta}^{(n)}$ \textit{across $1000$ trial replications}; we call this estimate $\widehat\Var(\hat{\theta}^{(n)})$. Note that while $\hat{s}^2/n$ is an approximate variance estimate based on just a \textit{single} trial, the estimate $\widehat\Var(\hat{\theta}^{(n)})$ is formed across many trial replications.

In Figure \ref{fig:reward-heatmap}, below we plot $\frac{\widehat\Var(\hat{\theta}^{(n)})}{\hat{s}^2/n}$ under Thompson sampling, $\epsilon$-greedy ($\epsilon=0.05$), and a non-adaptive algorithm that selects actions with a constant $0.5$ probability. Under the adaptive algorithms, $\frac{\widehat\Var(\hat{\theta}^{(n)})}{\hat{s}^2/n}$ is about $1$ when the value of $n$ is large relative to values of $|\Delta_1 |\not=0$. 
However, when $|\Delta_1|$ is small relative to the number of users $n$, the variance $\widehat{\Var}(\hat{\theta}^{(n)})$ is sometimes more than ten times greater than $\hat{s}^2/n$! 
%However, when $|\Delta_1 |$ is small relative to the sample size, $n$, the variance of $\hat{\theta}^{(n)}$ is $\geq 1$, over double. 
Notably, this phenomenon does not occur when using a non-adaptive data collection algorithm (selects actions with constant probability $0.5$). 

In summary, even when $\Delta_1$ is not exactly zero, the standard variance estimate $\hat{s}^2/n$ can \textit{severely} underestimate the true variance of $\hat{\theta}^{(n)}$ across trial replications. We expect variance estimation issue occur in this example whenever the signal-to-noise ratio $\frac{|\Delta_1|}{\sigma}$ is small relative to the sample size $n$. Note that small signal-to-noise ratios are small in practice because effect sizes and sample sizes are often small (e.g., $< 100$). 
}

\subsection{Misspecified linear contextual bandit} %%%%%%%%%%%%%%%%%%%%%%%%%%%%%%%%%%%%
\label{sec:linearExample}
\new{In the previous section, we discussed how non-replicability can occur in a nonstationary multi-armed bandit setting.} We now consider a \textit{stationary contextual} bandit setting in which the reward model used by the bandit algorithm is misspecified. This reflects the common practice in which simple RL/bandit algorithms are favored, to better trade-off bias and variance in these noisy settings (as discussed in Section \ref{sec:characteristics}). % we do not expect to know the true reward model. 
%In addition, 
We consider a post-trial inferential target which is a %often a
marginal loss minimizer; note in digital health, it is common to consider % such as a  
inferring causal effects \cite{boruvka2018assessing,nahum2024optimizinh,qian2019estimating} that are marginal over (a subset of) the features used in the bandit algorithm. 

For clarity, here we consider a toy version in which both the bandit algorithm and the post-trial inference use a model for the mean of the reward, conditional on features, that is not correctly specified for the underlying true conditional mean of the reward.
Specifically, in the post-trial analysis, we are interested in inference for $\thetastar$, which is the minimizer of a least squares criterion, i.e, $\thetastar$ solves:
\begin{align}
    \label{eqn:leastSquares}
    0 = \lim_{n \to \infty} \E^{(n)} \left[ \dot{\ell}_\theta(\HH_{i,T}) \right] \bigg|_{\theta = \thetastar},
\end{align}
where $\dot{\ell}_\theta$ is the derivative of $\ell_\theta$ with respect to $\theta = [\theta_0, \theta_1]$ where
\begin{multline}
    \ell_\theta(\HH_{i,T}) \\
    = \sum_{t=1}^T \big\{ R(Y_{i,t}) -\new{\phi_{\TN{inf}}(X_{i,t})^\top} \theta_0 - A_{i,t} \new{\phi_{\TN{inf}}(X_{i,t})^\top} \theta_1 \big\}^2.
    \label{eqn:ellDef}
\end{multline}
It is assumed for now that the limit in display \eqref{eqn:leastSquares} is well-defined. %; we prove this for the setting we consider below (see Proposition \ref{prop:nonrepMiss}). %\sam{note prior sentence.  }
Above, $\new{\phi_{\TN{inf}}(X_{i,t})} = [1, X_{i,t}]$ is a featurization of the covariates $X_{i,t}$ we use for inference. In the true data generating distribution, the mean reward will be a linear function of a different featurization of the state $\new{\phi_{\TN{gen}}(X_{i,t})}$; see display \eqref{eqn:phi0}.

The post-trial inference is based on the least squares estimator:
\begin{align}
    \label{eqn:leastSquaresMiss}
    \hat{\theta}^{(n)} = \argmin_{\theta_0, \theta_1 \in \real^2} ~ \frac{1}{n} \sum_{i=1}^n \ell_\theta(\HH_{i,T}).
\end{align}

\subsubsection{Data generating environment.} %%%%%%%%%%%%%%%%%%%%%%%%%%%
\label{sec:misspecifiedDGP}
Consider the following simple environment. Actions are binary, so  $\MC{A} = \{ 0, 1 \}$. The time-varying covariate features \\
$X_{i,t} \sim \TN{Uniform}[0,1]$ are i.i.d. over $i$ and $t$. The expected reward is linear in a featurization of the state $\new{\phi_{\TN{gen}}(X_{i,t})} = [1, X_{i,t}, X_{i,t}^2]$: % and the reward is linear in this featurization:
\begin{align}
    R(Y_{i,t}) = \new{\phi_{\TN{gen}}(X_{i,t})^\top} \alpha_0  + A_{i,t} \new{\phi_{\TN{gen}}(X_{i,t})^\top} \alpha_1 + \epsilon_{i,t},
    \label{eqn:phi0}
\end{align}
where noise $\epsilon_{i,t} \sim \N(0, 1)$ are i.i.d. over $i$ and $t$. Specifically, $\alpha_0 = [0.1, 0.1, 0]$ and $\alpha_1 = [ \frac{1}{3}, -2, 2]$. 
With these values of $\alpha_0,\alpha_1$ along with featurization vectors $\new{\phi_{\TN{gen}}}$ from \eqref{eqn:phi0}, the linear model used by the post-trial inference model, which uses $\new{\phi_{\TN{inf}}}$ from \eqref{eqn:ellDef}, is misspecified.

\subsubsection{Non-replicability under (contextual) 
\texorpdfstring{$\epsilon$}{epsilon}-greedy.} 
%%%%%%%%%%%%%%%%%%%%%%%%%%%
\label{sec:non_repl_e_greedy}
Suppose an $\epsilon$-greedy (or greedy) based contextual bandit algorithm is used to select treatments during the trial.
%; note when $\epsilon$ equals zero it is equivalent to an explore-then-commit algorithm. 
At the first decision time $(t=1)$, the algorithm randomizes with $\PP \big( A_{i,1} = 1 \mid X_{i,1} = x \big) = 0.5$ for all $x$. At the second decision time, the algorithm forms action selection probabilities as follows:
\begin{multline}
    \label{eqn:epsilonGreedy}
    \PP \big( A_{i,2} = 1 \mid \HH_{1:n,1}, X_{i,2} \big) \\
    = \begin{cases}
        1-\epsilon & \TN{if}~ \phi_{\TN{alg}}(X_{i,2})^\top \hat{\beta}_{1,1}^{(n)} > 0 \\
        \epsilon & \TN{otherwise}
    \end{cases}
\end{multline}
where $\hat\beta_{1,1}^{(n)}$ is the least squares estimator using the featurization function $\new{\phi_{\TN{alg}}(X_{i,t}) = \phi_{\TN{inf}}(X_{i,t})} = [1, X_{i,t}]$:
\begin{align*}
    \big[ \hat{\beta}_{1,0}^{(n)}, \hat{\beta}_{1,1}^{(n)} \big] = 
    \argmin_{\beta_0, \beta_1} ~ \frac{1}{n} \sum_{i=1}^n \tilde{\ell}_{(\beta_0, \beta_1)}(\HH_{i,1}),
\end{align*}
where $\tilde{\ell}_{(\beta_0, \beta_1)}(\HH_{i,1}) = \big\{ R(Y_{i,1}) - \new{\phi_{\TN{alg}}(X_{i,1})^\top} \beta_0 \\
- A_{i,1} \new{\phi_{\TN{alg}}(X_{i,1})^\top} \beta_1 \big\}^2$.  Recall that the above  model is misspecified since $\new{\phi_{\TN{alg}} \neq \phi_{\TN{gen}}}$. 

\begin{figure}[h]
    \centering
    \includegraphics[width=0.49\linewidth]{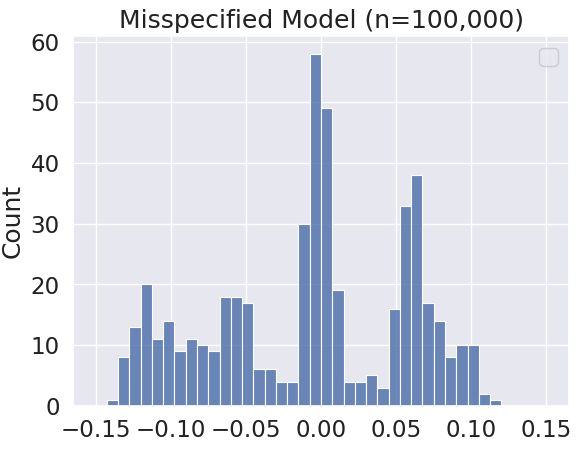}
    \includegraphics[width=0.49\linewidth]{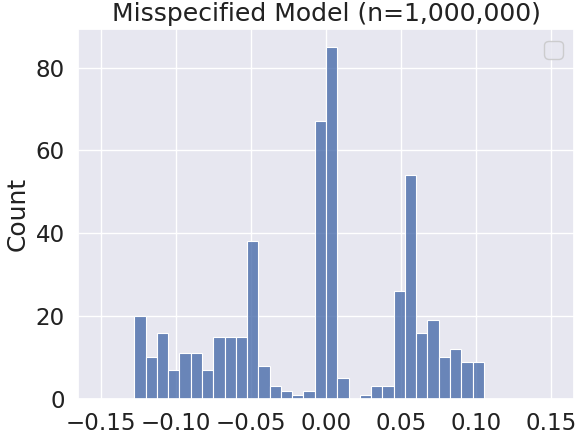}
    \caption{\bo{Inferential Non-replicability under Misspecified Linear Model.} Above we plot the empirical distribution of the first dimension of $\hat{\theta}_1^{(n)}$ for $n=100,000$ and $n=1,000,000$. Note that by increasing the sample size $10$-fold, the variance of $\hat{\theta}_1^{(n)}$ does not change much. Note that $n=1,000,000$ is an extremely large sample size given the noise variance in the environment is $1$.
    }\label{fig:twoFeaturizations}
\end{figure}

As seen in Figure \ref{fig:twoFeaturizations}, in this simple bandit setting, the non-replicability of the least squares estimator is severe. In this figure, we can see that the value of $\hat{\theta}_{1,1}^{(n)}$ (the first dimension of $\hat{\theta}_{1}^{(n)}$) varies significantly even with $100,000$ individuals in each replication trial. Furthermore, the non-replicability does not improve substantially by increasing the sample size $10$-fold to $1,000,000$ individuals.
Furthermore, we can prove that in this simple case even as $n \to \infty$, a constant fraction of replication trials will have $\hat{\theta}_{1,1}^{(n)} \geq \delta > 0$ and another constant fraction will have $\hat{\theta}_{1,1}^{(n)} \leq - \delta < 0$ for some $\delta > 0$ (see Proposition \ref{prop:nonrepMiss}).

\begin{prop}[Non-replicability under misspecified linear bandit]
    \label{prop:nonrepMiss}
    Under the contextual $\epsilon$-greedy algorithm (for any $\epsilon < 0.5$) in the environment described above, %$\thetastar$ is well defined. Moreover, 
    $\hat{\theta}^{(n)}$ defined in \eqref{eqn:leastSquaresMiss} (with the misspecified featurization $\phi_{\TN{gen}}$) does not converge in probability to a constant %$\thetastar$ 
    as $n \to \infty$. 
    
    Additionally, there exists a $\delta>0$ for which \\
    $\lim_{n\to\infty} \PP^{(n)} \big( \hat{\theta}_{1,1}^{(n)} > \delta\big)>0$ and \\
    $\lim_{n\to\infty} \PP^{(n)}\big(\hat{\theta}_{1,1}^{(n)} < \delta\big)>0$. We characterize the limiting distribution of $\hat{\theta}^{(n)}$ in Appendix \print{\ref{app:missLinear}}{A.2}.
\end{prop}

\new{
\begin{remark}[What if there is no misspecification?]
    In the case that $\phi_{\TN{gen}} = \phi_{\TN{inf}}$, there is no model misspecification. In this setting, there are no such replicability issues, and existing approaches for inference on adaptively collected data can be used for inference \citep{zhang2021mestimator}. Our non-replicability result shows that the results from \citet{zhang2021mestimator} can be rather fragile and sensitive to model misspecification.
\end{remark}
}

\begin{figure}[t]
    \centering
    \vspace{-3mm}
    \includegraphics[width=\linewidth]{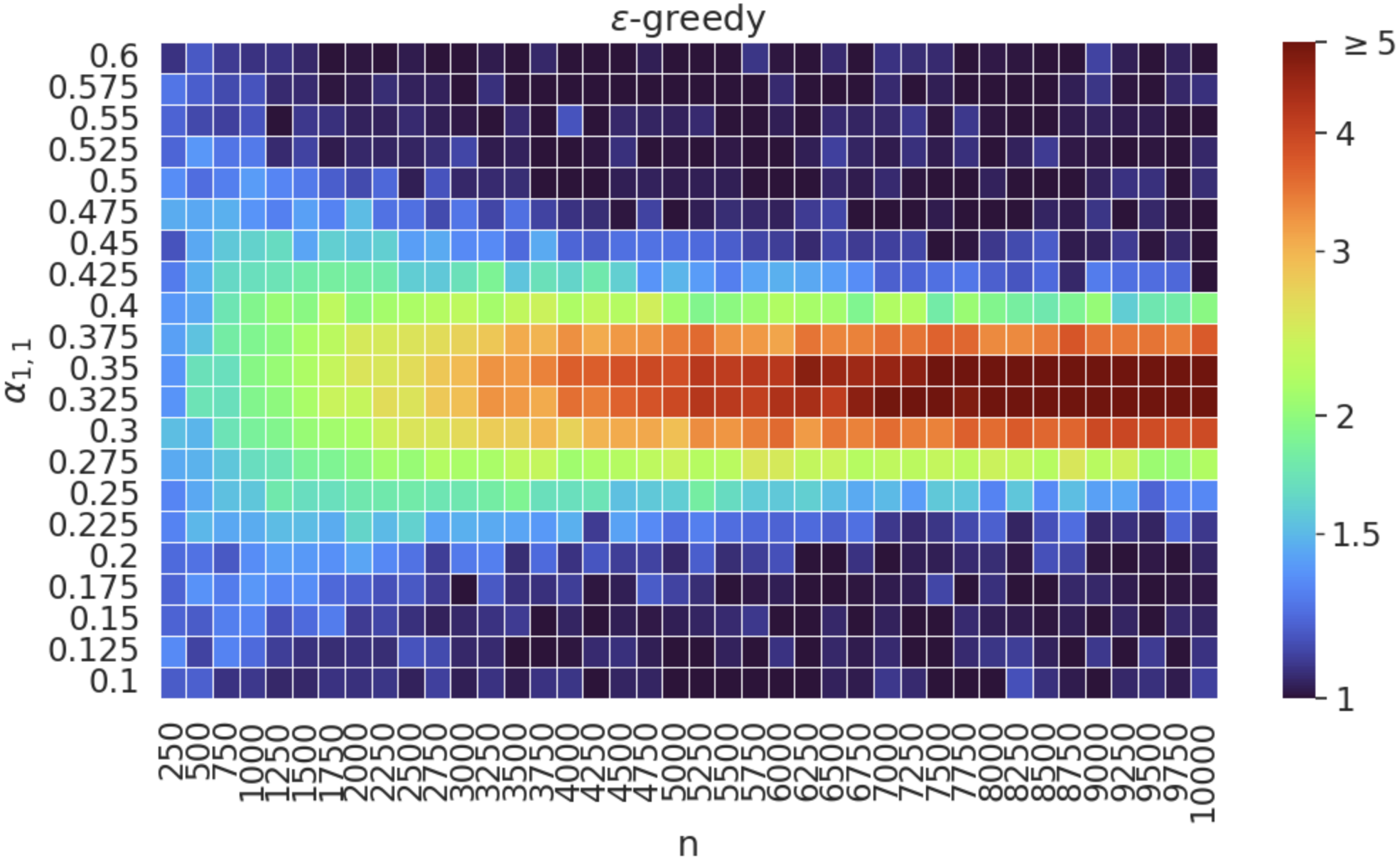}
    \vspace{-6mm}
    \caption{\new{\bo{Misspecified linear setting (Section \ref{sec:linearExample}): Robust standard errors underestimates the true variance under common bandit algorithms.} We plot the ratio of the estimated and the true empirical variance of $\hat\theta_{1,1}^{(n)}$, specifically $\frac{\Var(\hat{\theta}_{1,1}^{(n)})}{\hat{s}^2/n}$ under $\epsilon$-greedy across $1000$ replications for different values of $\alpha_{1,1}$ and $n$. Our heatmap uses a log-scale and rounds down all values greater than $5$. Note that $\hat{s}^2/n$ often severely underestimates $\Var(\hat{\theta}_{1,1}^{(n)})$, especially for small values of $|\alpha_{1,1}-1/3|$ and $n$. }}
    \label{fig:misspecified-heatmap}
\end{figure}

\new{\subsubsection{Standard robust standard errors underestimates the true variance under common bandit algorithms.} %%%%%%%%%%%%%%%%%%%%%%
\label{sec:heatmap-misspecified}
Building on the experimental setup from Section \ref{sec:misspecifiedDGP}, we now show that the issues that can occur regarding variance estimation in practice, even when $\alpha_{1,1}$ does not exactly equal $1/3$. We use the same contextual $\epsilon$-greedy algorithm, and keep most hyperparameters the same ($\alpha_0 = [0.1,0.1,0]$, $\alpha_1 = [\alpha_{1,1}, -2, 2]$, $T=2$). We only vary $\alpha_{1,1}$ and the number of individuals $n$ in each trial. 

In each simulated trial we compute $\hat{\theta}_{1,1}^{(n)}$, the first dimension of the least squares estimator $\hat\theta_1^{(n)}$. 
Also, we compute a standard variance estimator for $\hat{\theta}_{1,1}^{(n)}$, $\hat{s}^2$, using the sandwich variance estimator \citep[Chpt 5.3]{van2000asymptotic} (i.e., robust standard errors \citep{white1980heteroskedasticity}). Typically we expect that $\frac{\Var(\hat{\theta}_{1,1})}{\hat{s}^2/n} \approx 1$.
Note that across all our simulations ($1000$ replications for each value of $\alpha_{1,1}$ and $n$), the variance estimate $\hat{s}^2$ took values in the range $[9, 11]$. We used values $\alpha_{1,1} \in [0.1, 0.6]$ and $n \in [200, 10000]$. 
We compare $\hat{s}^2/n$ to the true variance $\Var(\hat{\theta}_{1,1}^{(n)})$. We form our approximate estimate $\widehat\Var(\hat{\theta}_{1,1}^{(n)})$ by taking the sample variance of $\hat{\theta}_{1,1}^{(n)}$ across $1000$ trial replications.

In Figure \ref{fig:misspecified-heatmap} we plot $\frac{\widehat\Var(\hat{\theta}_{1,1}^{(n)})}{\hat{s}^2/n}$ under contextual $\epsilon$-greedy ($\epsilon=0.05$). Notice that $\frac{\widehat\Var(\hat{\theta}_{1,1}^{(n)})}{\hat{s}^2/n}$ is about $1$ when the value of $n$ is large relative to values of $|\alpha_{1,1} - 1/3 |\not=0$. 
However, when $|\alpha_{1,1} - 1/3 |$ is small relative to the number of users $n$, the variance $\widehat{\Var}(\hat{\theta}_{1,1}^{(n)})$ is more than 5 times greater than $\hat{s}^2/n$!
In summary, even when $\alpha_{1,1}$ is not exactly $1/3$, the standard variance estimate $\hat{s}^2$ can \textit{severely} underestimate of the true variance of $\hat{\theta}_{1,1}^{(n)}$ across trial replications.
}

%%%%%%%%%%%%%%%%%%%%%%%%%%%%%%%%%%%%%%%%%%%%%%%%%%%%%%%
%%%%%%%%%%%%%%%%%%%%%%%%%%%%%%%%%%%%%%%%%%%%%%%%%%%%%%%
\section{Replicable Bandits}
\label{sec:replicableBandits}
\new{The non-replicable statistical analyses examples we presented in the previous section are fundamentally due to the non-replicability of the bandit algorithms used to make treatment decisions} (see the discussion below Proposition 
\ref{prop:nonrepNonstat} for additional intuition behind this). In this section, we show that if the bandit algorithm itself satisfies certain ``replicability'' properties, then a variety of statistical analyses after the experiment will be replicable \new{(we discuss the types of statistical analyses we are interested in detail later in Section \ref{sec:statisticalAnalyses})}. 
In other words, if the bandit algorithm is replicable, then the non-replicable statistical analyses phenomena from the previous section will no longer occur. These observations lead to a variety of open questions regarding how to design bandit algorithms with replicability in mind, which we list in Section \ref{sec:discussion}.

\subsection{Definition} %%%%%%%%%%%%%%%%%%%%%%%%%%%%%%%%%%%%
\label{sec:defReplicableBandit}

Informally, we say that a bandit algorithm is replicable if across repetitions of the trial, the learned policies become close in %expectation
with high probability
as the sample size, $n$, increases. 
Below, for any real-valued function of $x \in \MC{X}, a \in \MC{A}$ we use the notation $\| f \|_\infty \triangleq \sup_{x \in \MC{X}, a \in \MC{A}} | f(x, a) |$. In the definition below, we allow policies to depend on the entire collective history $\HH_{1:n,t-1}$. Note that this is a strict generalization of allowing the bandit algorithm to depend on particular statistics $\betahat{t-1}$ formed using $\HH_{1:n,t-1}$. Thus $\pi \big(x,a; \betahat{t-1} \big)$ is replaced by $\pi_t(x,a; \HH_{1:n,t-1})$ in the definition.

\begin{mydef}[Replicable bandit algorithm] 
    \label{def:replicableBandit}
    Let $\HH_{1:n,T}$ and $\tilde{\HH}_{1:n,T}$ be histories associated with independent replications of a trial. A bandit algorithm is replicable if %it uses a policy function class $\{ \pi(\cdotspace; \beta) \}_{\beta \in \real^{d_\beta}}$ %$\{ \pi_t \}_{t=1}^T$ 
    %such that each 
    for each $t \in [1 \colon T-1]$  and each $\epsilon > 0$, as $n \to \infty$
    \begin{multline}
        \label{eqn:replicableDef}
        %\lim_{n \to \infty} 
        \hspace{-5mm}\sup_{P \in \MC{P}} \PP_P^{(n)} \left( \big\| \pi_t \big(\cdotspace; \HH_{1:n,t-1} \big) - \pi_t \big(\cdotspace; \tilde{\HH}_{1:n,t-1} \, \big) \big\|_\infty > \epsilon \right) \\
        \to 0,
    \end{multline}
    Furthermore, we say that a bandit algorithm is replicable at a non-random, monotonically increasing rate $\{ r^{(n)} \}_{n \geq 1}$ if for each $t \in [1 \colon T-1]$ and each $\epsilon > 0$, as $n \to \infty$
    \begin{multline}
        \label{eqn:replicableDefRate}
        \hspace{-5mm}\sup_{P \in \MC{P}} \PP_P^{(n)} \left( r^{(n)} \big\| \pi_t \big(\cdotspace; \HH_{1:n,t-1} \big) 
        - \pi_t \big(\cdotspace; \tilde{\HH}_{1:n,t-1} \, \big) \big\|_\infty > \epsilon \right) \\
        \to 0.
    \end{multline}
\end{mydef}
\new{Above, $\PP_P^{(n)}$ averages over the randomness in forming the histories across replications $\HH_{1:n,t-1}$ and $\tilde{\HH}_{1:n,t-1}$; this randomness includes the draw of potential outcomes from $P \in \MC{P}$ as described in \eqref{eqn:OdistP} and any randomness in how the algorithm forms policies and selects actions. These definitions of replicable bandit algorithms require that the policies learned by the algorithm must be ``close'' with high probability across replications of the trial. Additionally, as the number of people in each trial $n$ grows, the policies learned across replications must get closer. 

The supremum over the class of outcome distributions $P \in \MC{P}$ ensures that replicability, as defined here, 
%In our definition, we take the supremum over the class of outcome distributions $P \in \MC{P}$.  The use of the supremum in the definition means that,   replicability as defined here,  
holds \textit{uniformly} across a range of possible data generating distributions, including the worst cases.  Further this definition of replicability assists in proving that convergence of the estimator is uniform in the underlying distribution (see Chapter 2.8 in \citet{van1996weak} for definition).
It is well-known in classical statistics that the use of asymptotic approximations with non-regular estimators whose convergence is not uniform in the underlying distribution can result in very poor statistical inference approximations \citep{romano2012uniform,laber2011adaptive,chakraborty2010inference,moodie2010estimating,luedtke2016statistical,hirano2012impossibility}.

In our earlier simulations from Section \ref{sec:nonstationary}, we found that the distribution of the mean reward estimator under Thompson sampling and $\epsilon$-greedy varied greatly with 
$\Delta_1$ the margin in the first timestep. % was close to zero (where ``close'' depends on the sample size $n$). 
This demonstrates non-uniform convergence since whether the mean reward estimator converges depends on the particular data generating environment $P$. %Intuitively, ensuring convergence, even for the worst case data generating distribution, ensures that the replicability of the policy does not depend on the particular environment or data generating distribution $P$ in which we are deploying the algorithm. 
Ensuring the algorithm is replicable uniformly over a range of data-generating distributions helps us ensure that the post-trial estimator $\hat\theta^{(n)}$ for the mean reward also converges uniformly in the underlying distribution.}

\subsection{Relationship to existing definitions of replicable decision-making.} %%%%%%%%%%%%%%%%%%%%%%%%%%%%%%%%%%
\label{sec:relatedReplicable}

In the literature on replicable bandits and reinforcement learning, there are other definitions of replicability. \citet{eaton2024replicable} and \citet{karbasi2024replicability} define a reinforcement learning algorithm as replicable if it learns the \textit{same} policy with high probability across repetitions. Formally, this means for some $\rho_n$ that converges to $1$ at some rate, %  \sam{does $\rho_n$ have to converge to $0$ as $n$ increases? }
\begin{align}
    \label{eqn:altDef}
    \PP \left( \big\| \pi_t \big( \cdotspace ; \HH_{1:n,t-1}, \xi \big) - \pi_t \big( \cdotspace ; \tilde{\HH}_{1:n,t-1}, \xi  \big) \|_\infty = 0 \right) 
    \geq \rho_n.
\end{align}
Above the policy also depends on noise $\xi$, which is independent of $\HH_{1:n,t-1}$, which allows for randomness in how the policy is formed.
%This random noise allows for randomness in how the policy is formed; 
Note that they explicitly assume that the random noise $\xi$ is the same across replicated trials.
Our definition differs in that we allow for policies formed to be close but not \textit{exactly} equal with high probability. %; note in practice many algorithms will not satisfy such \eqref{eqn:altDef} due to small changes in the learned policies between replicated trials. \sam{not sure I understand prior sentence!}
Additionally, the above formulation does not consider replicability that is uniform over the data-generating distribution $P \in \MC{P}$, i.e., for worst-case data-generating distributions.

A different work on replicable bandits aims to ensure that the \textit{entire sequence of actions} taken by the bandit algorithm across replications is the same with high probability \citep{esfandiari2022replicable}. However, their formulation holds constant across replications, any randomness in how the algorithm forms policies and selects actions. % are formed, constant across replications. %o the randomness is only due to differences in the draw of potential outcomes. 
This essentially means that their definition of replicability only encompasses the case in which replications are forced to use the same initial random seed. 

Finally, \citet{parmigiani2023defining} defines a notion of replicability of a decision-making rule with respect to a particular utility function, e.g., total expected reward. Their formulation of replicability allows for variation in the decision-making rule as long as it maximizes the utility function of interest. In contrast, we require a much stronger notion of replicability of the decision-making algorithm; our definition of replicability will allow replicability of estimators for a variety of quantities, rather than a single pre-specified utility function. \new{We discuss the types of post-trial statistical analyses we are interested in later in Section \ref{sec:statisticalAnalyses}.}

%%%%%%%%%%%%%%%%%%%%%%%%%%%%%%%%%%%%%%%%%%%%%%
%%%%%%%%%%%%%%%%%%%%%%%%%%%%%%%%%%%%%%%%%%%%%%
\new{\subsection{Example bandit algorithms that do not satisfy replicability}
As we saw in the non-replicability examples in Section \ref{sec:nonrepExamples}, $\epsilon$-greedy and Thompson sampling \citep{russo2018tutorial,kalvit2021closer} fail to satisfy Definition \ref{def:replicableBandit}.
%\eqref{eqn:replicableDef}. 
Many common bandit algorithms fail to satisfy the replicable algorithm condition when there is no unique optimal action (or the difference between the best action and the second best is close to zero or decreasing). 
This non-replicability is related to taking discontinuous functions (e.g., argmax) of the data to form the policy \citep{zhang2020inference,luedtke2016statistical}. This problem also occurs if the function used to form the policy is smooth but approaches a discontinuous function in the limit, e.g., a softmax function with a temperature that goes to zero as $n$ grows.
%The non-replicability is related to have a non-smooth limiting allocation function, i.e., the randomization probability in the limit is a non-smooth function of the underlying reward distribution (determined by $P$) \citep{zhang2020inference,luedtke2016statistical}. 
Note that discontinuous/non-smooth functions of data are problematic for asymptotic approximations in many areas of statistics \citep{laber2011adaptive,laber2014dynamic,leeb2005model,romano2012uniform}.

While the non-replicability of the bandit algorithm does not lead to inferential non-replicability when the post-trial inferential model is correctly specified \citep{zhang2021mestimator,zhang2022statistical}, once there is misspecification in the inferential model and/or bandit algorithm, non-replicability issues can become a severe problem.

%%%%%%%%%%%%%%%%%%%%%%%%%%%%%%%%%%%%%%%%%%%%%%
%%%%%%%%%%%%%%%%%%%%%%%%%%%%%%%%%%%%%%%%%%%%%%
\subsection{Example bandit algorithms that generally satisfy replicability}
\label{sec:satisfyAlgs}

In this section, we discuss several examples of bandit algorithms that will generally satisfy replicability from Definition \ref{def:replicableBandit}.

%%%%%%%%%%%%%%%%%%%%%%%%%%%%%%%%%%%%%%%
\subsubsection{Boltzmann sampling.} 
\label{sec:boltzmann}
Boltzmann sampling (or softmax exploration) \citep[Chpt 2.8]{sutton2018reinforcement} is a type of bandit algorithm that effectively is a ``softmax'' version of the $\epsilon$-greedy. Boltzmann sampling with (with a temperature parameter that does not go to zero as the sample size increases) satisfies the bandit replicability definition, and this type of algorithm has been used in several digital health trials \citep{yom2017encouraging,lauffenburger2021reinforcement}. Below, we describe a simple instantiation of Boltzmann sampling.

At each decision time, the algorithm forms $\betahat{t} = [\betahat{t,0}, \betahat{t,1}]$, which is the minimizer of a least squares criterion with a ridge regularization:
\begin{multline}
    \label{eqn:boltzmannRegression}
    \argmin_\beta ~ \lambda \| \beta \|_2^2 \\
    + \sum_{i=1}^n \sum_{t'=1}^t \left\{ R(Y_{i,t'}) - \big[ \phi_{\TN{alg}}(X_{i,t'}), A_{i,t'} \phi_{\TN{alg}}(X_{i,t}) \big]^\top \beta \right\}^2.
\end{multline}
$\epsilon$-greedy uses $\betahat{t}$ to form action selection probabilities using a thresholding function (see the formula from \eqref{eqn:epsilonGreedy}). In contrast, Boltzmann sampling forms treatment probabilities using a softmax function:
\begin{align}
    \label{eqn:boltzmannProb1}
    \PP \big( A_{i,t} = a \mid \betahat{t-1}, X_{i,t} \big) 
    = \frac{1}{1 + \exp(- s \cdot \phi_{\TN{alg}}(X_{i,t})^\top \betahat{t-1,1} )}.
\end{align}
Above the hyperparameter $s > 0$ changes the steepness or ``temperature'' of the softmax. We discuss considerations for choosing this hyperparameter later in Section \ref{sec:choosingAllocation}.

%%%%%%%%%%%%%%%%%%%%%%%%%%%%%%%%%%%%%%%
\subsubsection{Smoothed Thompson sampling.}
\label{sec:smoothed_thompson_sampling}
Several recent digital health intervention trials (including Oralytics) \citep{nahum2024optimizinh,ghosh2024rebandit,trella2024oralytics} use a modified version of Thompson sampling that ensures replicability of the algorithm in the sense of Definition \ref{def:replicableBandit}. We first discuss a simple instantiation of a standard Thompson sampling algorithm \citep{russo2018tutorial,thompson1933likelihood} and then describe the modified version.

A classical linear Thompson sampling algorithm posits a Bayesian model where a latent parameter $\alpha = [\alpha_1, \alpha_2]$ is sampled from a known, multivariate Gaussian prior distribution $\alpha \sim N(\mu, \Sigma)$. Then rewards are modeled as Gaussian conditional on $\alpha$ and the covariate $X_{i,t}$:
\begin{multline}
    \label{smooth_thompson_sampling_blr_reward_model}
    R(Y_{i,t}) \mid X_{i,t}, \alpha \\
    \sim N \big( \phi_{\TN{alg}}(X_{i,t})^\top \alpha_0 + A_{i,t} \phi_{\TN{alg}}(X_{i,t})^\top \alpha_1, \sigma^2 \big).
\end{multline}
Standard Thompson sampling selects action $1$ according to the posterior probability that action $1$ is optimal (i.e., the posterior probability that the treatment effect $X_{i,t}^\top \alpha_1$ is positive):
\begin{align*}
    &\PP \big( A_{i,t} = 1 \mid \betahat{t-1}, X_{i,t} \big) \\
    &= \PP \big( \phi_{\TN{alg}}(X_{i,t})^\top \alpha_1 > 0 \mid \HH_{1:n,t-1}, X_{i,t} \big) \nonumber \\
    &= \E \big[ \bs{1}_{ \phi_{\TN{alg}}(X_{i,t})^\top \alpha_1 > 0 } \mid \HH_{1:n,t-1}, X_{i,t} \big]. \nonumber 
\end{align*}
Above, the probability on the right-hand side averages over the randomness in the draw of $\alpha_1$ according to the posterior distribution given $\HH_{1:n,t-1}$. The second equality above holds because probabilities can be written as expectations of indicators.

In the smoothed version of Thompson sampling, we replace the indicator function with a smoothed function $\rho$:
\begin{multline}
    \label{sec:smoothTS}
    \PP \big( A_{i,t} = 1 \mid \betahat{t-1}, X_{i,t} \big) \\
    = \E \big[ \rho \big( \phi_{\TN{alg}}(X_{i,t})^\top \alpha_1 \big)  \mid \HH_{1:n,t-1}, X_{i,t} \big].
\end{multline}
For example, we can choose $\rho$ to be a softmax function $\rho(x) = \big( 1 + e^{-s \cdot x} \big)^{-1}$, where $s$ controls the temperature of the softmax.

%%%%%%%%%%%%%%%%%%%%%%%%%%%%%%%%%%%%%%%
\subsubsection{Covariate-Adjusted Response Adjusted randomization (CARA).} 
Many of the covariate-adjusted adaptive randomization algorithms from the adaptive clinical trial literature also satisfy our definition of replicability \citep{rosenberger2001covariate,hu2006theory,zhang2007asymptotic,atkinson2004adaptive}. Interestingly, many of the CARA algorithms described in the literature are essentially generalizations of the Boltzmann sampling from Section \ref{sec:boltzmann}, e.g., see Chapter 9 of \citet{rosenberger2001covariate}. Variants include fitting a logistic regression model instead of a linear regression model for binary rewards \citep{zhang2007asymptotic} and replacing the softmax function with other smooth functions like a Gaussian cumulative density function \citep{atkinson2004adaptive}.

\subsubsection{Choosing the allocation function.}
\label{sec:choosingAllocation}
The replicable algorithms we describe above require choosing an ``allocation function'', which maps a given estimated treatment effect to an action selection probability. For example, in Boltzmann sampling from \eqref{eqn:boltzmannProb1}, the allocation function is a softmax function with a temperature parameter $s$ that we need to choose. In smoothed Thompson sampling from \eqref{sec:smoothTS}, the allocation function is $\rho$.
In choosing the allocation function, there are multiple (competing) objectives, specifically, reward maximization and power maximization. See Chapter 2.2 of \citet{hu2006theory} and \citet{simchi2023multi} for further discussion of the tradeoff between reward maximization and power maximization.

If the environment is a bandit environment,\footnote{\new{In standard stochastic contextual bandit environments, the distribution of future contexts $X_{i,t+1}$ is independent of previous actions $A_{i,1:t}$ given previous contexts $X_{i,1:t}$; that is $X_{i,t+1} \indep A_{i,1:t} \mid X_{i,1:t}$. In contrast, in general Markov decision processes, the next context/state may depend on the past action.}} then in the limit, rewards are maximized if the allocation function approaches an indicator function, e.g., for smoothed Thompson sampling, this means $\rho$ will approach $\rho(x) = \bs{1}_{x > 0}$ and in Boltzmann sampling $s$ should approach positive infinity. Note, though, that if an indicator allocation function is used, the algorithm will fail to be replicable. In contrast, to maximize the power to detect a treatment effect, it is optimal for the allocation function to approach the Neyman allocation \citep{neyman1992two}; in simple bandit environments where the reward noise variance is the same for all actions, the Neyman allocation selects all actions with equal probability. 

In other words, the choice of the allocation function will simultaneously affect the algorithm's ability to maximize rewards and the power of statistical analyses to detect a treatment effect. Though we do not explore it thoroughly in this work, we believe that the allocation function can be chosen as a part of the clinical trial design by extending methods from standard power analyses \citep{cohen2013statistical,liao2016sample}, which specify that for a standardized effect size of interest (e.g., $0.1$ or $0.2$) the power must be sufficiently high (e.g., $\geq 80\%$). We discuss further open questions related to this point later in Section \ref{sec:discussion}.
}

\section{Statistical Analyses after using  Replicable Bandits}
\label{sec:statisticalAnalyses}

% \kwz{Talk about the tyeps of analyses we're interested in}

\new{In this section, we discuss post-trial statistical analyses of interest and present our theoretical guarantees for these statistical analyses on data collected by replicable bandit algorithms.
In most clinical trials, particularly behavioral health and digital health clinical trials, both the pre-specified and non-pre-specified data analyses concern inference for population parameters.  These parameters are defined via the population version of estimating functions or a loss criterion and estimated using Z or M estimators \citep{van1996weak}. Specifically, we are interested in constructing confidence intervals for a loss minimizer $\thetastar(P)$ that solves the following under the population-deployed policies\footnote{Recall we introduced population-deployed policies in Section \ref{sec:popDeployed}}:}
\begin{align}
    \label{eqn:lossDef}
    0 = \lim_{n \to \infty} \E_P^{(n)} \left[ \dot{\ell}_\theta \big( \HH_{i,T} \big) \right] \bigg|_{\theta = \thetastar(P)},
\end{align}
for a least squares criterion
\begin{align}
    \label{eqn:leastSquaresLinear}
    \ell_\theta \big( \HH_{i,T} \big) 
    = \sum_{t=1}^T \left\{ \Y_{i,t} - \phi_{\TN{inf}} \big( X_{i,t}, A_{i,t}, \HH_{i,t-1} \big)^\top \theta \right\}^2.
\end{align}
Above, we use $\Y_{i,t} \triangleq m(Y_{i,t}) \in \real$, where $m$ is a known mapping that takes $Y_{i,t}$ and outputs a scalar (e.g., $m$ could output a dimension of the vector $Y_{i,t}$ \new{or output the reward $R(Y_{i,t})$}). 
$\dot{\ell}_\theta$ is the derivative of $\ell_\theta$ with respect to $\theta$ and $\phi_{\TN{inf}}$ is a  vector of bounded feature mapping functions. 
We formalize that the limit in \eqref{eqn:lossDef} exists in Proposition \ref{prop:thetastarWellDefined}. 

\new{For different choices of $\MC{Y}_{i,t}$ and $\phi_{\TN{inf}}$, we can represent inference for different quantities of interest. For example, for $\MC{Y}_{i,t} = R(Y_{i,t})$ and $\phi_{\TN{inf}} = 1$, then $\theta^*(P)$ is the mean reward. Alternatively, if we were interested in inferring a treatment effect, we could consider choosing 
\begin{align*}
    \phi_{\TN{inf}}(X_{i,t}, A_{i,t}, \HH_{i,t-1}) = [g(X_{i,t}, \HH_{i,t-1}), A_{i,t} f(X_{i,t})]
\end{align*}
for some functions $g$ and $f$ that take values on $\real^{d_g}$ and $\real^{d_f}$ respectively. In this case, the last $\real^{d_f}$ dimensions of $\theta^*(P)$ will represent parameters in a treatment effect that is moderated by $X_{i,t}$. In the longitudinal digital health literature, these types of treatment effects are called \textit{causal excursion effects}. See \citet{boruvka2018assessing} and \citet{qian2021micro} for further discussion of causal excursion effects.}

To estimate $\theta^*(P)$ we consider an empirical loss minimizer $\hat\theta^{(n)}$ that solves
\begin{align}
    \label{eqn:thethatLS}
    0 = \frac{1}{n} \sum_{i=1}^n \dot{\ell}_\theta \big( \HH_{i,T} \big) \bigg|_{\theta = \hat\theta^{(n)}}.
\end{align}
\new{We saw earlier in Section \ref{sec:nonrepExamples} that if the bandit algorithm collecting the data is non-replicable, then standard estimators like $\thetahat$ can fail to be replicable. In this section, we assume the bandit algorithm is replicable according to Definition \ref{def:replicableBandit} and show that $\thetahat$ is replicable and that we can construct confidence intervals for $\thetastar(P)$.

The theoretical results we present in the following sections for this estimator on data collected by replicable bandit algorithms are two-fold. First, in Section \ref{sec:consistency} we show consistency, i.e., that the estimator $\hat\theta^{(n)}$ converges to $\theta^*(P)$ in probability. This ensures that for any replication of the trial, as the sample size $n$ grows, $\hat\theta^{(n)}$ will converge to the quantity of interest $\theta^*(P)$; this implies $\hat\theta^{(n)}$ is replicable in the sense of Definition \ref{def:replicableStat}. Then in Section \ref{sec:normality} we prove that the estimator $\hat\theta^{(n)}$ is asymptotically normal, i.e., $\sqrt{n} \big( \hat\theta^{(n)} - \thetastar(P) \big) \Dto N\big(0, \Sigma(P) \big)$ for some covariance matrix $\Sigma(P)$ which we can estimate; this will ensure we are able to characterize the variability of $\hat\theta^{(n)}$ across replications in the sense of Definition \ref{def:replicableVar}. With the asymptotic normality result, we are able to construct asymptotically valid confidence intervals (and confidence regions) for $\theta^*(P)$.}

\subsection{Consistency} 
\label{sec:consistency}
%%%%%%%%%%%%%%%%%%%%%%%%%%%%%%%%%%%%

\new{In this section, we show that the estimator $\hat\theta^{(n)}$ from \eqref{eqn:thethatLS} converges to the quantity of interest $\thetastar(P)$.} Formally, we show that for each $\epsilon > 0$,
\begin{align}
    \label{eqn:thetahatConsistency}
    \lim_{n \to \infty} \, \sup_{P \in \MC{P}} \, \PP_P^{(n)} \left( \big\| \hat\theta^{(n)} - \thetastar(P) \big\|_2 > \epsilon \right) = 0. %\quad \TN{for~each~} \epsilon > 0.
\end{align}
\new{The above ensures that the estimator $\thetahat$ is replicable, i.e., across trial replications, the proportion for which $\big\| \hat\theta^{(n)} - \thetastar(P) \big\|_2 > \epsilon$ goes to zero as the sample size in each trial grows. Note that the above ensures that $\hat\theta^{(n)}$ is replicable in the sense of Definition \ref{def:replicableStat}.
Above, the supremum over possible data generating distributions $P \in \MC{P}$ ensures that convergence is uniform in the underlying distribution (see Section \ref{sec:defReplicableBandit} for further discussion of the need for uniform convergence).} 

We now discuss the assumptions we use to ensure \eqref{eqn:thetahatConsistency}. The first, Assumption \ref{assump:limit}, below, helps ensure that $\thetastar$ from \eqref{eqn:lossDef} is well-defined (see Proposition \ref{prop:thetastarWellDefined}). Specifically, we assume there exists a population-deployed policy. This is the \textit{expected} policy that would be learned on average across trial replications as the number of individuals in each trial $n \to \infty$. Note that Assumption \ref{assump:limit} display \eqref{eqn:policyStabilizes} below is very mild and holds even for bandit algorithms that are not replicable in the sense of Definition \ref{def:replicableBandit}. This assumption rules out edge-case bandit algorithms that may continually oscillate, e.g., switch between two different policies indefinitely depending on whether $n$ is even or odd. Display \eqref{eqn:policyStabilizesRate} generalizes the assumption to hold at a particular rate.
\begin{assumption}[Population-deployed policies]
    \label{assump:limit}
    For each $t \in [2 \colon T]$, there exists a policy $\pi_t^*$ %$\betastar{t}$ 
    that may depend on
    %policy $\pi_t^*$ that may 
    $P \in \MC{P}$ such that %for each $a \in \MC{A}$ and $x \in \MC{X}$,
    \begin{align}
    \label{eqn:policyStabilizes}
    \lim_{n \to \infty} \, \sup_{P \in \MC{P}} \, \left\| \E_P^{(n)} \big[ \pi \big(\cdotspace; \betahat{t-1} \big) \big] - \pi_{t}^*(\cdotspace; P) \big) \right\|_\infty = 0.
    \end{align}
    Furthermore, we say this assumption holds for a non-random, monotonically increasing rate $\{ r^{(n)} \}_{n \geq 1}$ if %for each $a \in \MC{A}$ and $x \in \MC{X}$,
    \begin{align}
    \label{eqn:policyStabilizesRate}
    \lim_{n \to \infty} \, \sup_{P \in \MC{P}} \, r^{(n)} \left\| \E_P^{(n)} \big[ \pi \big(\cdotspace; \betahat{t-1} \big) \big] - \pi_{t}^*(\cdotspace; P) \big) \right\|_\infty = 0.
    \end{align}
\end{assumption}

Assumption \ref{assump:invertible} below concerns the second derivative of the population loss minimization criterion: $\ddot{\ell}_{\theta} \big( \HH_{i,T} \big) \triangleq \phi_{\TN{inf}} \big( X_{i,t}, A_{i,t}, \HH_{i,t-1} \big) \phi_{\TN{inf}} \big( X_{i,t}, A_{i,t}, \HH_{i,t-1} \big)^\top$. We assume that the expectation of this matrix is invertible. \new{This assumption is critical to ensuring that $\theta^*(P)$ is unique. Note that this type of assumption is standard in the literature, and similar assumptions are needed even for i.i.d. data settings \citep[Theorem 5.7]{van2000asymptotic}.}
%that the second derivative of the population loss minimization criterion is invertible. In particular, . % are unique and is standard in the literature \citep[Theorem 5.7]{van2000asymptotic}.
\begin{assumption}[Invertible]
    \label{assump:invertible}
    For some $\lambda > 0$ and all sufficiently large $n$, the minimum eigenvalue of the following matrix is bounded above $\lambda$ for each $P \in \MC{P}$:
    \begin{align*}
        \E_{P}^{(n)} \left[ \phi_{\TN{inf}} \big( X_{i,t}, A_{i,t}, \HH_{i,t-1} \big) \phi_{\TN{inf}} \big( X_{i,t}, A_{i,t}, \HH_{i,t-1} \big)^\top \right].
    \end{align*}
\end{assumption}
Assumption \ref{assump:moments} below ensures that the outcomes $Y_{i,t}$ have bounded fourth moments under any action sequence. \new{This assumption is quite mild, and a simple sufficient condition is that $|Y_{i,t}|$ is bounded.}
\begin{assumption}[Moments]
    \label{assump:moments}
    For each $P \in \MC{P}$, there exists a random variable $Y$ such that 
    %independent of $\HH_{1:n,T}$ given $P$ 
    \begin{align}
        \label{eqn:moments1}
        \max_{a_{1:T} \in \MC{A}^{1:T} }  \sum_{t=1}^T \big| \Y_{i,t}(a_{1:t}) \big| \leq Y, \quad P\TN{-almost surely}
    \end{align}
    and $\E_P \left[ | Y |^4 \right] < c_Y$ for a constant $c_Y < \infty$ that does not depend on $P$. 
\end{assumption}

Our first result below, Proposition \ref{prop:thetastarWellDefined}, ensures that the solution $\thetastar(P)$ from \eqref{eqn:replicableDef} is well-defined and unique (see Appendix \print{\ref{app:wellDefined}}{B.2}).
\begin{proposition}[$\thetastar(P)$ is well-defined]
    \label{prop:thetastarWellDefined}
    Suppose a bandit algorithm is replicable, i.e., satisfies \eqref{eqn:replicableDef}. If Assumption \ref{assump:limit} display \eqref{eqn:policyStabilizes}, Assumptions \ref{assump:invertible}-\ref{assump:moments} hold, then for each $P \in \MC{P}$ there is a unique $\thetastar(P)$ that solves \eqref{eqn:lossDef}.
\end{proposition}

\new{Before we state our consistency result for $\thetahat$, we introduce a final assumption that concerns the rate at which the bandit algorithm explores or randomizes treatments. Note that these types of exploration or minimum randomization conditions are common in the literature on data collected with bandit algorithms \citep{zhang2021mestimator,hadad2021confidence,boruvka2018assessing,qian2019estimating} and it is not directly related to replicability. Intuitively, if one is interested in inferring a treatment effect, each treatment must be sampled sufficiently often to ensure we can accurately estimate it. Ensuring the action selection probabilities do not converge to zero or one at too fast a rate guarantees that each treatment is sampled sufficiently often.}

We now define what we mean for an algorithm to have an exploration rate of $\{ \pi_{\min}^{(n)} \}_{n \geq 1}$. 
\begin{mydef}[Exploration rate]
    \label{def:exploration}
  Let $\{ \pi_{\min}^{(n)} \}_{n \geq 1}$ be a non-random, positive, monotonically decreasing sequence. We say that a bandit algorithm has an exploration rate of $\{ \pi_{\min}^{(n)} \}_{n \geq 1}$ if for any 
     $x_{1:T} \in \MC{X}^T$ and actions $a_{1:T} \in \MC{A}^T$,
    \begin{align*}
       \PP_P^{(n)} \bigg( \prod_{t=1}^T \pi \big(x_t,a_t; \betahat{t} \big)  \geq \pi_{\min}^{(n)} \bigg) = 1
    \end{align*}
    for each $t, n$ and $P \in \MC{P}$. 
\end{mydef}
Note that many digital health clinical trials use an exploration rate of $\pi_{\min}^{(n)} = \pi_{\min}$ for a constant $\pi_{\min} > 0$ in order to ensure sufficient power of a primary analyses \citep{liao2020personalized,nahum2024optimizinh,battalio2021sense2stop,figueroa2021adaptive,yom2017encouraging}. Our consistency results will only require that $\pi_{\min}^{(n)} n \to \infty$ as $n \to \infty$.

\begin{theorem}[Consistency]
    \label{thm:consistencyMiss}
    Suppose a bandit algorithm has an exploration rate $\{ \pi_{\min}^{(n)} \}_{n \geq 1}$ such that $\pi_{\min}^{(n)} n \to \infty$ as $n \to \infty$. Also let the bandit algorithm is replicable at rate $\big\{ \frac{1}{ \pi_{\min}^{(n)} } \big\}_{n \geq 1}$ according to Definition \ref{def:replicableBandit} display \eqref{eqn:replicableDefRate}.
    If Assumption \ref{assump:limit} display \eqref{eqn:policyStabilizesRate} for rate $\big\{ \frac{1}{ \pi_{\min}^{(n)} } \big\}_{n \geq 1}$, Assumptions \ref{assump:invertible}-\ref{assump:moments} hold, then $\thetahat$ converges in probability to $\thetastar(P)$ uniformly over $P$, i.e., \eqref{eqn:thetahatConsistency} holds.
\end{theorem}
 % \alt{Does Theorem 4.1 say if certain assumptions hold and your bandit algorithm is a replicable bandit, then Eq. 14 holds and the estimator, using the data collected by the bandit, is consistent?} 
\new{Note that for Theorem \ref{thm:consistencyMiss}, if the bandit algorithm has a constant exploration rate, $\pi_{\min} > 0$, then only Assumption \ref{assump:limit} display \eqref{eqn:policyStabilizes} is needed.} See Appendix \print{\ref{app:consistency}}{B.3} for the proof.

\subsection{Asymptotic normality} %%%%%%%%%%%%%%%%%%%%%%%%%%%%%%%%%%%%
\label{sec:normality}

In this section we discuss when %under replicable \textit{parametric} bandit algorithms, 
the estimator $\thetahat$ is asymptotically normal uniformly over $P \in \MC{P}$:
\begin{align}
    \sqrt{n} \big( \hat{\theta}^{(n)} - \thetastar \big) \Dto \N \left( 0, \ddot{L}_P^{-1} \Sigma_P \ddot{L}_P^{-1} \right). % ~~\TN{uniformly over } P \in \MC{P}.
    \label{eqn:normalityResult}
\end{align}

The limiting variance $\ddot{L}_P^{-1} \Sigma_P \ddot{L}_P^{-1}$ from \eqref{eqn:normalityResult} is called the ``adaptive'' sandwich variance estimator \citep{zhang2022statistical}. This is because it is an adjustment of the standard sandwich variance estimator \citep[Chpt 5.3]{van2000asymptotic}. \new{The variance can be estimated consistently and ensures we can characterize the variability of $\hat\theta^{(n)}$ across replications in the sense of Definition \ref{def:replicableVar}.}
We specify the formulas for the terms $\Sigma_P$ and $\ddot{L}_P$ in the limiting covariance matrix in Appendix \print{\ref{app:limitingVar}}{B.}. Note that the ideas in much of the following results are based on \citet{zhang2022statistical}.

We introduce two additional assumptions on the bandit algorithm that we use to prove the asymptotic normality result in \eqref{eqn:normalityResult}. The first is that the policy function $\pi(\cdotspace; \beta_{t-1})$ is differentiable with respect to $\beta_{t-1}$. \new{This assumption ensures that the algorithm's policies are smooth in the underlying $\beta$ parameters that it learns and helps us ensure that we can characterize in the limit how errors in estimating $\beta$ impact the decisions the algorithm makes.}

\begin{assumption}[Differentiable policies]
    %For each $t \in [2 \colon T]$, 
    The function $\pi(\cdotspace; \beta)$ is differentiable with respect to $\beta$, i.e., there exists a function $\dot{\pi} : \MC{X} \by \MC{A} \by \real^{d_\beta}\to \real^{d_\beta}$ such that for each $\beta \in \real^{d_\beta}, x \in \MC{X}, a \in \MC{A}$,
    \begin{align*}
        \lim_{h \to 0} \frac{ | \pi(x,a; \beta + h) - \pi(x,a; \beta) - \dot{\pi}(x,a; \beta)^\top  h| }{ \| h \|_2 } = 0.
    \end{align*}
    %Above, $h \in \real^{d_\beta}$.
    Also for each $\beta \in \real^{d_\beta}$, $\| \dot{\pi}(\cdotspace; \beta) \|_\infty < \infty$.\footnote{$\| \dot{\pi}(\cdotspace; \beta) \|_\infty =\sup_{(x,a)\in\MC{X} \by \MC{A},j\in[d_\beta]}|\dot{\pi}_j(x,a; \beta)|$} %\kwz{weaken?}
    %\dot{\ell}_{\thetastar}(\HH_{i,T})
    \label{assump:diffPolicy}
\end{assumption}

The final assumption, Assumption~\ref{assump:linearity},  
is that the vector of algorithm statistics 
$\betahat{1:T-1} = (\betahat1, \betahat2, \dots, \betahat{T-1})$ 
%converges to some fixed vector $\betastar{1:T} = (\betastar1, \betastar2, \dots, \betastar{T-1})$, that may depend on $P$. Moreover, $\betahat{1:T-1}$ 
is asymptotically linear \citep[Chpt 3]{tsiatis2006semiparametric}. \new{This assumption is commonly an intermediate step in asymptotic results for M and Z estimators \citep[Chapter 25.9]{van1996weak}. When $\beta_{1:T}$ are parameters in a linear model, this assumption is relatively straightforwardly satisfied if $\beta_{1:T}$.} 
% (see Lemma \ref{lemma:banditLinearity} for details). 
See \citet{zhang2022statistical} for the general sufficient conditions for bandit algorithms to satisfy Assumption \ref{assump:linearity}.
\begin{assumption}[Asymptotic linearity]
For some  vector $\betastar{1:T-1} = (\betastar1, \betastar2, \dots, \betastar{T-1})$ that  may depend on $P$, and 
%$((T-1)d_\beta) \times 1$ 
vector, $\psi_{\betastar{1:T-1}}(\HH_{1:n,T-1})$,
for which \\ 
    $\sup_{ P \in \MC{P} } \E_{P, \pi^*} \big[ \| \psi_{\betastar{1:T-1}}(\HH_{1:n,T-1}) \|_2^4 \big]< \infty$, we have
\begin{align*}
        &\frac{1}{\sqrt{n}} \sum_{i=1}^n \frac{\prod_{t=2}^T \pi_t^*(X_{i,t}, A_{i,t}; P)}{\prod_{t=2}^T {\pi}(X_{i,t}, A_{i,t};\betahat{t-1})} \psi_{\betastar{1:T-1}} (\HH_{i,T-1}) \\
        &= \sqrt{n} \left( \betahat{1:T-1} - \betastar{1:T-1} \right) + o_P(1) = O_P(1),
    \end{align*}
    uniformly over $P \in \MC{P}$. 
    Policies $\{ \pi_t^* \}_{t=2}^T$ were defined in Assumption \ref{assump:limit}. 
    \label{assump:linearity}
\end{assumption}

\begin{theorem}[Asymptotic normality]
    \label{thm:normality}
    Suppose a bandit algorithm is replicable, i.e., satisfies \eqref{eqn:replicableDef}, and has a constant exploration rate $\{ \pi_{\min}^{(n)} \}_{n \geq 1}$ where $\pi_{\min}^{(n)} = \pi_{\min}$ for a constant $\pi_{\min} > 0$. Under Assumption \ref{assump:limit} display \eqref{eqn:policyStabilizes} and Assumptions \ref{assump:invertible}-\ref{assump:linearity}, $\sqrt{n} \big( \thetahat - \thetastar \big)$ is asymptotically normal uniformly over $P \in \MC{P}$, i.e., \eqref{eqn:normalityResult} holds.
\end{theorem}
See Appendix \print{\ref{app:thmNormality}}{B.4} for the proof of the Theorem.

%%%%%%%%%%%%%%%%%%%%%%%%%%%%%%%%%%%%%%%%%%%%%%%%%%%%%%%
%%%%%%%%%%%%%%%%%%%%%%%%%%%%%%%%%%%%%%%%%%%%%%%%%%%%%%%
\section{Simulations}
\label{sec:simulations}
In this section, we examine replicability experimentally in a synthetic environment and an environment based on Oralytics \cite{nahum2024optimizinh, trella2024oralytics} (see the Example in Section \ref{ex:oralytics}). In both these settings, we are interested in the average outcome under the population-deployed policy (similar to \eqref{eqn:diffAveReward}): %For simplicity, we focus on inferring first term in this difference: 
\begin{align*}
   \thetastar = \lim_{n \to \infty} \E^{(n)} \left[ \frac{1}{T} \sum_{t=1}^T \Y_{i,t} \right].
\end{align*}
We estimate $\thetastar$ above with the empirical average outcome:
\begin{align*}
    \thetahat = \frac{1}{n} \sum_{i=1}^n \frac{1}{T} \sum_{t=1}^T \Y_{i,t}.
\end{align*}

In each of our two experimental settings, we have binary actions ($\MC{A} = \{ 0, 1\}$) and \new{we consider the smoothed Thompson sampling algorithm and Boltzmann sampling from Section \ref{sec:satisfyAlgs}, as well as an $\epsilon$-greedy contextual bandit algorithm. See Appendix \print{\ref{app:simulations}}{C} for full details on algorithms. Note that smoothed Thompson sampling and Boltzmann sampling satisfy our replicable bandit definition, i.e., \eqref{eqn:replicableDef}, and $\epsilon$-greedy does not. As a result, we find in our simulations that $\epsilon$-greedy can fail to be replicable in some environments, while Boltzmann sampling is consistently replicable across various environments.}

\subsection{Synthetic simulations} 
\label{sec:synthetic}

%%%%%%%%%%%%%%%%%%%%%%%%%%%%%%%%%%%%%%%%%%%%%%%%%%%%%%%
\subsubsection{Simulation environment.}
We consider the synthetic simulation environment that was used in \citet{zhang2022statistical}. In this environment, $T = 50$ and the mean reward is dependent on the recent ``dosage'', i.e., an exponentially discounted average of the number of previous treatments. In other words, in this setting, the sequence of previous treatments can impact an individual's future responsiveness. Specifically, the outcome of interest $\Y_{i,t} = Y_{i,t}$ where
\begin{align}
    Y_{i,t} = \alpha_0 + \alpha_1 A_{i,t} \frac{1}{c_\gamma} \sum_{t'=1}^{t-1} \gamma^{t-1-t'} A_{i,t'} + \alpha_2 A_{i,t} + \epsilon_{i,t}.
    \label{eqn:Ysynthetic}
\end{align}
Above, we use $\gamma = 0.95$ and use $c_\gamma = 1/(1-\gamma)$ to normalize the exponentially discounted sum to be in the range $[0,1]$. The errors $\epsilon_{i,t}$ are correlated across $t$ for each individual $i$. See Appendix \print{\ref{app:syntheticEnv}}{C.1} for more details.   

\subsubsection{Contextual bandit algorithms.}
The Boltzmann sampling, Thompson sampling, and $\epsilon$-greedy algorithms use the reward $R(Y_{i,t})=Y_{i,t}$ and use previous decision outcomes, $Y_{i,t-1}$, as the context.
Specifically, the algorithms use the following working model for the expected reward:
\begin{align*}
    \E[ R(Y_{i,t}) \mid X_{i,t}, A_{i,t} ] 
    = \beta_0^\top \phi_{\TN{alg}}(X_{i,t}) + \beta_1^\top A_{i,t} \phi_{\TN{alg}}(X_{i,t})
\end{align*}
where $\phi_{\TN{alg}}(X_{i,t}) = [1, R(Y_{i,t-1})]$. \new{The Boltzmann sampling and smoothed Thompson sampling (s-TS) algorithms use a steepness of $s=5$; see displays \eqref{eqn:boltzmannProb1} and \eqref{sec:smoothTS} respectively. The smoothed Thompson sampling algorithm use a multivariate Gaussian prior with mean zero and an identity covariance matrix. For $\epsilon$-greedy, we use $\epsilon = 0.1$; see display \eqref{eqn:epsilonGreedy}.}

\subsubsection{Results.}
We vary the number of individuals in the trial $n$.   Recall that $
    \thetahat = \frac{1}{n} \sum_{i=1}^n \frac{1}{T} \sum_{t=1}^T Y_{i,t}$. 
Further note that under a baseline algorithm $\pi^{\TN{baseline}}$ which selects $\PP(A_{i,t} = 1 \mid \HH_{1:n,t-1}) = 0.5$,
the average expected outcome is $\E_{\pi^{\TN{baseline}}} \big[ \frac{1}{T} \sum_{t=1}^T\ Y_{i,t} \big] = 0.0492$.

As seen in Table \ref{table:synthetic}, the average outcome under Boltzmann sampling ($0.303$), Smoothed Thompson sampling ($0.301$), and $\epsilon$-greedy ($0.427$) both greatly exceed the average outcome under the baseline algorithm ($0.0492$). The average outcome under Boltzmann sampling and smoothed Thompson sampling  (s-TS) is less than that under $\epsilon$-greedy. However, under the replicable Boltzmann sampling algorithm, we can construct accurate confidence intervals for the average outcome, which is not the case in our simulations under the non-replicable $\epsilon$-greedy algorithm.

\begin{figure}
    
    \centering
    \includegraphics[width=0.49\linewidth]{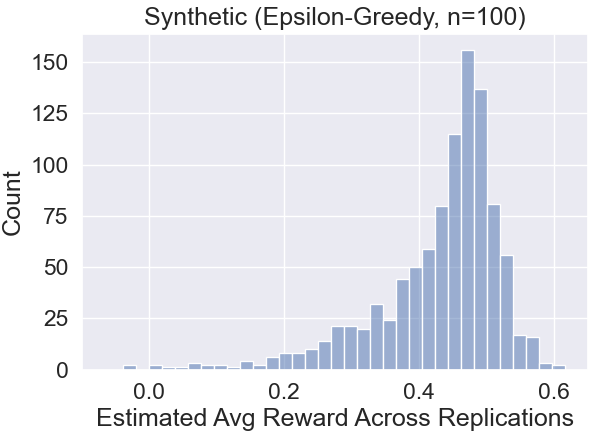}
   \includegraphics[width=0.49\linewidth]{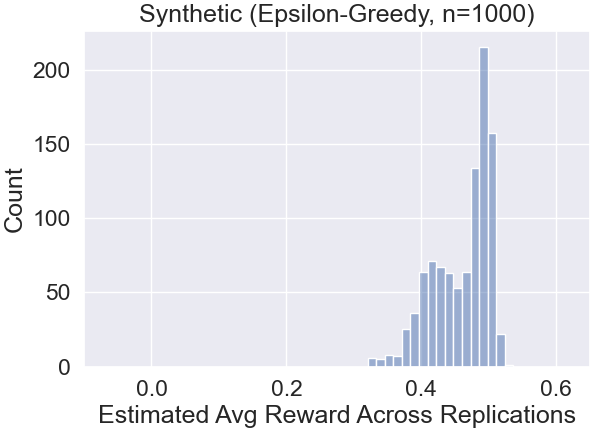}
    \caption{\bo{Average outcome estimator $\thetahat$ across replications under $\epsilon$-greedy in synthetic environment.} We consider $n=100$ (left) and $n=1000$ (right). Results are across $1000$ trial replications.} \label{figure:synthetic}
\end{figure}

We construct confidence intervals for the average outcome using a normal approximation with an estimator of the adaptive sandwich limiting variance from \eqref{eqn:normalityResult} (denoted AS). We also compare to the standard sandwich variance estimator (denoted S). Under $\epsilon$-greedy, the distribution of the average outcome estimator is far from normally distributed (see Figure \ref{figure:synthetic}). Moreover, in this setting under $\epsilon$-greedy, the naive standard sandwich variance estimator severely underestimates the true variance. In fact, for $n=1000$ under $\epsilon$-greedy, the standard variance estimator underestimates the true variance by a factor of $9$! Also, naively using a normal approximation with such a variance estimator leads to confidence intervals that \textit{severely} undercover (e.g., a confidence interval designed to cover $95\%$ of the time only has an empirical coverage of $58.3\%$ for $n=100$ and $23.1\%$ for $n=1000$).

\begin{table}
\caption{\bo{Synthetic simulation results.} We average over $1000$ Monte Carlo repetitions. We use (AS) to refer to variances estimated using the adaptive sandwich variance estimator from display \eqref{eqn:normalityResult}. We use (S) to refer to variances estimated using the standard sandwich variance estimator. Note that the (AS) and (S) variance estimates reported are means across all 1000 repetitions.} 

\begin{tabular}{lcccccl}
\toprule 
 $n=100$ & Boltzmann & \new{s-TS} & $\epsilon$-Greedy \\
\midrule
Average $\thetahat$ & $0.303$ & \new{$0.301$} & $0.427$ \\
Empirical $\thetahat$ Variance & $0.00212$ & \new{$0.00204$} & $0.009$ \\
Estimated $\thetahat$ Variance (AS) & $0.00499$ & \new{$0.00337$} & N/A \\
Estimated $\thetahat$ Variance (S) & $0.00114$ & \new{$0.00114$} & $0.00109$   \\
Coverage ($95\%$ Interval, AS) & $97.1\%$ & \new{$95.1\%$} & N/A \\
Coverage ($95\%$ Interval, S) & $84.7\%$ & \new{$85.6\%$} & $58.3\%$ \\
\bottomrule
\end{tabular}

\medskip
\label{table:synthetic}
\begin{tabular}{lcccccl}
\toprule 
 $n=1000$ & Boltzmann & \new{s-TS} & $\epsilon$-Greedy \\
\midrule
Average $\thetahat$ & $0.305$ & \new{$0.304$}  & $0.461$\\
Empirical $\thetahat$ Variance & $0.00025$ & \new{$0.00022$} & $0.002$ \\
Estimated $\thetahat$ Variance (AS) & $0.00025$ & \new{$0.00025$} & N/A \\
Estimated $\thetahat$ Variance (S) & $0.00012$ & \new{$0.00012$} & $0.00011$   \\
Coverage ($95\%$ Interval, AS) & $93.5\%$ & \new{$95.4\%$} & N/A \\
Coverage ($95\%$ Interval, S) & $82.0\%$ & \new{$84.7\%$} & $23.1\%$ \\
\bottomrule
\end{tabular}

\end{table}

\subsection{Oralytics simulations} 
%%%%%%%%%%%%%%%%%%%%%%%%%%%%%%%%%%%%%%%%%%%%%%%%%%%%%%%

Oralytics involves a series of oral health clinical trials, each using a contextual bandit algorithm to personalize the delivery of intervention prompts encouraging oral self-care behaviors for individuals at-risk for dental disease. For each individual, the Oralytics algorithm makes decisions every day at two decision points (i.e., once in the morning and once in the evening). The action space is binary, representing sending an intervention prompt ($A_{i, t} = 1$) or not ($A_{i, t} = 0$). In the upcoming Oralytics trial \citep{nahum2024optimizinh}, individuals are each in the trial for $10$ weeks, which means a total of $T=140$ decision times. We use $n=100$ in our simulations.

\subsubsection{Simulation environment.}
To build the simulation environment, we use data from the pilot phase of the Oralytics study \citep{nahum2024optimizinh,trella2024oralytics}. %\footnote{Oralytics pilot data is publicly accessible in GitHub \href{https://github.com/StatisticalReinforcementLearningLab/oralytics_pilot_data}{here}.} 
The pilot phase ran in spring 2023 for 35 days. All individuals were recruited and started the study at the same time (no incremental recruitment). For each individual in the pilot study, we fit an outcome model, which is used to simulate rewards. % model to take in the individual's time varying covariates and the current action, to output a distribution the subsequent brushing time. 
To make the simulation environment more realistic, we also modeled the negative delayed effects of actions and individual app opening behavior. See Appendix~\print{\ref{sec_sim_env}}{C.2.1} for full details.

In Oralytics, a variety of brushing outcomes are collected passively via the blue-tooth enabled toothbrushes. These include $B_{i,t}$, the number of seconds the individual brushes, and $P_{i,t}$, the number of seconds the user brushes with too much pressure, following the $t^{\TN{th}}$ decision time. Brushing quality is the true behavioral outcome that the scientific team aims to improve for each individual to achieve long-term oral health. % the main outcome of interest $\Y_{i,t}$ is a measure of individual's brushing quality, which 
Our working definition of brushing quality is the number of seconds the individual brushes total minus the number of seconds of over-pressure brushing:
\begin{align}
    \Y_{i,t} = \max(180, B_{i,t} - P_{i,t})
    \label{eqn:qualityDef}
\end{align}
Note above that we truncate by $180$ to avoid optimizing for over-brushing. 

Although the main outcome of interest $\Y_{i,t}$ is the brushing quality, the reward is a design choice that can be modified in order to improve the algorithm's learning. In Oralytics we use $R(Y_{i,t}) = \Y_{i,t} - C_{i, t}$, where $C_{i , t}$ is a cost term that represents a crude approximation of the delayed effects of actions on future contexts and rewards. This allows us to approximate a full reinforcement learning algorithm with a contextual bandit algorithm. 
%approximation for the value of next state. 
%Using this reward design, we can generalize the contextual bandit to consider the negative delayed effect of actions, which is beyond the standard framework for bandit algorithms. 
For more details on the reward design see Appendix~\print{\ref{sec_app_reward}}{C.2.2} and \citet{trella2022reward}.
The featurization used by the contextual bandit algorithms, $\phi_{\TN{inf}}(X_{i,t}) \in \real^5$, consists of an intercept, time of day (morning vs. evening), brush times over the past week, ``dosage'' (i.e., number of promps sent) in the past week, and a prior day app engagement indicator.

\new{\subsubsection{Contextual bandit algorithms.}
The Boltzmann sampling, Thompson sampling, and $\epsilon$-greedy algorithms all use the same reward definition and context features. The Boltzmann sampling and smoothed Thompson sampling (s-TS) algorithms use a steepness of $s=0.025$; see displays \eqref{eqn:boltzmannProb1} and \eqref{sec:smoothTS} respectively. We set $s=0.025 \approx \frac{1}{38.83}$, where $38.83$ is the standard deviation of the reward noise \citep{trella2024oralytics}. The smoothed Thompson sampling algorithm uses a prior based on that used in the Oralytics trial \citep{trella2024oralytics}. For $\epsilon$-greedy, we use $\epsilon = 0.2$; see display \eqref{eqn:epsilonGreedy}.}

\subsubsection{Results.}
As seen in Table \ref{table:oralytics}, the expected outcome under Boltzmann sampling, smoothed Thompson sampling (s-TS), and $\epsilon$-greedy in the Oralytics environment is comparable. Similar to the synthetic simulation setting, we find that under Boltzmann sampling and smoothed Thompson sampling, using a normal approximation for $\thetahat$ with the adaptive sandwich variance estimator of the limiting variance leads to accurate confidence intervals. However, in contrast to the synthetic simulation setting, we find that under $\epsilon$-greedy, the naive approach of using a normal approximation for $\thetahat$ and estimating the limiting variance using the standard sandwich variance estimators empirically works well in this setting. 

These simulation results support the idea that Boltzmann sampling can better ensure inferential replicability across a variety of possible data-generating environments $P \in \MC{P}$, while the inferential replicability under $\epsilon$-greedy may depend on properties of the environment itself, properties that may not be known a priori to deploying the algorithm.

\begin{table}
\label{table:oralytics}
\caption{\bo{Oralytics simulations.} We average over $1000$ Monte Carlo repetitions. We use (AS) to refer to variances estimated using the adaptive sandwich variance estimator from display \eqref{eqn:normalityResult}. We use (S) to refer to variances estimated using the standard sandwich variance estimator. Note that the (AS) and (S) variance estimates reported are means across all 1000 repetitions.}
\begin{tabular}{lcccccl}
\toprule 
  & Boltzmann & \new{s-TS} & $\epsilon$-Greedy \\
\midrule
Average $\thetahat$ & $91.66$  & \new{$91.62$} & $91.45$ \\
Empirical $\thetahat$ Variance & $7.56$ & \new{$7.77$} & $8.61$ \\
Estimated $\thetahat$ Variance (AS) & $9.37$ & \new{$9.32$} & N/A \\
Estimated $\thetahat$ Variance (S) & $7.71$ & \new{$7.72$}  & $7.81$ \\
Coverage ($95\%$ Interval, AS) & $95.8\%$ & \new{$95.5\%$} & N/A  \\
Coverage ($95\%$ Interval, S) & $94.1\%$ & \new{$93.7\%$} & $92.3\%$ \\
\bottomrule
\end{tabular}
\end{table}

%%%%%%%%%%%%%%%%%%%%%%%%%%%%%%%%%%%%%%%%%%%%%%%%%%%%%%%
%%%%%%%%%%%%%%%%%%%%%%%%%%%%%%%%%%%%%%%%%%%%%%%%%%%%%%%
\section{Discussion and Open Questions}
\label{sec:discussion}

In this work, we connect inferential replicability to replicability of the underlying bandit algorithm used to make treatment decisions. We show that in a variety of misspecified environments (non-stationary and misspecified reward model), when common bandit algorithms are used to assign treatments, standard estimators can be inconsistent and thus non-replicable across trials. We discuss how this non-replicability relates to the non-replicability of the underlying policies formed by the bandit algorithm across trials.
We then provide a definition of what it means for a bandit algorithm to be replicable. We prove that under replicable bandit algorithms, a variety of standard estimators are consistent and asymptotically normal.

There are several open questions given this work, which we summarize below: \\

\noindent\bo{Replicability and regret minimization:}
%\begin{itemize}[leftmargin=15pt]
%    \item  
    What are the consequences of ensuring replicability of the bandit algorithm on regret \citep{kalvit2021closer}? Is it possible to ensure that a bandit algorithm is both replicable and minimizes regret optimally? We found in our experiments in Section \ref{sec:synthetic} that the algorithm that was not replicable lead to a bit higher rewards than an algorithm that was replicable. What about algorithms that are designed to optimally balance regret minimization and the power of a particular pre-specified post-trial primary analysis \citep{erraqabi2017trading,hu2006theory,simchi2023multi,qin2024optimizing,krishnamurthy2023proportional}? \\

\noindent\bo{Replicability and robustness:} Several papers have found that algorithms that minimize regret aggressively can be less robust to model misspecification \citep{fan2021fragility,eysenbach2021maximum,xu2024fallacy}. % and have regret distributions with fat-tails \citep{simchi2023regret}. 
    Could there be a relationship between replicability and ensuring the algorithm minimizes regret robustly across many environments?

%%%%%%%%%%%%%%%%%%%%%%%%%%%%%%%%%%%%%%%%%%%%%%
\begin{supplement}

\stitle{Appendix \print{\ref{app:nonreplicability}}{A}: Proofs for Section \ref{sec:nonrepExamples}}
\sdescription{This appendix contains the proofs for the results in Section \ref{sec:nonrepExamples} (Two Examples of Non-Replicability).}
\end{supplement}

\begin{supplement}
\stitle{Appendix \print{\ref{app:replicable}}{B}: Proofs for Section \ref{sec:replicableBandits}}
\sdescription{This appendix contains the proofs for the results in Section \ref{sec:replicableBandits} (Replicable Bandits).}
\end{supplement}

\begin{supplement}
\stitle{Appendix \print{\ref{app:simulations}}{C}: Additional Simulation Details}
\sdescription{This appendix contains additional details on
%and expanded versions of 
the simulation results from Section \ref{sec:simulations}.}
\end{supplement}

\section*{Acknowledgements}
\noindent This work is supported by grants NIH/NIDA P50DA054039, NIH/NIDCR UH3DE028723, NIH/NHLBI R01HL125440
NIH/NIBIB, OD P41EB028242, and 5P30AG073107-03 GY3 Pilots.
Susan Murphy holds concurrent appointments as a Professor of Statistics and Computer Science at Harvard University and as an Amazon Scholar. This paper describes her work performed at Harvard University and is not associated with Amazon.

\bibliographystyle{plainnat}
%% if your bibliography is in bibtex format, uncomment commands:
%\bibliographystyle{imsart-number} % Style BST file (imsart-number.bst or imsart-nameyear.bst)
\bibliography{reference}       % Bibliography file (usually '*.bib')

\clearpage
\appendix
\onecolumn

%%%%%%%%%%%%%%%%%%%%%%%%%%%%%%%%%%%%%%%%%%%%%%%%%%%%%%%%%%
%%%%%%%%%%%%%%%%%%%%%%%%%%%%%%%%%%%%%%%%%%%%%%%%%%%%%%%%%%
\section{Proofs for Section \print{\ref{sec:nonrepExamples}}{3} (Two Examples of Non-Replicability)}
\label{app:nonreplicability}

%%%%%%%%%%%%%%%%%%%%%%%%%%%%%%%%%%%%%%%%%%%%%%%%%%%%%%%%%%
\subsection{Proof of Proposition \print{\ref{prop:nonrepNonstat}}{3.1} (Non-Replicability under Non-Stationary)}
\label{app:nonstationary}

\begin{lemma}
    \label{lemma:nonstationaryBandit}
    Under the environment described in Section \print{\ref{sec:nonstationaryDGP}}{3.3.1}, for any bandit algorithm that uses treatment assignment probabilities $\pi_1(1) = \PP( A_{i,1} = 1 ) = 0.5$ and $\hat{\pi}_2(1) = \PP( A_{i,2} = 1 \mid \HH_{1:n,1})$,
    \begin{align*}
        \frac{1}{n} \sum_{i=1}^n \left( R(Y_{i,1}) + R(Y_{i,2}) \right) - \Delta_2 \hat{\pi}_2(1) \Pto 0.
    \end{align*}
\end{lemma}

\begin{proof}
Since $\{ A_{1,i} \}_{i=1}^n \iidsim \TN{Bernoulli}(0.5)$ and $\{ \epsilon_{1,i} \} \sim \N(0,1)$ are independent of $\{ A_{1,i} \}_{i=1}^n$, by the Weak Law of Large Numbers for i.i.d. random variables, 
\begin{align}
    \label{eqn:WLLNnonstationary}
    \frac{1}{n} \sum_{i=1}^n R(Y_{i,1}) = \frac{1}{n} \sum_{i=1}^n \mu_0 + \Delta_1 A_{i,1} + \epsilon_{i,1}
    \Pto \mu_0 + \Delta_1 \pi_1(1) = 0.
\end{align}
The last equality above holds because in the environment $\mu_0 = 0$ and $\Delta_1 = 0$. \\

We now show that $\frac{1}{n} \sum_{i=1}^n R(Y_{i,2}) - \Delta_2 \hat{\pi}_2(1) \Pto 0$. By Slutsky's Theorem, together with \eqref{eqn:WLLNnonstationary}, will be sufficient to prove this Lemma.
Note that for any $\alpha > 0$,
\begin{align*}
    \PP \left( \left| \frac{1}{n} \sum_{i=1}^n R(Y_{i,2}) - \Delta_2 \hat{\pi}_2(1) \right| > \alpha \right)
    &\underbrace{\leq}_{(a)} \frac{1}{\alpha^2} \E^{(n)} \left[ \left( \frac{1}{n} \sum_{i=1}^n R(Y_{i,2}) - \Delta_2 \hat{\pi}_2(1) \right)^2 \right] \\
    &= \frac{1}{\alpha^2} \E^{(n)} \left[ \left( \frac{1}{n} \sum_{i=1}^n \left\{ \mu_0 + \Delta_2 A_{i,2} + \epsilon_{i,2} \right\} - \Delta_2 \hat{\pi}_2(1) \right)^2 \right] \\
    &= \frac{1}{\alpha^2 n^2} \E^{(n)} \left[ \left( \sum_{i=1}^n \big\{ \Delta_2 \big( A_{i,2} - \hat{\pi}_2(1) \big) + \epsilon_{i,2} \big\} \right)^2 \right] \\
    &\underbrace{=}_{(b)} \frac{1}{\alpha^2 n^2} \E^{(n)} \left[ \sum_{i=1}^n \big\{ \Delta_2 \cdot \big( A_{i,2} - \hat{\pi}_2(1) \big) + \epsilon_{i,2} \big\}^2 \right] \\
    &\underbrace{=}_{(c)} \frac{1}{\alpha^2 n^2} \E^{(n)} \left[ \sum_{i=1}^n \big\{ \Delta_2^2 \cdot \big( A_{i,2} - \hat{\pi}_2(1) \big)^2 + \epsilon_{i,2}^2 \big\} \right] \\
    &\underbrace{\leq}_{(d)} \frac{1}{\alpha^2 n^2} n \left( \Delta_2^2 \cdot 0.25 + 1 \right) \to 0.
\end{align*}
Above, 
\begin{itemize}
    \item (a) holds by Markov inequality (Chebychev inequality).
    \item (b) holds because for $i \not= j$,
    \begin{align*}
        &\E^{(n)} \left[ \big\{ \Delta_2 \cdot \big( A_{i,2} - \hat{\pi}_2(1) \big) + \epsilon_{i,2} \big\} \big\{ \Delta_2 \cdot \big( A_{j,2} - \hat{\pi}_2(1) \big) + \epsilon_{j,2} \big\} \right] \\
        &= \E^{(n)} \left[ \E^{(n)} \left[ \big\{ \Delta_2 \cdot \big( A_{i,2} - \hat{\pi}_2(1) \big) + \epsilon_{i,2} \big\} \big\{ \Delta_2 \cdot \big( A_{j,2} - \hat{\pi}_2(1) \big) + \epsilon_{j,2} \big\} \mid \HH_{1:n,1} \right] \right] = 0.
        %&= \E^{(n)} \left[ \E^{(n)} \left[ \big\{ \Delta_2 \cdot \big( A_{i,2} - \hat{\pi}_2(1) \big) + \epsilon_{i,2} \big\} \Delta_2 \cdot \big( A_{j,2} - \hat{\pi}_2(1) \big) \mid \HH_{1:n,1} \right] \right] \\
        %&+ \E^{(n)} \left[ \E^{(n)} \left[ \big\{ \Delta_2 \cdot \big( A_{i,2} - \hat{\pi}_2(1) \big) + \epsilon_{i,2} \big\} \epsilon_{j,2} \mid \HH_{1:n,1} \right] \right] = 0.
    \end{align*}
    The last equality above holds because (i) $\epsilon_{i,2}, \epsilon_{j,2}$ are independent of $A_{i,2}, A_{j,2}, \HH_{1:n,1}$ and have expectation zero, and (2) $A_{i,2}, A_{j,2}$ are independent conditional on $\HH_{1:n,1}$ and $\E^{(n)} \left[ A_{i,2} - \hat{\pi}_2 \mid \HH_{1:n,1} \right] = 0$.
    \item (c) holds because $\E^{(n)} \left[ \Delta_2 \cdot \big( A_{i,2} - \hat{\pi}_2(1) \big) \epsilon_{i,2} \right] = \E^{(n)} \left[ \E^{(n)} \left[ \Delta_2 \cdot \big( A_{i,2} - \hat{\pi}_2(1) \big) \epsilon_{i,2} \mid \HH_{1:n,1}, A_{i,2} \right] \right] \\
    = \E^{(n)} \left[ \Delta_2 \cdot \big( A_{i,2} - \hat{\pi}_2(1) \big) \E^{(n)} \left[ \epsilon_{i,2} \mid \HH_{1:n,1}, A_{i,2} \right] \right]
    = 0$.
    \item (d) holds because the variance of a binary random variable ($A_{i,2}$) is bounded above by $0.25$ and $\Var(\epsilon_{i,t}) = 1$.
\end{itemize}
\end{proof}

\subsubsection{\texorpdfstring{$\epsilon$}{Epsilon}-Greedy.}
By Proposition 2 of \print{\citet{zhang2020inference}}{Zhang et al. [104]}, under $\epsilon$-greedy
\begin{align*}
    \hat{\pi}_2(1) \Dto Z \quad \quad \TN{where} \quad Z = \begin{cases}
        \epsilon / 2 & \TN{w.p.} ~ 1/2 \\
        1-\epsilon / 2 & \TN{w.p.} ~ 1/2 
    \end{cases}.
\end{align*}
Thus, by Lemma \ref{lemma:nonstationaryBandit} and continuous mapping theorem,
\begin{align*}
    \thetahat = \frac{1}{n} \sum_{i=1}^n \frac{1}{2} 
 \left( R(Y_{i,1}) + R(Y_{i,2}) \right) 
    = \Delta_2 \hat{\pi}_2(1) + o_P(1)
    \Dto \Delta_2 / 2 \cdot Z.
\end{align*}
Note by the result above and Dominated Convergence Theorem that
\begin{align*}
    \theta^* = \lim_{n \to \infty} \E^{(n)} \left[ \frac{1}{2} \sum_{t=1}^2 R(Y_{i,t}) \right]
    = \E[ \Delta_2 / 2 \cdot Z ] 
    = \Delta \cdot 0.25.
\end{align*}

\subsubsection{Thompson Sampling.}
By Proposition 1 of \print{\citet{zhang2020inference}}{Zhang et al. [104]}, under Gaussian-Gaussian Thompson sampling (standard normal priors and $\N(0,1)$ rewards),
\begin{align*}
    \hat{\pi}_2(1) \Dto \TN{Uniform}[0,1].
\end{align*}
Thus, by Lemma \ref{lemma:nonstationaryBandit} and continuous mapping theorem,
\begin{align*}
    \frac{1}{n} \sum_{i=1}^n \frac{1}{2} \left( R(Y_{i,1}) + R(Y_{i,2}) \right) 
    = \Delta_2 / 2 \cdot \hat{\pi}_2(1) + o_P(1)
    \Dto \Delta_2 / 2 \cdot Z, \quad \TN{where}~ Z \sim \TN{Uniform}[0,1].
\end{align*}

Note by the result above and by the definition of convergence in distribution,
\begin{align*}
    \theta^* = \lim_{n \to \infty} \E^{(n)} \left[ \frac{1}{2} \sum_{t=1}^2 R(Y_{i,t}) \right]
    = \E[ \Delta_2 / 2 \cdot Z ] 
    = \Delta \cdot 0.25.
\end{align*}

\medskip

%%%%%%%%%%%%%%%%%%%%%%%%%%%%%%%%%%%%%%%%%%%%%%%%%%%%%%%%%%
\subsection{Proof of Proposition \print{\ref{prop:nonrepMiss}}{3.2} (Non-Replicability under Misspecifed Linear Bandit)}
\label{app:missLinear}

\begin{lemma}
    \label{lemma:misspecHelp}
    Assume the conditions of Proposition \print{\ref{prop:nonrepMiss}}{3.2}.  Let $k$ be a finite positive constant.   Then we have that for any function $f$ such that
    $\max_{a \in \{0,1\}} \E[ f (Y_{i,2}, X_{i,2}, a)^2 |X_{i,2}, A_{i,2}=a] \leq \sigma^2\ a.s.$, we have that  
    \begin{align*}
        \frac{1}{n} \sum_{i=1}^n f(Y_{i,2}, X_{i,2}, A_{i,2}) - \E \big[  f(Y_{i,2}, X_{i,2}, A_{i,2}) \mid \betahat1 \big] \Pto 0.
    \end{align*}
\end{lemma}

\begin{proof}
Proof of Lemma \ref{lemma:misspecHelp}.

Let $\delta > 0$.
\begin{align*}
    &\PP^{(n)} \left( \left| \frac{1}{n} \sum_{i=1}^n f(Y_{i,2}, X_{i,2}, A_{i,2}) - \E \big[  f(Y_{i,2}, X_{i,2}, A_{i,2}) \mid \betahat1 \big] \right| > \delta \right) \\
    &\underbrace{\leq}_{(a)} \frac{1}{\delta^2} \E^{(n)} \left[ \left( \frac{1}{n} \sum_{i=1}^n  f(Y_{i,2}, X_{i,2}, A_{i,2}) - \E \big[  f(Y_{i,2}, X_{i,2}, A_{i,2}) \mid \betahat1 \big] \right)^2 \right] \\
    &\underbrace{=}_{(b)} \frac{1}{\delta^2 n^2} \E^{(n)} \left[ \sum_{i=1}^n \left(  f(Y_{i,2}, X_{i,2}, A_{i,2}) - \E \big[  f(Y_{i,2}, X_{i,2}, A_{i,2}) \mid \betahat1 \big] \right)^2 \right] \\
    &= \frac{1}{\delta^2 n^2} \sum_{i=1}^n \E^{(n)} \left[ \E^{(n)} \left[ \left(  f(Y_{i,2}, X_{i,2}, A_{i,2}) - \E \big[  f(Y_{i,2}, X_{i,2}, A_{i,2}) \mid \betahat1 \big] \right)^2 \bigg| X_{i,2}, \betahat1 \right] \right] \\
    &= \frac{1}{\delta^2 n^2} \sum_{i=1}^n \E^{(n)} \left[ \E^{(n)} \left[ f(Y_{i,2}, X_{i,2}, A_{i,2})^2 \mid X_{i,2}, \betahat1 \right] - \E \big[  f(Y_{i,2}, X_{i,2}, A_{i,2}) \mid \betahat1 \big]^2 \right] \\
    &\leq \frac{2}{\delta^2 n^2} \sum_{i=1}^n \E^{(n)} \left[ \E^{(n)} \left[ f(Y_{i,2}, X_{i,2}, A_{i,2})^2 \mid X_{i,2}, \betahat1 \right] \right] \\
    &= \frac{2}{\delta^2 n^2} \sum_{i=1}^n \E^{(n)} \bigg[ \pi(X_{i,2}, 1; \betahat1) \E^{(n)} \left[ f(Y_{i,2}, X_{i,2}, A_{i,2})^2 \mid X_{i,2}, A_{i,2} = 1 \right] \\
    &+ \pi(X_{i,2}, 0; \betahat1) \E^{(n)} \left[ f(Y_{i,2}, X_{i,2}, A_{i,2})^2 \mid X_{i,2}, A_{i,2} = 0 \right] \bigg] \\
    &\underbrace{\leq}_{(c)} \frac{2}{\delta^2 n^2} \sum_{i=1}^n \E^{(n)} \left[ \pi(X_{i,2}, 1; \betahat1) \sigma^2 + \pi(X_{i,2}, 0; \betahat1) \sigma^2 \right] 
    \underbrace{=}_{(d)} \frac{2}{\delta^2 n^2} n \sigma^2 \to 0.
\end{align*}
Above, 
\begin{itemize}
    \item (a) holds by Markov inequality (Chebychev inequality).
    \item (b) holds because for $i \not= j$,
    \begin{align*}
        &\E^{(n)} \left[ \big\{ f(Y_{i,2}, X_{i,2}, A_{i,2}) - \E \big[  f(Y_{i,2}, X_{i,2}, A_{i,2}) \mid \betahat1 \big] \big\} \big\{ f(Y_{2,j}, X_{2,j}, A_{2,j}) - \E \big[  f(Y_{2,j}, X_{2,j}, A_{2,j}) \mid \betahat1 \big] \big\} \right] = 0.
    \end{align*}
    The above holds because (i) $\E^{(n)} \left[ \big\{ f(Y_{2,j}, X_{2,j}, A_{2,j}) - \E \big[  f(Y_{2,j}, X_{2,j}, A_{2,j}) \mid \betahat1 \big] \mid \HH_{1:n,1} \right] = 0$, and (ii) the tuples $(Y_{2,j}, X_{2,j}, A_{2,j}), (Y_{i,2}, X_{i,2}, A_{i,2})$ are independent conditional on $\HH_{1:n,1}$.
    \item (c) holds by our assumption that $\max_{a \in \{0,1\}} \E[ f (Y_{i,2}, X_{i,2}, a)^2 |X_{i,2}, A_{i,2}=a] \leq \sigma^2$ a.s. 
    \item (d) holds since $\pi(X_{i,2}, 1; \betahat1) + \pi(X_{i,2}, 0; \betahat1) = 1$.
\end{itemize}
\end{proof}

\begin{proof}
Proof of Proposition \print{\ref{prop:nonrepMiss}}{3.2}.

\bo{Part 1: Convergence of $\betahat1$.}
We first show that $\betahat1$ converges in probability to a constant. Using the definition of $\betahat1$ we can derive that:
\begin{align*}
    \betahat1 = \bigg( \frac{1}{n} \sum_{i=1}^n \tilde{\phi}(X_{i,1}, A_{i,1}) \tilde{\phi}(X_{i,1}, A_{i,1})^\top \bigg)^{-1} \frac{1}{n} \sum_{i=1}^n \tilde{\phi}(X_{i,1}, A_{i,1}) R(Y_{i,1})
\end{align*}
where  $\tilde{\phi}(X_{i,1}, A_{i,1}) \triangleq [ \phi_{\TN{inf}}(X_{i,1}), A_{i,1} \phi_{\TN{inf}}(X_{i,1}) \big]^\top$.

By the Weak Law of Large Numbers for i.i.d. random variables,
\begin{align}
    \label{eqn:misspec1}
    \frac{1}{n} \sum_{i=1}^n \tilde{\phi}(X_{i,1}, A_{i,1}) \tilde{\phi}(X_{i,1}, A_{i,1})^\top 
    \Pto \E \left[ \tilde{\phi}(X_{i,1}, A_{i,1}) \tilde{\phi}(X_{i,1}, A_{i,1})^\top \right] 
    \underbrace{=}_{(a)} \begin{bmatrix}
        1 & 0.5 ~ & 0.5 ~ & 0.25 \\
        0.5 & \frac{1}{3} & 0.25 & \frac{1}{6} \\
        0.5 & 0.25 & 0.5 & 0.25 \\
        0.25 & \frac{1}{6} & 0.25 & \frac{1}{6}
    \end{bmatrix} 
\end{align}
and
\begin{align}
    \label{eqn:misspec2}
    &\frac{1}{n} \sum_{i=1}^n \tilde{\phi}(X_{i,1}, A_{i,1}) R(Y_{i,1}) \Pto \E \left[ \tilde{\phi}(X_{i,1}, A_{i,1}) R(Y_{i,1}) \right] \\
    &\underbrace{=}_{(b)} \E \left[ \begin{pmatrix}
        1 \\
        U_{1,i} \\
        A_{i,1} \\
        A_{i,1} U_{1,i}
    \end{pmatrix} \left\{ 0.1+0.1 \cdot U_{1,i} + A_{i,1} \left[ \frac{1}{3} - 2 \cdot U_{1,i} + 2 \cdot U_{1,i}^2 \right] + \epsilon_{i,1} \right\} \right] \\
    &= \E \left[ \begin{pmatrix}
        0.1+0.1 \cdot U_{1,i} + 0.5 \left[ \frac{1}{3} - 2 \cdot U_{1,i} + 2 \cdot U_{1,i}^2 \right] \\
        U_{1,i} \left\{ 0.1+0.1 \cdot U_{1,i} + 0.5 \cdot \left[ \frac{1}{3} - 2 \cdot U_{1,i} + 2 \cdot U_{1,i}^2 \right] \right\} \\
        0.5 \cdot \left\{ 0.1+0.1 \cdot U_{1,i} + \frac{1}{3} - 2 \cdot U_{1,i} + 2 \cdot U_{1,i}^2 \right\} \\
        0.5 \cdot U_{1,i} \left\{ 0.1+0.1 \cdot U_{1,i} + \frac{1}{3} - 2 \cdot U_{1,i} + 2 \cdot U_{1,i}^2 \right\}
    \end{pmatrix} \right] \\
    &= \begin{pmatrix}
        0.15 + 0.5 \left[ \frac{1}{3} - 1 + \frac{2}{3} \right] \\
         0.05 + 0.1 \cdot \frac{1}{3} + 0.5 \cdot \left\{ \frac{1}{6} - \frac{1}{3} + 0.5 \right\} \\
        0.5 \cdot \left\{ 0.15 + 0.5 \left[ \frac{1}{3} - 1 + \frac{2}{3} \right] \right\} \\
        0.5 \left\{ 0.05 + 0.1 \cdot \frac{1}{3} + 0.5 \cdot \left( \frac{1}{6} - \frac{1}{3} + 0.5 \right) \right\}
    \end{pmatrix} 
    = \begin{pmatrix}
        0.15 \\
        0.05 + 0.1 \cdot \frac{1}{3} \\
        0.15/2 \\
        0.05/2 + 0.1 \cdot \frac{1}{2} / 2
    \end{pmatrix}
\end{align}
Above, (a) holds since $\tilde{\phi}(X_{i,1}, A_{i,1}) = \big[ 1, \, U_{1,i}, \, A_{i,1}, \, A_{i,1} U_{1,i} \big]^\top$, $U_{1,i} \sim \TN{Uniform}[0,1]$, and $A_{i,1} \mid U_{1,i} \sim \TN{Bernoulli}(0.5)$.
Above (b) also uses the definition of $R(Y_{i,1})$ from \print{\eqref{eqn:phi0}}{(10)}.

By continuous mapping theorem,
\begin{align*}
    \betahat1 \Pto \E \left[ \tilde{\phi}(X_{i,1}, A_{i,1}) \tilde{\phi}(X_{i,1}, A_{i,1})^\top \right]^{-1} \E \left[ \tilde{\phi}(X_{i,1}, A_{i,1}) R(Y_{i,1}) \right]
    = \betastar1 = \begin{pmatrix}
        0.1 \\
        0.1 \\
        0 \\
        0 
    \end{pmatrix}
\end{align*}

Also, note that by standard results for the asymptotic normality of least squares estimators on i.i.d. data \print{\citep[Chpt 5.3]{van2000asymptotic}}{[90, Chpt 5.3]}
\begin{align}
    \label{eqn:misspecNormality}
    \sqrt{n} \left( \betahat1 - \betastar1 \right) \Dto \N \left( 0, B^{-1} \Sigma B^{-1} \right),
\end{align}
where $B = \E \left[ \tilde{\phi}(X_{i,1}, A_{i,1}) \tilde{\phi}(X_{i,1}, A_{i,1})^\top \right]$ and $\Sigma = \E \left[ \big\{ R(Y_{i,1}) - \tilde{\phi}(X_{i,1}, A_{i,1})^\top \betastar1 \big\}^2 \tilde{\phi}(X_{i,1}, A_{i,1}) \tilde{\phi}(X_{i,1}, A_{i,1})^\top \right]$. \\

\medskip

\medskip 
%%%%%%%%%%%%%%%%%%%%%%%%%%%%%%%%%%%%%%%
\bo{Part 2: Deriving $\thetahat$ distribution.}
Using the definition of $\thetahat$ we can derive that:
\begin{align*}
    \thetahat = \left( \frac{1}{n} \sum_{i=1}^n \sum_{t=1}^2 \tilde\phi_{\TN{inf}}(X_{t,i}, A_{t,i}) \tilde\phi_{\TN{inf}}(X_{t,i}, A_{t,i})^\top \right)^{-1} \frac{1}{n} \sum_{i=1}^n \sum_{t=1}^2 \tilde\phi_{\TN{inf}}(X_{t,i}, A_{t,i}) R(Y_{i,t})
\end{align*}
where  $\tilde{\phi}(X_{i,1}, A_{i,1}) \triangleq [ \phi_{\TN{inf}}(X_{i,1}), A_{i,1} \phi_{\TN{inf}}(X_{i,1}) \big]^\top$.
By Lemma \ref{lemma:misspecHelp} we have that
\begin{align}
    \label{eqn:misspec3}
    \frac{1}{n} \sum_{i=1}^n \tilde\phi_{\TN{inf}}(X_{i,2}, A_{i,2}) \tilde\phi_{\TN{inf}}(X_{i,2}, A_{i,2})^\top - \E^{(n)} \big[ \tilde\phi_{\TN{inf}}(X_{i,2}, A_{i,2}) \tilde\phi_{\TN{inf}}(X_{i,2}, A_{i,2})^\top \mid \betahat{1,1} \big] \Pto 0
\end{align}
and 
\begin{align}
    \label{eqn:misspec4}
    \frac{1}{n} \sum_{i=1}^n \tilde\phi_{\TN{inf}}(X_{i,2}, A_{i,2}) R(Y_{i,2}) - \E^{(n)} \big[ \tilde\phi_{\TN{inf}}(X_{i,2}, A_{i,2}) R(Y_{i,2}) \mid \betahat{1,1} \big] \Pto 0.
\end{align}
Above the conditional expectations use $\betahat{1,1}$ rather than $\betahat1$ because the $\epsilon$-greedy action selection policy only uses $\betahat{1,1}$ to select actions for $\betahat1 = [\betahat{1,0}, \betahat{1,1}]$.

Let
\begin{align*}
    &g_0(\beta) \triangleq \E^{(n)} \big[ \tilde\phi_{\TN{inf}}(X_{i,2}, A_{i,2}) \tilde\phi_{\TN{inf}}(X_{i,2}, A_{i,2})^\top \mid \betahat{1,1} = \beta \big] \\
    &= \E \big[ \pi(X_{i,2}, 1; \beta ) \tilde\phi_{\TN{inf}}(X_{i,2}, 1) \tilde\phi_{\TN{inf}}(X_{i,2}, 1)^\top \big]
    + \E \big[ \pi(X_{i,2}, 0; \beta ) \tilde\phi_{\TN{inf}}(X_{i,2}, 0) \tilde\phi_{\TN{inf}}(X_{i,2}, 0)^\top \big]
\end{align*}
and let
\begin{align*}
    &g_1(\beta) \triangleq \E^{(n)} \big[ \tilde\phi_{\TN{inf}}(X_{i,2}, A_{i,2}) R(Y_{i,2}) \mid \betahat{1,1} = \beta \big] \\
    &= \E \big[ \pi(X_{i,2}, 1; \beta ) \tilde\phi_{\TN{inf}}(X_{i,2}, 1) R(Y_{i,2}(1)) \big]
    + \E \big[ \pi(X_{i,2}, 0; \beta ) \tilde\phi_{\TN{inf}}(X_{i,2}, 0) R(Y_{i,2}(0))^\top \big]
\end{align*}

By \eqref{eqn:misspec1}, \eqref{eqn:misspec2}, \eqref{eqn:misspec3}, \eqref{eqn:misspec4}, and continuous mapping theorem, we have that
\begin{align*}
    \thetahat 
    = \left( \E^{(n)} \left[ \tilde{\phi}(X_{i,1}, A_{i,1})^{\otimes 2} \right] + g_0(\betahat{1,1}) \right)^{-1} 
    \left\{ \E^{(n)} \big[ \tilde\phi_{\TN{inf}}(X_{i,1}, A_{i,1}) R(Y_{i,1}) \big] + g_1(\betahat{1,1}) \right\} + o_P(1).
\end{align*}

Recall that
\begin{align*}
    \PP \big( A_{i,2} = 1 \mid X_{i,2}, \HH_{1:n,1} \big)
    = \pi(X_{i,2}, 1; \betahat{1,1} )
    = \begin{cases}
        1-\epsilon/2 & \TN{if}~ X_{i,2}^\top \hat{\beta}_{1,1}^{(n)} > 0 \\
        \epsilon/2 & \TN{otherwise}
    \end{cases}.
\end{align*}
By the above, we have that $g_0(\sqrt{n} \beta) = g_0(\beta)$ and $g_1(\sqrt{n} \beta) = g_1(\beta)$. 
Thus,
\begin{align*}
    \thetahat 
    = \left( \E \left[ \tilde{\phi}(X_{i,1}, A_{i,1})^{\otimes 2} \right] + g_0(\sqrt{n}\betahat{1,1}) \right)^{-1} 
    \left\{ \E \big[ \tilde\phi_{\TN{inf}}(X_{i,1}, A_{i,1}) R(Y_{i,1}) \big] + g_1(\sqrt{n}\betahat{1,1}) \right\} + o_P(1).
\end{align*}

Furthermore, even though $\pi(\cdotspace; \beta)$ is discontinuous in $\beta$, since because $X_{i,2} \sim \TN{Uniform}[0,1]$ and $R(Y_{i,2}(a)) \sim \alpha_0 \phi_{\TN{gen}}(X_{i,t})^\top + a \phi_{\TN{gen}}(X_{i,t})^\top \alpha_1 + \epsilon_{i,t}$ it can be shown that $g_0$ and $g_1$ are continuous in $\beta$. Recall from \eqref{eqn:misspecNormality} that $\betahat1 = [\betahat{1,0}, \betahat{1,1}]$ is such that $\betahat{1,1} \Pto [0,0]$ and $\sqrt{n} \betahat{1,1} \Dto \N(0, S)$, for a covariance matrix $S$ equal to the lower right block of the matrix $B^{-1} \Sigma B^{-1}$. By continuous mapping theorem we have that
\begin{align*}
    \thetahat = \left( \E \left[ \tilde{\phi}(X_{i,1}, A_{i,1})^{\otimes 2} \right] + g_0( \tilde{\beta}) \right)^{-1} 
    \left\{ \E \big[ \tilde\phi_{\TN{inf}}(X_{i,1}, A_{i,1}) R(Y_{i,1}) \big] + g_1( \tilde{\beta}) \right\} + o_P(1)
    \quad \TN{where} \quad \tilde{\beta} \sim \N(0, S).
\end{align*}
One can simulate from this distribution to verify that it is not a constant. Moreover, one can verify that there exists a $\delta>0$ for which $\lim_{n\to\infty} \PP^{(n)} \big( \hat{\theta}_{1,1}^{(n)} > -0.0625 + \delta\big)>0$ and $\lim_{n\to\infty} \PP^{(n)}\big(\hat{\theta}_{1,1}^{(n)} < -0.0625 + \delta\big)>0$.
\end{proof}

\clearpage
%%%%%%%%%%%%%%%%%%%%%%%%%%%%%%%%%%%%%%%%%%%%%%%%%%%%%%%%%%
%%%%%%%%%%%%%%%%%%%%%%%%%%%%%%%%%%%%%%%%%%%%%%%%%%%%%%%%%%
\section{Proofs for Section \print{\ref{sec:replicableBandits}}{4} (Replicable Bandits)}
\label{app:replicable}

\subsection{Notation}
\label{app:notation}
In Sections \ref{app:notation}-\ref{sec:mainproof}, we allow policies to depend on the entire collective history $\HH_{1:n,t-1}$. Note that this is a strict generalization of allowing the bandit algorithm to depend on particular statistics $\betahat{t-1}$ formed using $\HH_{1:n,t-1}$. Thus $\pi \big(x,a; \betahat{t-1} \big)$ is replaced by $\pi_t(x,a; \HH_{1:n,t-1})$ in the definition and assumptions.

For convenience in this Appendix, we use the following notation:
\begin{align}
    \label{eqn:pinotation}
    \pi_{i,1:T}^*(P) \triangleq \prod_{t=1}^T \pi_t^*(X_{i,t}, A_{i,t}; P)
    \quad \quad \TN{and} \quad \quad
    \hat{\pi}_{i,1:T}^{(n)} \triangleq \prod_{t=1}^T \pi_t(X_{i,t}, A_{i,t}; \HH_{1:n,t-1}).
\end{align}

\subsubsection{Formalizing Uniform Convergence.}
~ \\

\bo{Convergence in probability.} When we say $Z^{(n)}$ converges in probability to $z$ uniformly over $P \in \MC{P}$ we mean that for any $\epsilon > 0$,
\begin{align*}
    \lim_{n \to \infty} \sup_{P \in \MC{P}} \PP_{P}^{(n)} \left( \big| Z^{(n)} -z \big| > \epsilon \right) = 0.
\end{align*}
We also use the notation $Z^{(n)} - z = \sup_{P \in \MC{P}} o_P(1)$. \\

\bo{Convergence in Distribution.} Similarly, when we say $Z^{(n)}$ converges in distribution to $Z$ uniformly over $P \in \MC{P}$ we mean that
\begin{align*}
    \lim_{n \to \infty} \sup_{P \in \MC{P}} \sup_{f \in BL} \left\{ \E_{P}^{(n)} \left[ f(Z^{(n)}) \right] - \E_{P}^{(n)} \left[ f(Z) \right] \right\} = 0,
\end{align*}
where we use $BL$ to denote all bounded Lipschitz functions. \\

\bo{Bounded in probability.} Finally, we use $Z^{(n)} = \sup_{P \in \MC{P}} O_P(1)$ to mean that for any $\epsilon > 0$, there exists some $M < \infty$ and $N \in \mathbb{N}$ such that for all $n \geq N$,
\begin{align*}
    \sup_{P \in \MC{P}} \PP_{P}^{(n)} \left( \big| Z^{(n)} \big| > M \right) < \epsilon.
\end{align*}

%%%%%%%%%%%%%%%%%%%%%%%%%%%%%%%%%%%%%%%%%%%%%%%%%%%%%%%%%%
\subsection{\texorpdfstring{$\thetastar(P)$}{Theta Star} is Well-Defined: Proof of Proposition \print{\ref{prop:thetastarWellDefined}}{1}}
\label{app:wellDefined}

\subsubsection{Helpful Lemma.}

\begin{lemma}
Suppose a bandit algorithm is replicable at a non-random, monotonic rate $\{ r^{(n)} \}_{n \geq 1}$ according to Definition \print{\ref{def:replicableBandit}}{3} display \print{\eqref{eqn:replicableDefRate}}{(13)} and Assumption \print{\ref{assump:limit}}{1} display \print{\eqref{eqn:policyStabilizesRate}}{(24)} holds at rate $\{ r^{(n)} \}_{n \geq 1}$, then
\begin{align}
    \sup_{P \in \MC{P}} \E_P^{(n)} \left[ r^{(n)} \left\| \pi_t \big(\cdotspace; \HH_{1:n,t-1} \big) - \pi_t^*(\cdotspace; P) \right\|_\infty  \right]
    \to 0.
    \label{eqn:limitingPistar}
\end{align}
where $\pi_t^*(x,a;P ) = \lim_{n \to \infty} \E_P^{(n)} \left[ \pi_t \big(x,a; \HH_{1:n,t-1} \big) \right]$; this latter limit exists by Assumption \print{\ref{assump:limit}}{1}.
\label{lemma:limitingPistar}
\end{lemma}

Note that a special case of Lemma \ref{lemma:limitingPistar} above is the setting in which there is a constant rate where $r^{(n)} = 1$ for all $n$.

\begin{proof}

We first show \eqref{eqn:limitingPistar}. Note that for any $P$ and $n$,
\begin{align*}
     \E_P^{(n)} \left[ \left\| \pi_t \big(\cdotspace; \HH_{1:n,t-1} \big) - \pi_t \big(\cdotspace; \tilde{\HH}_{1:n,t-1} \big) \right\|_\infty \right] 
    &= \E_P^{(n)} \left[ \E_P^{(n)} \left[ \left\| \pi_t \big(\cdotspace; \HH_{1:n,t-1} \big) - \pi_t \big(\cdotspace; \tilde{\HH}_{1:n,t-1} \big) \right\|_\infty \, \bigg| \, \HH_{1:n,t-1} \right] \right] \\
    &\geq \E_P^{(n)} \left[ \left\| \E_P^{(n)} \left[\pi_t \big(\cdotspace; \HH_{1:n,t-1} \big) - \pi_t \big(\cdotspace; \tilde{\HH}_{1:n,t-1} \big) \, \bigg| \, \HH_{1:n,t-1} \right] \right\|_\infty  \right] \\
    &= \E_P^{(n)} \left[ \left\| \pi_t \big(\cdotspace; \HH_{1:n,t-1} \big) - \E_P^{(n)} \left[ \pi_t \big(\cdotspace; \tilde{\HH}_{1:n,t-1} \big) \mid \HH_{1:n,t-1} \right] \right\|_\infty  \right] \\
    &= \E_P^{(n)} \left[ \left\| \pi_t \big(\cdotspace; \HH_{1:n,t-1} \big) - \E_P^{(n)} \left[ \pi_t \big(\cdotspace; \HH_{1:n,t-1} \big) \right] \right\|_\infty  \right].
\end{align*}
Above the inequality holds by Jensen's inequality. The final equality holds by independence of $\HH_{1:n,t-1}$ and $\tilde\HH_{1:n,t-1}$. 

By the above result we have that
\begin{align}
    &\sup_{P \in \MC{P}} \E_P^{(n)} \left[ r^{(n)} \left\| \pi_t \big(\cdotspace; \HH_{1:n,t-1} \big) - \E_P^{(n)} \left[ \pi_t \big(\cdotspace; \HH_{1:n,t-1} \big) \right] \right\|_\infty  \right] \\
    &\leq \sup_{P \in \MC{P}} \E_P^{(n)} \left[ r^{(n)} \left\| \pi_t \big(\cdotspace; \HH_{1:n,t-1} \big) - \pi_t \big(\cdotspace; \tilde{\HH}_{1:n,t-1} \big) \right\|_\infty \right] \to 0,
    \label{eqn:existenceHelp1}
\end{align}
as $n \to \infty$. The final limit above holds because for any $\delta > 0$,
\begin{align*}
    &\sup_{P \in \MC{P}} \E_P^{(n)} \left[ r^{(n)} \left\| \pi_t \big(\cdotspace; \HH_{1:n,t-1} \big) - \pi_t \big(\cdotspace; \tilde{\HH}_{1:n,t-1} \big) \right\|_\infty \right] \\
    &= \sup_{P \in \MC{P}} \E_P^{(n)} \left[ r^{(n)} \left\| \pi_t (\cdotspace; \HH_{1:n,t-1} ) - \pi_t (\cdotspace; \tilde{\HH}_{1:n,t-1} ) \right\|_\infty \II_{r^{(n)} \| \pi_t (\cdotspace; \HH_{1:n,t-1} ) - \pi_t (\cdotspace; \tilde{\HH}_{1:n,t-1} ) \|_\infty \leq \delta} \right] \\
    &+ \sup_{P \in \MC{P}} \E_P^{(n)} \left[ r^{(n)} \left\| \pi_t (\cdotspace; \HH_{1:n,t-1} ) - \pi_t (\cdotspace; \tilde{\HH}_{1:n,t-1} ) \right\|_\infty \II_{r^{(n)} \| \pi_t (\cdotspace; \HH_{1:n,t-1} ) - \pi_t (\cdotspace; \tilde{\HH}_{1:n,t-1} ) \|_\infty > \delta} \right] \\
    &\leq \delta + \sup_{P \in \MC{P}} \E_P^{(n)} \left[ 2 \cdot \II_{r^{(n)} \| \pi_t (\cdotspace; \HH_{1:n,t-1} ) - \pi_t (\cdotspace; \tilde{\HH}_{1:n,t-1} ) \|_\infty > \delta} \right] \\
    &= \delta + 2 \sup_{P \in \MC{P}} \PP_P^{(n)} \left( r^{(n)} \big\| \pi_t (\cdotspace; \HH_{1:n,t-1} ) - \pi_t (\cdotspace; \tilde{\HH}_{1:n,t-1} ) \big\|_\infty > \delta \right) \to \delta,
\end{align*}
as $n \to \infty$. The final limit above holds by our assumption that the bandit algorithm is replicable at rate $\{ r^{(n)} \}_{n \geq 1}$ according to Definition \print{\ref{def:replicableBandit}}{3} display \print{\eqref{eqn:replicableDefRate}}{(13)}
(recall $\pi \big(x,a; \betahat{t-1} \big)$ is replaced by $\pi_t(x,a; \HH_{1:n,t-1})$ in the definition and assumptions). Then by taking $\delta \to 0$ we have that \eqref{eqn:existenceHelp1} holds. \\

Note that by triangle inequality,
\begin{align*}
    &\sup_{P \in \MC{P}}  \E_P^{(n)} \left[ r^{(n)} \left\| \pi_t \big(\cdotspace; \HH_{1:n,t-1} \big) - \pi_t^*(\cdotspace; P) \right\|_\infty  \right] \\
    &\leq \underbrace{ \sup_{P \in \MC{P}}  \E_P^{(n)} \left[ r^{(n)} \left\| \pi_t \big(\cdotspace; \HH_{1:n,t-1} \big) - \E_P^{(n)} \left[ \pi_t \big(\cdotspace; \HH_{1:n,t-1} \big) \right] \right\|_\infty  \right] }_{(a)}
    + \underbrace{ \sup_{P \in \MC{P}} r^{(n)} \left\| \E_P^{(n)} \left[ \pi_t \big(\cdotspace; \HH_{1:n,t-1} \big) \right]  - \pi_t^*(\cdotspace; P) \right\|_\infty }_{(b)}
\end{align*}
Above (a) converges to 0 as $n \to \infty$ by \eqref{eqn:existenceHelp1}. Term (b) above goes to zero by Assumption \print{\ref{assump:limit}}{1} display \print{\eqref{eqn:policyStabilizesRate}}{(24)}, which we assume holds for rate $\{ r^{(n)} \}_{n \geq 1}$. 
\end{proof}

\medskip
\begin{corollary}
    \label{corollary:pihatConverges}
   Suppose a bandit algorithm has an exploration rate $\{ \pi_{\min}^{(n)} \}_{n \geq 1}$ such that $\pi_{\min}^{(n)} n \to \infty$ as $n \to \infty$. The bandit algorithm is also replicable at rate $\big\{ \frac{1}{ \pi_{\min}^{(n)} } \big\}_{n \geq 1}$ (according to Definition \print{\ref{def:replicableBandit}}{3}) and Assumption \print{\ref{assump:limit}}{1}  display \print{\eqref{eqn:policyStabilizesRate}}{(24)} holds at rate $\big\{ \frac{1}{ \pi_{\min}^{(n)} } \big\}_{n \geq 1}$. Then for any $t \in [2 \colon T]$ and any $\epsilon > 0$, 
    \begin{align*}
        \lim_{n \to \infty} \sup_{P \in \MC{P}} \PP_P \left( (\pi_{\min}^{(n)})^{-1} \left\| \prod_{t=1}^T \pi_t \big( \cdotspace; \HH_{1:n,t-1} \big) - \prod_{t=1}^T \pi_t^* \big( \cdotspace; P\big) \right\|_\infty > \epsilon \right) = 0.
    \end{align*}
    Above, $\pi_t^*(x,a;P ) = \lim_{n \to \infty} \E_P^{(n)} \left[ \pi_t \big(x,a; \HH_{1:n,t-1} \big) \right]$; this limit exists by Assumption \print{\ref{assump:limit}}{1}.
\end{corollary}

\begin{proof}
Proof of Corollary \ref{corollary:pihatConverges}. \\

Note that for any $x_{1:T} \in \MC{X}^T$ and $a_{1:T} \in \MC{A}^{1:T}$,
\begin{align*}
    \left| \prod_{t=1}^T \hat{\pi}_t^{(n)}(x_t, a_t) - \prod_{t=1}^T \pi_t^*(x_t, a_t; P) \right|
    &= \sum_{\tau=1}^T \bigg\{ \prod_{t=1}^{\tau-1} \hat{\pi}_t^{(n)}(x_t, a_t) \bigg\} \left| \hat{\pi}_\tau^{(n)}(x_\tau, a_\tau) - \pi_\tau^*(x_\tau, a_\tau; P) \right| \bigg\{ \prod_{t=\tau+1}^T \pi_t^*(x_t, a_t; P) \bigg\} \\
    &\leq \sum_{\tau=1}^T \left| \hat{\pi}_\tau^{(n)}(x_\tau, a_\tau) - \pi_\tau^*(x_\tau, a_\tau; P) \right| 
\end{align*}

By Slutsky's theorem it is sufficient to show that for any $t \in [2 \colon T]$, the following limit goes to zero for each $\epsilon > 0$:
\begin{align*}
    \lim_{n \to \infty} \sup_{P \in \MC{P}} \, \PP_P^{(n)} \left( (\pi_{\min}^{(n)})^{-1} \left\| \hat{\pi}_t^{(n)}(\cdotspace) - \pi_t^*(\cdotspace; P) \right\|_\infty > \epsilon \right)
    \leq \lim_{n \to \infty} \frac{1}{\epsilon} \sup_{P \in \MC{P}} \, \E_P^{(n)} \left[ (\pi_{\min}^{(n)})^{-1} \left\| \hat{\pi}_t^{(n)}(\cdotspace) - \pi_t^*(\cdotspace; P) \right\|_\infty \right] = 0.
\end{align*}
The inequality above holds by Markov inequality. The final limit above holds by Lemma \ref{lemma:limitingPistar} (which uses that Assumption \print{\ref{assump:limit}}{1}  display \print{\eqref{eqn:policyStabilizesRate}}{(24)} holds for rate $\big\{ \frac{1}{ \pi_{\min}^{(n)} } \big\}_{n \geq 1}$ and that the bandit algorithm is replicable at rate $\big\{ \frac{1}{ \pi_{\min}^{(n)} } \big\}_{n \geq 1}$).
\end{proof}

\subsubsection{Main Proof of Proposition \print{\ref{prop:thetastarWellDefined}}{1}.}

\begin{proof}
Recall from \print{\eqref{eqn:lossDef}}{(19)} that $\thetastar(P)$ solves $0 = \lim_{n \to \infty} \E_P^{(n)} \left[ \dot{\ell}_\theta(\HH_{i,T}) \right] \bigg|_{\theta = \thetastar(P)}$,
where $\dot{\ell}_\theta$ is the derivative of $\ell_\theta$ with respect to $\theta$ for $\ell_\theta \big( \HH_{i,T} \big) 
= \sum_{t=1}^T \left\{ \Y_{i,t} - \phi \big( X_{i,t}, A_{i,t}, \HH_{i,t-1} \big)^\top \theta \right\}^2$.   We show that $\thetastar(P)$ is well-defined.

Define $\theta^{*,(n)}(P)$ as the solution to $0 = \E_P^{(n)} \left[ \dot{\ell}_\theta(\HH_{i,T}) \right] \big|_{\theta = \theta^{*,(n)}(P)}$ which is given by
\begin{align*}
    \theta^{*,(n)}(P) = \E_P^{(n)} \bigg[ \sum_{t=1}^T \phi_{i,t} \phi_{i,t}^\top \bigg]^{-1}  \E_P^{(n)} \bigg[ \sum_{t=1}^T \phi_{i,t} \Y_{i,t} \bigg],
\end{align*}
where $\phi_{i,t} = \phi \big( X_{i,t}, A_{i,t}, \HH_{i,t-1} \big)$. The matrix $\E_P^{(n)} \left[ \sum_{t=1}^T \phi_{i,t} \phi_{i,t}^\top \right]^{-1}$ above is invertible by Assumption \print{\ref{assump:invertible}}{2}.

If we can show the following two results:
\begin{align}
    \label{eqn:wellDefinedPt1}
    \lim_{n \to \infty} \E_P^{(n)} \left[ \sum_{t=1}^T \phi_{i,t} \phi_{i,t}^\top \right] 
    = \E_{P, \pi^*} \left[ \sum_{t=1}^T \phi_{i,t} \phi_{i,t}^\top \right] 
\end{align}
and
\begin{align}
    \label{eqn:wellDefinedPt2}
    \lim_{n \to \infty} \E_P^{(n)} \left[ \sum_{t=1}^T \phi_{i,t} \Y_{i,t} \right] 
    = \E_{P, \pi^*} \left[ \sum_{t=1}^T \phi_{i,t} \Y_{i,t} \right] 
\end{align}
then we can conclude that
\begin{align*}
    \thetastar = \lim_{n \to \infty} \theta^{*,(n)}(P) = \E_{P, \pi^*} \left[ \sum_{t=1}^T \phi_{i,t} \phi_{i,t}^\top \right]^{-1}  \E_{P, \pi^*} \left[ \sum_{t=1}^T \phi_{i,t} \phi_{i,t}^\top \right].
\end{align*}

We now show that \eqref{eqn:wellDefinedPt2} holds. An analogous argument can be used to show \eqref{eqn:wellDefinedPt1}. For convenience, denote $f \big( \HH_{T,i} \big) \triangleq \sum_{t=1}^T \phi_{i,t} \Y_{i,t}$.

By telescoping sums,
\begin{align*}
    \E_P^{(n)} \left[ f \big( \HH_{T,i} \big) \right]
    - \E_{P, \pi^*} \left[ f \big( \HH_{T,i} \big) \right]
    = \sum_{t=3}^{T+1} \left\{ \E_{P, \hat{\pi}_{2:t-1}^{(n)}, \pi_{t:T}^*}^{(n)} \left[ f \big( \HH_{T,i} \big) \right] - \E_{P, \hat{\pi}_{2:t-2}^{(n)}, \pi_{t-1:T}^*}^{(n)} \left[ f \big( \HH_{T,i} \big) \right] \right\}
\end{align*}
Above we use the subscript $\hat{\pi}_{2:t-1}^{(n)}, \pi_{t:T}^*$ to denote that the actions $A_{i,2}, A_{i,3}, \dots, A_{i,t-1}$ are selected using the estimated policies $\hat\pi_{2:t-1}^{(n)}$ and the actions $A_{i,t}, A_{i,t+1}, \dots, A_{i,T}$ decisions are selected using policies $\pi_{t:T}^*(P)$. Note further that $\E_P^{(n)}\equiv \E_{P,\hat{\pi}_{2:T}^{(n)}}^{(n)}$.  Recall that $\pi_t^*(x,a;P ) = \lim_{n \to \infty} \E_P^{(n)} \left[ \pi_t \big(x,a; \HH_{1:n,t-1} \big) \right]$; this limit exists by Assumption \print{\ref{assump:limit}}{1} display \print{\eqref{eqn:policyStabilizes}}{(23)}.

Pick any $t \in [3 \colon T+1]$. By law of iterated expectations and rearranging terms:
\begin{align*}
    &\E_{P, \hat{\pi}_{2:t-1}^{(n)}, \pi_{t:T}^*}^{(n)} \left[ f \big( \HH_{T,i} \big) \right] - \E_{P, \hat{\pi}_{2:t-2}^{(n)}, \pi_{t-1:T}^*}^{(n)} \left[ f \big( \HH_{T,i} \big) \right] \\
    &= \E_P^{(n)} \left[ \sum_{a \in \MC{A}} \pi_{t-1} \big( X_{i,t-1}, a; \HH_{1:n,t-2} \big) \, \E_{P,\pi^*} \big[ f \big( \HH_{T,i} \big) \mid \HH_{1:n,t-2}, X_{i,t-1}, A_{i,t-1}=a \big] \right] \\
    &- \E_P^{(n)} \left[ \sum_{a \in \MC{A}} \pi_{t-1}^* \big( X_{i,t-1}, a \big) \, \E_{P,\pi^*} \big[ f \big( \HH_{T,i} \big) \mid \HH_{1:n,t-2}, X_{i,t-1}, A_{i,t-1}=a \big] \right] \\
    &= \sum_{a \in \MC{A}} \E_P^{(n)} \left[ \left\{  \pi_{t-1}\big( X_{i,t-1}, a; \HH_{1:n,t-2} \big) - \pi_{t-1}^* \big( X_{i,t-1}, a ; P\big) \right\} \, \E_{P,\pi^*} \big[ f \big( \HH_{T,i} \big) \mid \HH_{1:n,t-2}, X_{i,t-1}, A_{i,t-1}=a \big] \right].
\end{align*}

Thus, we have that\footnote{$||z||_1$ for a vector, $z$,  is the sum of the absolute values of the entries in the vector $z$.}
\begin{align*}
    &\left\| \E_P^{(n)} \left[ f \big( \HH_{T,i} \big) \right]
    - \E_{P, \pi^*} \left[ f \big( \HH_{T,i} \big) \right] \right\|_1 \\
    &= \left\| \sum_{t=2}^T \sum_{a \in \MC{A}} \E_P^{(n)} \left[ \left\{ \pi_t \big( X_{i,t}, a; \HH_{1:n,t-1} \big) - \pi_t^* \big( X_{i,t}, a ; P\big) \right\} \, \E_{P,\pi^*} \big[ f \big( \HH_{T,i} \big) \mid \HH_{1:n,t-1}, X_{i,t}, A_{i,t}=a \big] \right] \right\|_1 \\
    &\leq \sum_{t=2}^T \sum_{a \in \MC{A}} \E_P^{(n)} \left[ \left| \pi_t \big( X_{i,t}, a; \HH_{1:n,t-1} \big) - \pi_t^* \big( X_{i,t}, a ; P\big) \right| \, \left\| \E_{P,\pi^*} \big[ f \big( \HH_{T,i} \big) \mid \HH_{1:n,t-1}, X_{i,t}, A_{i,t}=a \big] \right\|_1 \right] \\
    &\leq \sum_{t=2}^T \sum_{a \in \MC{A}} \E_P^{(n)} \left[ \left\| \pi_t \big( \cdotspace; \HH_{1:n,t-1} \big) - \pi_t^* \big( \cdotspace; P\big) \right\|_\infty \, \left\| \E_{P,\pi^*} \big[ f \big( \HH_{T,i} \big) \mid \HH_{1:n,t-1}, X_{i,t}, A_{i,t}=a \big] \right\|_1 \right] \\
    &\leq 2 \sum_{t=2}^T \sum_{a \in \MC{A}} \E_P^{(n)} \left[  \left\| \E_{P,\pi^*} \big[ f \big( \HH_{T,i} \big) \mid \HH_{1:n,t-1}, X_{i,t}, A_{i,t}=a \big] \right\|_1 \right] 
    \to 0,
\end{align*}
as $n \to \infty$. We now show that the final limit above holds by Dominated Convergence Theorem.
\begin{itemize}
    %%%%%%%%%%%%%%%%%%%%%%%%%%%%%%
    \item Since $\| \phi \|_\infty < \infty$, we have that
    \begin{align*}
        \left\| \E_{P,\pi^*} \big[ f \big( \HH_{T,i} \big) \mid \HH_{1:n,t-1}, X_{i,t}, A_{i,t}=a \big] \right\|_1
        &= \bigg\| \E_{P,\pi^*} \bigg[ \sum_{j=1}^T \phi_{i,j} \Y_{i,j} \bigg| \HH_{1:n,t-1}, X_{i,t}, A_{i,t}=a \bigg] \bigg\|_1 \\
        &\leq \|\phi\|_\infty \E_{P,\pi^*} \bigg[ \sum_{j=1}^T \big| \Y_{i,j} \big| \bigg| \HH_{1:n,t-1}, X_{i,t}, A_{i,t}=a \bigg] \\
        &\leq \|\phi\|_\infty \E_{P,\pi^*} \left[ Y \mid \HH_{1:n,t-1}, X_{i,t}, A_{i,t}=a \right] 
        = O_P(1).
    \end{align*}
    The final limit above holds because for any $\epsilon > 0$ (below let $q$ be a constant we specify later), 
    \begin{align*}
        \PP_P^{(n)} \left( \big| \E_{P,\pi^*} \left[ Y \mid \HH_{1:n,t-1}, X_{i,t}, A_{i,t}=a \right] \big| > q \right) 
        &\underbrace{\leq}_{(i)} \frac{1}{q^2} \E_P^{(n)} \left[ \E_{P,\pi^*} \left[ Y \mid \HH_{1:n,t-1}, X_{i,t}, A_{i,t}=a \right]^2 \right] \\
        &\underbrace{\leq}_{(ii)} \frac{1}{q^2} \E_P^{(n)} \left[ \E_{P,\pi^*} \left[ Y^2 \mid \HH_{1:n,t-1}, X_{i,t}, A_{i,t}=a \right] \right] \\
        &\underbrace{\leq}_{(iii)} \frac{1}{q^2} c_Y^{1/2}.
    \end{align*}
    (i) holds by by Chebychev inequality. (ii) holds by Jensen's inequality. (iii) holds by Assumption \print{\ref{assump:moments}}{3}. By choosing $q = (\epsilon^{1/2} \cdot c_Y^{1/4})^{-1}$, we have shown the desired result.
\end{itemize}
\end{proof}

%%%%%%%%%%%%%%%%%%%%%%%%%%%%%%%%%%%%%%%%%%%%%%%%%%%%%%%%%%
\subsection{Proof of Theorem \print{\ref{thm:consistencyMiss}}{5.1} (Consistency)}
\label{app:consistency}

\subsubsection{Helpful Lemmas.}
\begin{lemma}
    Suppose a bandit algorithm 
    has an exploration rate $\{ \pi_{\min}^{(n)} \}_{n \geq 1}$ such that $\pi_{\min}^{(n)} n \to \infty$ as $n \to \infty$.
    Suppose also that Assumption \print{\ref{assump:limit}}{1} display \print{\eqref{eqn:policyStabilizes}}{(23)} holds. 
    Let $f$ be a real-valued function of $\HH_{i,T}$ such that 
    $\sup_{P \in \MC{P}} \E_{P, \pi^*} \big[ f(\HH_{i,T})^2 \big] < \infty$. 
    Then for each $\epsilon > 0$,
    \begin{align*}
        \lim_{n \to 0} \, \sup_{P \in \MC{P}} \, \PP_P^{(n)} \left( \bigg| \frac{1}{n} \sum_{i=1}^n \frac{\pi_{i,1:T}^*(P)}{ \hat{\pi}_{i,1:T}^{(n)} } f( \HH_{i,T} ) - \E_{P,\pi^*} \left[ f( \HH_{i,T} ) \right] \bigg| > \epsilon \right) = 0.
    \end{align*}
    Above $\pi_t^*(x,a;P ) = \lim_{n \to \infty} \E_P^{(n)} \left[ \pi_t \big(x,a; \HH_{1:n,t-1} \big) \right]$; this limit exists by Assumption \print{\ref{assump:limit}}{1} display \print{\eqref{eqn:policyStabilizes}}{(23)}.
    \label{lemma:uniformMWLLN}
\end{lemma}

\begin{proof}
Proof of Lemma \ref{lemma:uniformMWLLN}. \\

Using the notation from \eqref{eqn:pinotation}, note that by telescoping series: 
\begin{multline*}
    \frac{1}{n} \sum_{i=1}^n \frac{\pi_{i,1:T}^*(P)}{ \hat{\pi}_{i,1:T}^{(n)} } f( \HH_{i,T} ) - \E_{P, \pi^*} \left[ f( \HH_{i,T} ) \right] \\
    = \sum_{t=1}^T \frac{1}{n} \sum_{i=1}^n\bigg\{ \E_P^{(n)} \bigg[  \frac{\pi_{i,1:T}^*(P)}{ \hat{\pi}_{i,1:T}^{(n)} } f( \HH_{i,T} ) \bigg| \HH_{1:n,t} \bigg] 
    - \E_P^{(n)} \bigg[ \frac{\pi_{i,1:T}^*(P)}{ \hat{\pi}_{i,1:T}^{(n)} } f( \HH_{i,T} ) \bigg| \HH_{1:n,t-1} \bigg] \bigg\}.
\end{multline*}
Note that 
$\E_P^{(n)} \left[ \frac{\pi_{i,1:T}^*(P)}{ \hat{\pi}_{i,1:T}^{(n)} } f( \HH_{i,T} ) \big| \HH_{1:n,0} \right]=\E_P^{(n)} \bigg[ \frac{\pi_{i,1:T}^*(P)}{ \hat{\pi}_{i,1:T}^{(n)} } f( \HH_{i,T} ) \bigg]=\E_{P, \pi^*} \left[ f( \HH_{i,T} ) \right]  $.
Since $T$ is finite, to prove the Lemma, it is sufficient to show that for any $t \in [1 \colon T]$:
\begin{align*}
    \lim_{n \to 0} \, \sup_{P \in \MC{P}} \, \PP_P^{(n)} \bigg( \bigg| \frac{1}{n} \sum_{i=1}^n\bigg\{ \E_P^{(n)} \bigg[  \frac{\pi_{i,1:T}^*(P)}{ \hat{\pi}_{i,1:T}^{(n)} } f( \HH_{i,T} ) \bigg| \HH_{1:n,t} \bigg] 
    - \E_P^{(n)} \bigg[ \frac{\pi_{i,1:T}^*(P)}{ \hat{\pi}_{i,1:T}^{(n)} } f( \HH_{i,T} ) \bigg| \HH_{1:n,t-1} \bigg] \bigg\} \bigg| > \epsilon \bigg) = 0.
\end{align*}
Note that
\begin{align*}
    &\sup_{P \in \MC{P}} \, \PP_P^{(n)} \bigg( \bigg| \frac{1}{n} \sum_{i=1}^n\bigg\{ \E_P^{(n)} \bigg[  \frac{\pi_{i,1:T}^*(P)}{ \hat{\pi}_{i,1:T}^{(n)} } f( \HH_{i,T} ) \bigg| \HH_{1:n,t} \bigg] 
    - \E_P^{(n)} \bigg[ \frac{\pi_{i,1:T}^*(P)}{ \hat{\pi}_{i,1:T}^{(n)} } f( \HH_{i,T} ) \bigg| \HH_{1:n,t-1} \bigg] \bigg\} \bigg| > \epsilon \bigg) \\
    &\underbrace{\leq}_{(a)} \sup_{P \in \MC{P}} \, \frac{1}{ n^2 \epsilon^2}\E_P^{(n)} \bigg[ \bigg( \sum_{i=1}^n\bigg\{ \E_P^{(n)} \bigg[  \frac{\pi_{i,1:T}^*(P)}{ \hat{\pi}_{i,1:T}^{(n)} } f( \HH_{i,T} ) \bigg| \HH_{1:n,t} \bigg] 
    - \E_P^{(n)} \bigg[ \frac{\pi_{i,1:T}^*(P)}{ \hat{\pi}_{i,1:T}^{(n)} } f( \HH_{i,T} ) \bigg| \HH_{1:n,t-1} \bigg] \bigg\} \bigg)^2 \bigg] \\
    &= \sup_{P \in \MC{P}} \, \frac{1}{ n^2 \epsilon^2}\E_P^{(n)} \bigg[ \bigg( \sum_{i=1}^n\bigg\{ \frac{\pi_{i,1:t}^*(P)}{ \hat{\pi}_{i,1:t}^{(n)} } \E_{P, \pi^*} \left[ f( \HH_{i,T} ) \big| \HH_{1:n,t} \right] 
    - \frac{\pi_{i,1:t-1}^*(P)}{ \hat{\pi}_{i,1:t-1}^{(n)} } \E_{P, \pi^*} \left[ f( \HH_{i,T} ) \big| \HH_{1:n,t-1} \right] \bigg\} \bigg)^2 \bigg] \\
    &\underbrace{=}_{(b)} \sup_{P \in \MC{P}} \, \frac{1}{ n^2 \epsilon^2}\E_P^{(n)} \bigg[ \bigg( \sum_{i=1}^n \frac{\pi_{i,1:t-1}^*(P)}{ \hat{\pi}_{i,1:t-1}^{(n)} } \bigg\{ \frac{\pi_{i,t}^*(P)}{ \hat{\pi}_{i,t}^{(n)} } \E_{P, \pi^*} \left[ f( \HH_{i,T} ) \big| \HH_{i,t} \right] 
    - \E_{P, \pi^*} \left[ f( \HH_{i,T} ) \big| \HH_{i,t-1} \right] \bigg\} \bigg)^2 \bigg] \\
    &\underbrace{=}_{(c)} \sup_{P \in \MC{P}} \, \frac{1}{ n^2 \epsilon^2} \sum_{i=1}^n\E_P^{(n)} \bigg[  \bigg( \frac{\pi_{i,1:t-1}^*(P)}{ \hat{\pi}_{i,1:t-1}^{(n)} } \bigg)^2 \bigg\{ \frac{\pi_{i,t}^*(P)}{ \hat{\pi}_{i,t}^{(n)} } \E_{P, \pi^*} \left[ f( \HH_{i,T} ) \big| \HH_{i,t} \right] 
    - \E_{P, \pi^*} \left[ f( \HH_{i,T} ) \big| \HH_{i,t-1} \right] \bigg\}^2 \bigg] \\
    &\underbrace{\leq}_{(d)} \sup_{P \in \MC{P}} \, \frac{3 }{ n \epsilon^2} \bigg\{\E_P^{(n)} \bigg[ \bigg( \frac{\pi_{i,1:t}^*(P)}{ \hat{\pi}_{i,1:t}^{(n)} } \E_{P, \pi^*} \left[ f( \HH_{i,T} ) \big| \HH_{i,t} \right]  \bigg)^2 \bigg] 
    +\E_P^{(n)} \bigg[ \bigg( \frac{\pi_{i,1:t-1}^*(P)}{ \hat{\pi}_{i,1:t-1}^{(n)} } \E_{P, \pi^*} \left[ f( \HH_{i,T} ) \big| \HH_{i,t-1} \right] \bigg)^2 \bigg] \bigg\} \\
    &\underbrace{\leq}_{(e)} \sup_{P \in \MC{P}} \, \frac{3}{ n \epsilon^2 \pi_{\min}^{(n)} } \bigg\{ \E_P^{(n)} \bigg[ \frac{\pi_{i,1:t}^*(P)}{ \hat{\pi}_{i,1:t}^{(n)} } \E_{P, \pi^*} \left[ f( \HH_{i,T} ) \big| \HH_{i,t} \right]^2 \bigg] + \E_P^{(n)} \bigg[ \frac{\pi_{i,1:t-1}^*(P)}{ \hat{\pi}_{i,1:t-1}^{(n)} } \E_{P, \pi^*} \left[ f( \HH_{i,T} ) \big| \HH_{i,t-1} \right]^2 \bigg] \bigg\} \\
    &\underbrace{\leq}_{(f)} \sup_{P \in \MC{P}} \, \frac{6}{ n \epsilon^2 \pi_{\min}^{(n)} } \E_{P, \pi^*} \left[ f( \HH_{i,T} )^2 \right] 
    \underbrace{\to}_{(g)} 0.
\end{align*}
Above
\begin{itemize}
    \item (a) holds by Markov/Chebychev inequality.
    \item (b) holds since $\E_{P, \pi^*} \left[ f( \HH_{i,T} ) \big| \HH_{1:n,t} \right] = \E_{P, \pi^*} \left[ f( \HH_{i,T} ) \big| \HH_{i,t} \right]$, because when the fixed policy $\pi_t^*(\cdotspace; P)$ is used to select actions, the distribution of $\HH_{i,T}$ no longer depends on the histories of other individuals given $\HH_{i,t}$.
    \item (c) since the cross terms have expectation zero by law of iterated expectations since (i)
    \begin{align*}
        &\E_P^{(n)} \left[ \frac{\pi_{i,1:t-1}^*(P)}{ \hat{\pi}_{i,1:t-1}^{(n)} } \bigg\{ \frac{\pi_{i,t}^*(P)}{ \hat{\pi}_{i,t}^{(n)} } \E_{P, \pi^*} \left[ f( \HH_{i,T} ) \big| \HH_{i,t} \right] 
        - \E_{P, \pi^*} \left[ f( \HH_{i,T} ) \big| \HH_{i,t-1} \right] \bigg\} \bigg| \HH_{1:n,t-1} \right] \\
        &= \E_P^{(n)} \left[ \frac{\pi_{i,1:t-1}^*(P)}{ \hat{\pi}_{i,1:t-1}^{(n)} } \bigg\{ \frac{\pi_{i,t}^*(P)}{ \hat{\pi}_{i,t}^{(n)} } \E_{P, \pi^*} \left[ f( \HH_{i,T} ) \big| \HH_{i,t} \right] 
        - \E_{P, \pi^*} \left[ f( \HH_{i,T} ) \big| \HH_{i,t-1} \right] \bigg\} \bigg| \HH_{1:n,t-1} \right] = 0
    \end{align*}
    and (ii) for any $i \not= j$,
    \begin{multline*}
        \E_P^{(n)} \bigg[ \frac{\pi_{i,1:t-1}^*(P)}{ \hat{\pi}_{i,1:t-1}^{(n)} } \bigg\{ \frac{\pi_{i,t}^*(P)}{ \hat{\pi}_{i,t}^{(n)} } \E_{P, \pi^*} \left[ f( \HH_{i,T} ) \big| \HH_{i,t} \right] 
        - \E_{P, \pi^*} \left[ f( \HH_{i,T} ) \big| \HH_{i,t-1} \right] \bigg\} \\
        \frac{\pi_{j,1:t-1}^*(P)}{ \hat{\pi}_{j,1:t-1}^{(n)} } \bigg\{ \frac{\pi_{j,t}^*(P)}{ \hat{\pi}_{j,t}^{(n)} } \E_{P, \pi^*} \left[ f( \HH_{j,T} ) \big| \HH_{j,t} \right] 
        - \E_{P, \pi^*} \left[ f( \HH_{j,T} ) \big| \HH_{j,t-1} \right] \bigg\} \bigg| \HH_{1:n,t-1} \bigg]
    \end{multline*}
    \begin{multline*}
        = \frac{\pi_{i,1:t-1}^*(P)}{ \hat{\pi}_{i,1:t-1}^{(n)} } \frac{\pi_{j,1:t-1}^*(P)}{ \hat{\pi}_{j,1:t-1}^{(n)} } \E_P^{(n)} \bigg[ \bigg\{ \frac{\pi_{i,t}^*(P)}{ \hat{\pi}_{i,t}^{(n)} } \E_{P, \pi^*} \left[ f( \HH_{i,T} ) \big| \HH_{i,t} \right] 
        - \E_{P, \pi^*} \left[ f( \HH_{i,T} ) \big| \HH_{i,t-1} \right] \bigg\} \\
        \bigg\{ \frac{\pi_{j,t}^*(P)}{ \hat{\pi}_{j,t}^{(n)} } \E_{P, \pi^*} \left[ f( \HH_{j,T} ) \big| \HH_{j,t} \right] 
        - \E_{P, \pi^*} \left[ f( \HH_{j,T} ) \big| \HH_{j,t-1} \right] \bigg\} \bigg| \HH_{1:n,t-1} \bigg] 
    \end{multline*}
    \begin{multline*}
        = \frac{\pi_{i,1:t-1}^*(P)}{ \hat{\pi}_{i,1:t-1}^{(n)} } \frac{\pi_{j,1:t-1}^*(P)}{ \hat{\pi}_{j,1:t-1}^{(n)} } \bigg\{ \E_P^{(n)} \bigg[ \frac{\pi_{i,t}^*(P)}{ \hat{\pi}_{i,t}^{(n)} } \E_{P, \pi^*} \left[ f( \HH_{i,T} ) \big| \HH_{i,t} \right] 
        - \E_{P, \pi^*} \left[ f( \HH_{i,T} ) \big| \HH_{i,t-1} \right] \bigg| \HH_{1:n,t-1} \bigg]  \\
        \E_P^{(n)} \bigg[  \frac{\pi_{j,t}^*(P)}{ \hat{\pi}_{j,t}^{(n)} } \E_{P, \pi^*} \left[ f( \HH_{j,T} ) \big| \HH_{j,t} \right] 
        - \E_{P, \pi^*} \left[ f( \HH_{j,T} ) \big| \HH_{j,t-1} \right] \bigg| \HH_{1:n,t-1} \bigg] \bigg\}
    \end{multline*}
    The final equality holds because given $\HH_{1:n,t-1}$ the policy $\hat{\pi}_t^{(n)}$ is non-random, and individual histories $\HH_{i,t}$ and $\HH_{j,t}$ are independent conditional on $\HH_{1:n,t-1}$.
    \item (d) holds because $(a-b)^2 = a^2 - 2ab + b^2 \leq 3 (a^2 + b^2)$.
    \item (e) and (g) hold because the bandit algorithm has an exploration rate $\{ \pi_{\min}^{(n)} \}_{n \geq 1}$ such that $\pi_{\min}^{(n)} n \to \infty$ as $n \to \infty$.
    \item (f) holds by Jensen's inequality.
\end{itemize}
\end{proof}

\begin{lemma}
    Suppose a bandit algorithm has an exploration rate $\{ \pi_{\min}^{(n)} \}_{n \geq 1}$ such that $\pi_{\min}^{(n)} n \to \infty$ as $n \to \infty$. Also let the bandit algorithm is replicable at rate $\big\{ \frac{1}{ \pi_{\min}^{(n)} } \big\}_{n \geq 1}$ according to Definition \print{\ref{def:replicableBandit}}{3} display \print{\eqref{eqn:replicableDefRate}}{(13)} and Assumption \print{\ref{assump:limit}}{1} display \print{\eqref{eqn:policyStabilizesRate}}{(24)} holds for rate $\big\{ \frac{1}{ \pi_{\min}^{(n)} } \big\}_{n \geq 1}$. 
    Let $f$ be a real-valued function of $\HH_{i,T}$ such that $\sup_{P \in \MC{P}} \E_{P, \pi^*} \big[ f(\HH_{i,T})^2 \big] < \infty$. 
    Then for each $\epsilon > 0$,
    \begin{align*}
        \lim_{n \to 0} \, \sup_{P \in \MC{P}} \, \PP_P^{(n)} \left( \bigg| \frac{1}{n} \sum_{i=1}^n f( \HH_{i,T} ) - \E_{P,\pi^*} \left[ f( \HH_{i,T} ) \right] \bigg| > \epsilon \right) = 0.
    \end{align*}
    Above $\pi_t^*(x,a;P ) = \lim_{n \to \infty} \E_P^{(n)} \left[ \pi_t \big(x,a; \HH_{1:n,t-1} \big) \right]$; this limit exists by Assumption \print{\ref{assump:limit}}{1}.
    \label{lemma:noweightsWLLN}
\end{lemma}

\begin{proof}
Proof of Lemma \ref{lemma:noweightsWLLN}. \\

Using the notation from \eqref{eqn:pinotation}, note the following decomposition:
\begin{align*}
    \frac{1}{n} \sum_{i=1}^n f( \HH_{i,T} ) 
    = \underbrace{ \frac{1}{n} \sum_{i=1}^n \frac{\pi_{i,1:T}^*(P)}{ \hat{\pi}_{i,1:T}^{(n)} } f( \HH_{i,T} ) }_{(a)} 
    - \underbrace{ \frac{1}{n} \sum_{i=1}^n \frac{\pi_{i,1:T}^*(P) - \hat{\pi}_{i,1:T}^{(n)}}{ \hat{\pi}_{i,1:T}^{(n)} } f( \HH_{i,T} ) }_{(b)}
\end{align*}

\bo{Regarding term (a) above:} 
By Lemma \ref{lemma:uniformMWLLN} we have that for each $\epsilon > 0$:
\begin{align*}
     \lim_{n \to 0} \, \sup_{P \in \MC{P}} \, \PP_P^{(n)} \bigg( \bigg| \frac{1}{n} \sum_{i=1}^n \frac{\pi_{i,1:T}^*(P)}{ \hat{\pi}_{i,1:T}^{(n)} } f( \HH_{i,T} ) - \E_{P,\pi^*} \left[ f( \HH_{i,T} ) \right] \bigg| > \epsilon \bigg) = 0.
\end{align*}
We are able to apply Lemma \ref{lemma:uniformMWLLN} due to Assumption \print{\ref{assump:limit}}{1}, $\sup_{P \in \MC{P}} \E_{P, \pi^*} \big[ f(\HH_{i,T})^2 \big] < \infty$, and since the bandit algorithm has an exploration rate $\{ \pi_{\min}^{(n)} \}_{n \geq 1}$ such that $\lim_{n \to \infty} \pi_{\min}^{(n)} n \to \infty$. \\

\bo{Regarding term (b) above:} 
Note that for any $P$ and $n$ the following inequalities hold almost surely:
\begin{align*}
    & \left| \frac{1}{n} \sum_{i=1}^n \frac{\pi_{i,1:T}^*(P) - \hat{\pi}_{i,1:T}^{(n)}}{ \hat{\pi}_{i,1:T}^{(n)} } f( \HH_{i,T} ) \right| 
    \leq \frac{1}{n} \sum_{i=1}^n \frac{ \big| \pi_{i,1:T}^*(P) - \hat{\pi}_{i,1:T}^{(n)} \big|}{ \hat{\pi}_{i,1:T}^{(n)} } \left| f( \HH_{i,T} ) \right| \\
    &\leq \bigg( \frac{1}{n} \sum_{i=1}^n \frac{ \left| f( \HH_{i,T} ) \right| }{ \hat{\pi}_{i,1:T}^{(n)} } \bigg) \sup_{x_{1:T}, a_{1:T}} \left| \prod_{t=1}^T \hat{\pi}_t^{(n)}(x_t, a_t) - \prod_{t=1}^T \pi_t^*(x_t, a_t; P) \right| \\
    &\leq \underbrace{\bigg( \frac{1}{n} \sum_{i=1}^n \frac{ \pi^*_{1:T,i} }{ \hat{\pi}_{i,1:T}^{(n)} } \left| f( \HH_{i,T} ) \right| \bigg) }_{(i)} 
    \cdot \underbrace{ (\pi_{\min}^{(n)})^{-1} \cdot \sup_{x_{1:T}, a_{1:T}} \left| \prod_{t=1}^T \hat{\pi}_t^{(n)}(x_t, a_t) - \prod_{t=1}^T \pi_t^*(x_t, a_t; P) \right| }_{(ii)}
    %&= O_P(1) o_P(1)
\end{align*}
Note that the term (i) equals $\E_{P, \pi^*} \left[ \left| f( \HH_{i,T} ) \right| \right] + \sup_{P \in \MC{P}} o_P(1)$ by Lemma \ref{lemma:uniformMWLLN}. We are able to apply Lemma \ref{lemma:uniformMWLLN} due to Assumption \print{\ref{assump:limit}}{1}, $\sup_{P \in \MC{P}} \E_{P, \pi^*} \big[ f(\HH_{i,T})^2 \big] < \infty$, and since the bandit algorithm has an exploration rate $\{ \pi_{\min}^{(n)} \}_{n \geq 1}$ such that $\lim_{n \to \infty} \pi_{\min}^{(n)} n \to \infty$.

Regarding term (ii), note that by Corollary \ref{corollary:pihatConverges}, 
\begin{align*}
    \sup_{x_{1:T}, a_{1:T}} (\pi_{\min}^{(n)})^{-1} \left| \prod_{t=1}^T \hat{\pi}_t^{(n)}(x_t, a_t) - \prod_{t=1}^T \pi_t^*(x_t, a_t; P) \right| = \sup_{P \in \MC{P}} o_P(1).
\end{align*}
We are able to apply Corollary \ref{corollary:pihatConverges} as we do above because the bandit algorithm has an exploration rate $\{ \pi_{\min}^{(n)} \}_{n \geq 1}$ such that $\pi_{\min}^{(n)} n \to \infty$ as $n \to \infty$; Assumption \print{\ref{assump:limit}}{1} display \print{\eqref{eqn:policyStabilizesRate}}{(24)} holds for rate $\big\{ \frac{1}{\pi_{\min}^{(n)}} \big\}_{n \geq 1}$; and the bandit algorithm is replicable at rate $\big\{ \frac{1}{\pi_{\min}^{(n)}} \big\}_{n \geq 1}$.
\end{proof}

\subsubsection{Main Proof.}
\label{sec:mainproof}
\begin{proof}
Proof of Theorem \print{\ref{thm:consistencyMiss}}{5.1}. \\

Recall that:
\begin{align*}
    \thetastar(P) = \bigg( \E_{P, \pi^*} \bigg[ \sum_{t=1}^T \phi_{i,t} \phi_{i,t}^\top \bigg] \bigg)^{-1} \E_{P, \pi^*} \bigg[ \sum_{t=1}^T \phi_{i,t} \Y_{i,t}\bigg]
\end{align*}
Above, the matrix $\E_{P, \pi^*} \big[ \sum_{t=1}^T \phi_{i,t} \phi_{i,t}^\top \big]$ is invertible by Assumption \print{\ref{assump:invertible}}{2}.

By Lemma \ref{lemma:noweightsWLLN}, for each $\epsilon > 0$,
\begin{align}
    \label{eqn:covarianceConsistency}
    \lim_{n \to 0} \, \sup_{P \in \MC{P}} \, \PP_P^{(n)} \bigg( \bigg| 
    \frac{1}{n} \sum_{i=1}^n \sum_{t=1}^T \phi_{i,t} \phi_{i,t}^\top 
    - \E_{P, \pi^*} \left[ \sum_{t=1}^T \phi_{i,t} \phi_{i,t}^\top \right]
    \bigg| > \epsilon \bigg) = 0
\end{align}
We are able to apply Lemma \ref{lemma:noweightsWLLN} because $\|\phi\|_\infty < \infty$, the bandit algorithm has an exploration rate $\{ \pi_{\min}^{(n)} \}_{n \geq 1}$ such that $\pi_{\min}^{(n)} n \to \infty$ as $n \to \infty$, Assumption \print{\ref{assump:limit}}{1} display \print{\eqref{eqn:policyStabilizesRate}}{(24)} holds for rate $\big\{ \frac{1}{ \pi_{\min}^{(n)} } \big\}_{n \geq 1}$, and the bandit is replicable at rate $\big\{ \frac{1}{ \pi_{\min}^{(n)} } \big\}_{n \geq 1}$.

By \eqref{eqn:covarianceConsistency} and Assumption \print{\ref{assump:invertible}}{2} we have that the probability that $\frac{1}{n} \sum_{i=1}^n \sum_{t=1}^T \phi_{i,t} \phi_{i,t}^\top$ has eigenvalue greater than a constant $\lambda > 0$ goes to $1$ uniformly over $P \in \MC{P}$. Thus,
\begin{align}
    \label{eqn:invCovarianceConsistency}
    \lim_{n \to 0} \, \sup_{P \in \MC{P}} \, \PP_P^{(n)} \bigg( \bigg| 
    \bigg( \frac{1}{n} \sum_{i=1}^n \sum_{t=1}^T \phi_{i,t} \phi_{i,t}^\top \bigg)^{-1} 
    - \E_{P, \pi^*} \left[ \sum_{t=1}^T \phi_{i,t} \phi_{i,t}^\top \right]^{-1}
    \bigg| > \epsilon \bigg) = 0
\end{align}

By Lemma \ref{lemma:noweightsWLLN}, for each $\epsilon > 0$,
\begin{align}
    \label{eqn:numeratorConsistency}
    \lim_{n \to 0} \, \sup_{P \in \MC{P}} \, \PP_P^{(n)} \bigg( \bigg| 
    \frac{1}{n} \sum_{i=1}^n \sum_{t=1}^T \phi_{i,t} \Y_{i,t}
    - \E_{P, \pi^*} \left[ \sum_{t=1}^T \phi_{i,t} \Y_{i,t} \right]
    \bigg| > \epsilon \bigg) = 0
\end{align}
Again, we are able to apply Lemma \ref{lemma:noweightsWLLN} because $\|\phi\|_\infty < \infty$, Assumption \print{\ref{assump:moments}}{3} holds, the bandit algorithm has an exploration rate $\{ \pi_{\min}^{(n)} \}_{n \geq 1}$ such that $\pi_{\min}^{(n)} n \to \infty$ as $n \to \infty$, Assumption \print{\ref{assump:limit}}{1} display \print{\eqref{eqn:policyStabilizesRate}}{(24)} holds for rate $\big\{ \frac{1}{ \pi_{\min}^{(n)} } \big\}_{n \geq 1}$, and the bandit is replicable at rate $\big\{ \frac{1}{ \pi_{\min}^{(n)} } \big\}_{n \geq 1}$.

Using the definition of $\thetahat$ we can derive that:
\begin{align*}
    \thetahat = \left( \frac{1}{n} \sum_{i=1}^n \sum_{t=1}^T \phi_{i,t} \phi_{i,t}^\top \right)^{-1} \frac{1}{n} \sum_{i=1}^n \sum_{t=1}^T \phi_{i,t} \Y_{i,t}
\end{align*}
Above, $\phi_{i,t} \triangleq \phi \big(X_{t,i}, A_{t,i}, \HH_{i,t-1} \big)$. 

By continuous mapping theorem, and \eqref{eqn:numeratorConsistency} and \eqref{eqn:invCovarianceConsistency}, for each $\epsilon > 0$,
\begin{align*}
    \lim_{n \to \infty} \, \sup_{P \in \MC{P}} \, \PP_P^{(n)} \left( \big\| \hat\theta^{(n)} - \thetastar(P) \big\|_2 > \epsilon \right) = 0. 
\end{align*}
\end{proof}

%%%%%%%%%%%%%%%%%%%%%%%%%%%%%%%%%%%%%%%%%%%%%%%%%%%%%%%%%%
\subsection{Proof of Theorem \print{\ref{thm:normality}}{5.2} (Asymptotic Normality)}
\label{app:thmNormality}
In this section, we restrict the action sampling by the bandit algorithm to depend on  statistics $\betahat{t-1}$ formed using $\HH_{1:n,t-1}$ as in the main text.   

\subsubsection{Formula for Limiting Variance.}
\label{app:limitingVar}

We now write the formulas for the limiting variance from \print{\eqref{eqn:normalityResult}}{(26)}:
\begin{align*}
    \ddot{L}_P \triangleq \E_{P, \pi^*} \left[ \sum_{t=1}^T \phi_{i,t} \phi_{i,t}^\top \right]
\end{align*}
\begin{align*}
    \Sigma_P \triangleq \E_{P, \pi^*} \left[ \left( \dot{\ell}_{\thetastar}(\HH_{i,T}) - Q_{P,2:T} \psi_{\betastar{1:T-1}} (\HH_{i,T-1}) \right) \left( \dot{\ell}_{\thetastar}(\HH_{i,T}) - Q_{P,2:T} \psi_{\betastar{1:T-1}} (\HH_{i,T-1}) \right)^\top \right]
\end{align*}
\begin{align*}
    Q_{P,2:T} \triangleq 
    \left(Q_{P,2},Q_{P,3},\ldots, Q_{P,T}\right)
    \quad \TN{where} \quad 
    Q_{P,t} \triangleq \E_{P,\pi^*} \left[ \frac{ \dot{\ell}_{\thetastar}(\HH_{i,T}) }{\pi_{i,t}^*} \dot{\pi}_t \big(X_{i,t}, A_{i,t}; \betastar{t-1} \big)^\top \right].
\end{align*}
Recall that $\betastar{1:T-1}$ may depend on $P$.

\subsubsection{Main Proof of Theorem \print{\ref{thm:normality}}{5.2}.}
\begin{proof}
Recall  that:
\begin{align}
    \label{eqn:thetahatNormality}
    \thetahat = \left( \frac{1}{n} \sum_{i=1}^n \sum_{t=1}^T \phi_{i,t} \phi_{i,t}^\top \right)^{-1} \frac{1}{n} \sum_{i=1}^n \sum_{t=1}^T \phi_{i,t} \Y_{i,t}
\end{align}
where $\phi_{i,t} \triangleq \phi \big(X_{t,i}, A_{t,i}, \HH_{i,t-1} \big)$. 

Note that the probability that $\frac{1}{n} \sum_{i=1}^n \sum_{t=1}^T \phi_{i,t} \phi_{i,t}^\top$ has eigenvalue greater than a constant $\lambda > 0$ goes to $1$ uniformly over $P \in \MC{P}$. This is because by Lemma \ref{lemma:noweightsWLLN}, for each $\epsilon > 0$,
\begin{align}
    \label{eqn:covarianceConsistencyhi}
    \lim_{n \to 0} \, \sup_{P \in \MC{P}} \, \PP_P^{(n)} \bigg( \bigg| 
    \frac{1}{n} \sum_{i=1}^n \sum_{t=1}^T \phi_{i,t} \phi_{i,t}^\top 
    - \E_{P, \pi^*} \left[ \sum_{t=1}^T \phi_{i,t} \phi_{i,t}^\top \right]
    \bigg| > \epsilon \bigg) = 0
\end{align}
We are able to apply Lemma \ref{lemma:noweightsWLLN} because $\|\phi\|_\infty < \infty$, the bandit algorithm has a constant exploration rate, Assumption \print{\ref{assump:limit}}{1} display \print{\eqref{eqn:policyStabilizes}}{(23)} holds, and the bandit is replicable. By \eqref{eqn:covarianceConsistencyhi} and Assumption \print{\ref{assump:invertible}}{2} we have that the probability that $\frac{1}{n} \sum_{i=1}^n \sum_{t=1}^T \phi_{i,t} \phi_{i,t}^\top$ has eigenvalue greater than a constant $\lambda > 0$ goes to $1$ uniformly over $P \in \MC{P}$.

Note that by rearranging terms in the equality \eqref{eqn:thetahatNormality} we get that
\begin{align}
    &\sqrt{n} \big( \thetahat - \thetastar \big) 
    = \left( \frac{1}{n} \sum_{i=1}^n \sum_{t=1}^T \phi_{i,t} \phi_{i,t}^\top \right)^{-1} \frac{1}{\sqrt{n}} \sum_{i=1}^n \sum_{t=1}^T \phi_{i,t} \big( \Y_{i,t} - \phi_{i,t}^\top \thetastar \big) \\
    &\underbrace{=}_{(a)} - \left( \E_{P, \pi^*} \left[ \sum_{t=1}^T \phi_{i,t} \phi_{i,t}^\top \right] + \sup_{P \in \MC{P}} o_P(1) \right)^{-1} \frac{1}{\sqrt{n}} \sum_{i=1}^n \dot{\ell}_{\thetastar}(\HH_{i,T})  \\
    &= - \left( \E_{P, \pi^*} \left[ \sum_{t=1}^T \phi_{i,t} \phi_{i,t}^\top \right]^{-1} + \sup_{P \in \MC{P}} o_P(1) \right) \frac{1}{\sqrt{n}} \sum_{i=1}^n \frac{ \hat{\pi}_{i,2:T}^{(n)} }{ \hat{\pi}_{i,2:T}^{(n)} } \dot{\ell}_{\thetastar}(\HH_{i,T}) \\
    &\underbrace{=}_{(b)} - \left( \ddot{L}_{P}^{-1} + \sup_{P \in \MC{P}} o_P(1) \right) \bigg[ \underbrace{ \frac{1}{\sqrt{n}} \sum_{i=1}^n \frac{ \pi_{i,2:T}^* }{ \hat{\pi}_{i,2:T}^{(n)} } \dot{\ell}_{\thetastar}(\HH_{i,T}) }_{(i)}
    - \underbrace{ \frac{1}{\sqrt{n}} \sum_{i=1}^n \frac{ \hat{\pi}_{i,2:T}^{(n)} - \pi_{i,2:T}^* }{ \hat{\pi}_{i,2:T}^{(n)} } \dot{\ell}_{\thetastar}(\HH_{i,T}) }_{(ii)} \bigg]
    \label{eqn:thetahatDist}
\end{align}
Above, equality (a) uses \eqref{eqn:covarianceConsistencyhi} and that $\dot{\ell}_{\theta}(\HH_{i,T}) = - \sum_{t=1}^T \phi_{i,t} \big( \Y_{i,t} - \phi_{i,t}^\top \thetastar \big)$ by \print{\eqref{eqn:ellDef}}{(8)}. \\
Equality (b) uses that $\ddot{L}_{P} \triangleq \E_{P, \pi^*} \left[ \sum_{t=1}^T \phi_{i,t} \phi_{i,t}^\top \right]$ by definition. \\

\bo{Simplifying (ii) above.} 
\begin{align*}
    &\frac{1}{\sqrt{n}} \sum_{i=1}^n \frac{ \hat{\pi}_{i,2:T}^{(n)} - \pi_{i,2:T}^* }{ \hat{\pi}_{i,2:T}^{(n)} } \dot{\ell}_{\thetastar}(\HH_{i,T}) 
    \underbrace{=}_{(a)} \frac{1}{\sqrt{n}} \sum_{i=1}^n \frac{ \dot{\ell}_{\thetastar}(\HH_{i,T})  }{ \hat{\pi}_{i,2:T}^{(n)} } \sum_{t=2}^T \hat{\pi}_{i,2:t-1}^{(n)}  \cdot \big\{ \hat{\pi}_{i,t}^{(n)} - \pi_{i,t}^*(P) \big\} \cdot  \pi_{i,t+1:T}^*(P) \\
    &= \sum_{t=2}^T \frac{1}{n} \sum_{i=1}^n \frac{ \dot{\ell}_{\thetastar}(\HH_{i,T})  }{ \hat{\pi}_{i,2:T}^{(n)} } \hat\pi_{i,2:t-1}^{(n)} \cdot \pi_{i,t+1:T}^*(P) \cdot \sqrt{n} \big\{  \hat{\pi}_{i,t}^{(n)} - \pi_{i,t}^*(P) \big\} \\
    &\underbrace{=}_{(b)} \sum_{t=2}^T \frac{1}{n} \sum_{i=1}^n \frac{ \dot{\ell}_{\thetastar}(\HH_{i,T})  }{ \hat{\pi}_{i,2:T}^{(n)}} \hat\pi_{i,2:t-1}^{(n)} \cdot \pi_{i,t+1:T}^*(P) 
    \bigg\{ \dot{\pi} \big(X_{i,t}, A_{i,t}; \betastar{t-1}(P) \big)^\top \sqrt{n}\big( \betahat{t-1} - \betastar{t-1}(P) \big) \\
    &\quad \quad \quad \quad \quad \quad \quad \quad \quad \quad \quad \quad \quad \quad \quad \quad \quad \quad \quad \quad \quad \quad \quad \quad \quad \quad \quad \quad + \sqrt{n} \sup_{P \in \MC{P}}  o_P \big( \big\| \betahat{t-1} - \betastar{t-1}(P) \big\|_2 \big) \bigg\} \\
    &\underbrace{=}_{(c)} \sum_{t=2}^T \bigg[ \frac{1}{n} \sum_{i=1}^n \frac{ \dot{\ell}_{\thetastar}(\HH_{i,T}) }{ \hat{\pi}_{i,2:T}^{(n)} } \hat\pi_{i,2:t-1}^{(n)} \cdot \pi_{i,t+1:T}^*(P) \cdot \dot{\pi} \big(X_{i,t}, A_{i,t}; \betastar{t-1}(P) \big)^\top \bigg] \sqrt{n} \big( \betahat{t-1} - \betastar{t-1}(P) \big) \\
    &+ \sum_{t=2}^T \bigg[ \frac{1}{n} \sum_{i=1}^n \frac{ \dot{\ell}_{\thetastar}(\HH_{i,T})  }{ \hat{\pi}_{i,2:T}^{(n)} } \hat\pi_{i,2:t-1}^{(n)} \cdot \pi_{i,t+1:T}^*(P)\cdot \dot{\pi} \big(X_{i,t}, A_{i,t}; \betastar{t-1}(P) \big)^\top \bigg] \sup_{P \in \MC{P}} o_P (1) \\
    &\underbrace{=}_{(d)} \sum_{t=2}^T \bigg[ \frac{1}{n} \sum_{i=1}^n \frac{ \dot{\ell}_{\thetastar}(\HH_{i,T}) }{ \hat{\pi}_{i,2:T}^{(n)} } \left( \pi_{i,2:t-1}^*(P) + \sup_{P \in \MC{P}} o_P(1) \right) \cdot \pi_{i,t+1:T}^*(P) \cdot \dot{\pi} \big(X_{i,t}, A_{i,t}; \betastar{t-1}(P) \big)^\top \bigg] \sqrt{n} \big( \betahat{t-1} - \betastar{t-1}(P) \big) \\
    &+ \sum_{t=2}^T \bigg[ \frac{1}{n} \sum_{i=1}^n \frac{ \dot{\ell}_{\thetastar}(\HH_{i,T})  }{ \hat{\pi}_{i,2:T}^{(n)} } \left( \pi_{i,2:t-1}^*(P) + \sup_{P \in \MC{P}} o_P(1) \right) \cdot \pi_{i,t+1:T}^*(P) \cdot \dot{\pi} \big(X_{i,t}, A_{i,t}; \betastar{t-1}(P) \big)^\top\bigg] \sup_{P \in \MC{P}} o_P (1) \\
    &\underbrace{=}_{(e)} \sum_{t=2}^T \left[ Q_{P,t} + \sup_{P \in \MC{P}} o_P(1) \right] \sqrt{n} \big( \betahat{t-1} - \betastar{t-1}(P) \big)
    + \left\{ \sup_{P \in \MC{P}} O_P (1) \right\} \left\{ \sup_{P \in \MC{P}} o_P (1) \right\} \\
    &= \sum_{t=2}^T Q_{P,t} \sqrt{n} \big( \betahat{t-1} - \betastar{t-1}(P) \big) + \sup_{P \in \MC{P}} o_P (1) \\
    &
    =Q_{P,2:T} \sqrt{n} \big( \betahat{1:T-1} - \betastar{1:T-1}(P) \big) + \sup_{P \in \MC{P}} o_P (1) \\
    &\underbrace{=}_{(f)} Q_{P,2:T} \frac{1}{\sqrt{n}} \sum_{i=1}^n \frac{ \pi_{i,2:T}^* }{\hat\pi_{i,2:T}^{(n)}} \psi_{\betastar{1:T-1}} (\HH_{i,T-1}) + \sup_{P \in \MC{P}} o_P (1).
\end{align*}
Above,
\begin{itemize}
    \item (a) holds since by telescoping sums
    \begin{align*}
        \pi_{i,2:T}^*(P) - \hat{\pi}_{i,2:T}^{(n)}
        = \sum_{t=2}^T \pi_{i,2:t-1}^*(P) \cdot \big\{ \pi_{i,t}^*(P) - \hat{\pi}_{i,t} \big\} \cdot \hat{\pi}_{i,t+1:T}^{(n)}
    \end{align*}
    Above, by slight abuse of notation we use ``out of bounds'' indexing to refer to one, i.e., $\pi_{i,2:1}^*(P) \triangleq 1$ and $\hat{\pi}_{i,T+1:T}^{(n)} \triangleq 1$.
    %%%%%%%%%%%%%%%%%%%%%%%%%%
    \item (b) holds since by Assumption \print{\ref{assump:diffPolicy}}{4},
    \begin{align*}
        &\hat{\pi}_{i,t}^{(n)} - \pi_{i,t}^*(P)
        = \pi \big(X_{i,t}, A_{i,t}; \betahat{t-1} \big) - \pi \big(X_{i,t}, A_{i,t}; \betastar{t-1}(P) \big) \\
        &= \dot{\pi} \big(X_{i,t}, A_{i,t}; \betastar{t-1}(P) \big) \big( \betahat{t-1} - \betastar{t-1}(P) \big)
        + \sup_{P \in \MC{P}} o_P \big( \big\| \betahat{t-1} - \betastar{t-1}(P) \big\|_2 \big)
    \end{align*}
    %%%%%%%%%%%%%%%%%%%%%%%%%%
    \item (c) holds since by Assumption \print{\ref{assump:linearity}}{5},
    \begin{align*}
        \sup_{P \in \MC{P}}  o_P \big( \sqrt{n} \big\| \betahat{t-1} - \betastar{t-1}(P) \big\|_2 \big)
        = \sup_{P \in \MC{P}} o_P (1)
    \end{align*}
    %%%%%%%%%%%%%%%%%%%%%%%%%%
    \item (d) holds since
    \begin{align*}
        \hat\pi_{i,2:t-1}^{(n)} = \left( \hat\pi_{i,2:t-1}^{(n)} - \pi_{i,2:t-1}^*(P) \right) + \pi_{i,2:t-1}^*(P) 
        = \pi_{i,2:t-1}^*(P) + \sup_{P \in \MC{P}} o_P(1)
    \end{align*}
    The final equality above holds by Corollary \ref{corollary:pihatConverges} (which uses Assumption \print{\ref{assump:limit}}{1}, as well as the constant minimum exploration rate and the bandit replicability assumptions).
    %%%%%%%%%%%%%%%%%%%%%%%%%%
    \item (e) holds because of two results. The first is that by Lemma \ref{lemma:uniformMWLLN},
    \begin{align*}
        &\frac{1}{n} \sum_{i=1}^n \frac{ \dot{\ell}_{\thetastar}(\HH_{i,T}) }{ \hat{\pi}_{i,2:T}^{(n)} } \pi_{i,2:t-1}^*(P) \cdot \pi_{i,t+1:T}^*(P) \cdot \dot{\pi} \big(X_{i,t}, A_{i,t}; \betastar{t-1}(P) \big)^\top \\
        &= \frac{1}{n} \sum_{i=1}^n \frac{ \pi_{i,2:T}^*(P) }{ \hat{\pi}_{i,2:T}^{(n)} } \frac{ \dot{\ell}_{\thetastar}(\HH_{i,T}) }{ \pi_{i,t}^* } \cdot \dot{\pi} \big(X_{i,t}, A_{i,t}; \betastar{t-1}(P) \big)^\top \\
        &= \E_{P,\pi^*} \left[ \frac{ \dot{\ell}_{\thetastar}(\HH_{i,T}) }{\pi_{i,t}^*} \dot{\pi} \big(X_{i,t}, A_{i,t}; \betastar{t-1}(P) \big)^\top \right] + \sup_{P \in \MC{P}} o_P(1) = Q_{P,t} + \sup_{P \in \MC{P}} o_P(1)
    \end{align*}
     We are able to apply Lemma \ref{lemma:uniformMWLLN} due to Assumption \print{\ref{assump:limit}}{1} and since each  $(j,k)$ entry,
     \begin{align*}
         &\sup_{P \in \MC{P}} \E_{P, \pi^*} \bigg[ \bigg( \frac{ \dot{\ell}_{\thetastar}(\HH_{i,T})_{[k]} }{ \pi_{i,t}^* } \cdot \dot{\pi} \big(X_{i,t}, A_{i,t}; \betastar{t-1}(P) \big)_{[j]} \bigg)^2 \bigg] \\
         &\leq \frac{1}{\pi_{\min}^2} \sup_{ P \in \MC{P} } \E_{P, \pi^*} \bigg[  \dot{\ell}_{\thetastar}(\HH_{i,T})_{[k]}^2 \dot{\pi} \big(X_{i,t}, A_{i,t}; \betastar{t-1}(P) \big)_{[j]}^2 \bigg] \\
         &\leq \frac{1}{\pi_{\min}^2} \sup_{ P \in \MC{P} } \E_{P, \pi^*} \bigg[ \dot{\ell}_{\thetastar}(\HH_{i,T})_{[k]}^4 + \dot{\pi} \big(X_{i,t}, A_{i,t}; \betastar{t-1}(P) \big)_{[j]}^4 \bigg] < \infty
     \end{align*}
     Above the first inequality holds by our constant exploration assumption. The second inequality holds because $|a|^2 |b|^2 \leq |a|^4 + |b|^4$. The final inequality holds by Assumptions \print{\ref{assump:moments}}{3} and \print{\ref{assump:diffPolicy}}{4}. \\

     The second result is that
     \begin{align*}
        &\frac{1}{n} \sum_{i=1}^n \frac{ \dot{\ell}_{\thetastar}(\HH_{i,T}) }{ \hat{\pi}_{i,2:T}^{(n)} } \pi_{i,t+1:T}^*(P) \cdot \dot{\pi} \big(X_{i,t}, A_{i,t}; \betastar{t-1}(P) \big)^\top = \sup_{P \in \MC{P}} O_P(1),
    \end{align*}
    which follows by an analogous argument (by applying Lemma \ref{lemma:uniformMWLLN}).
    %%%%%%%%%%%%%%%%%%%%%%%%%%
    \item (f) holds by Assumption \print{\ref{assump:linearity}}{5} and continuous mapping theorem.
\end{itemize}

~\\
\bo{Consolidating Results.}
 Using \eqref{eqn:thetahatDist} and our simplification of the term (ii) above, we have that
\begin{align}
    \label{eqn:mNormalityApplyAlmost}
    &\sqrt{n} \big( \thetahat - \thetastar \big) \\
    &= -\left( \ddot{L}_{P}^{-1} + \sup_{P \in \MC{P}} o_P(1) \right) \frac{1}{\sqrt{n}} \sum_{i=1}^n \frac{ \pi_{i,2:T}^* }{ \hat{\pi}_{i,2:T}^{(n)} } \left( \dot{\ell}_{\thetastar}(\HH_{i,T}) - Q_{P,2:T} \psi_{\betastar{1:T-1}} (\HH_{i,T-1}) \right) + \sup_{P \in \MC{P}} o_P (1) 
\end{align}

Finally, we apply Theorem D.4 in \print{\citet{zhang2022statistical}}{Zhang et al. [106]} to get that
\begin{multline}
    \label{eqn:mNormalityApply}
    \frac{1}{\sqrt{n}} \sum_{i=1}^n \frac{ \pi_{i,2:T}^* }{ \hat{\pi}_{i,2:T}^{(n)} } \left( \dot{\ell}_{\thetastar}(\HH_{i,T}) + Q_{P,2:T} \psi_{\betastar{1:T-1}} (\HH_{i,T-1}) \right) \\
    \Dto \N \left( 0, \E_{P, \pi^*} \left[ \left( \dot{\ell}_{\thetastar}(\HH_{i,T}) - Q_{P,2:T} \psi_{\betastar{1:T-1}} (\HH_{i,T-1}) \right) \left( \dot{\ell}_{\thetastar}(\HH_{i,T}) - Q_{P,2:T} \psi_{\betastar{1:T-1}} (\HH_{i,T-1}) \right)^\top \right] \right)
\end{multline}
uniformly over $P \in \MC{P}$. We now list and verify the conditions for Theorem D.4 in \print{\citet{zhang2022statistical}}{Zhang et al. [106]} hold uniformly over $P \in \MC{P}$:
\begin{enumerate}
    \item \bo{Bounded Moments:} For each dimension $k$ for some $\alpha > 0$,
    \begin{align*}
        \sup_{P \in \MC{P}} \E_{P, \pi^*} \left[ \left( \dot{\ell}_{\thetastar}(\HH_{i,T}) + Q_{P,2:T} \psi_{\betastar{1:T-1}} (\HH_{i,T-1}) \right)_{[k]}^{2+\alpha} \right] < \infty.
    \end{align*}
    \item \bo{Minimum Exploration:} For a constant $\pi_{\min}$,
    \begin{align*}
        \inf_{\beta \in \real^{d_\beta}} \pi(a, x; \beta) > \pi_{\min}.
    \end{align*}
    \item \bo{Lipschitz Policy Function:} There exists a non-negative real-valued function $\dot{\pi}(a, s; P) \in \real^{d_\beta}$ such that $\sup_{P \in \MC{P}} \E_{P, \pi^*} \left[ \| \dot{\pi}(a, s; P) \|_1^{2+\alpha} \right] < \infty$ for some $\alpha > 0$ and for any $\beta, \beta' \in \real^{d_\beta}$
    \begin{align*}
        \sup_{P \in \MC{P}} \PP_P \bigg( | \pi(A_{i,t}, X_{i,t}; \beta) - \pi(A_{i,t}, X_{i,t}; \beta') | \leq \dot{\pi}(A_{i,t}, X_{i,t}; P) \| \beta - \beta' \|_2 \bigg) = 1.
    \end{align*}
    \item \bo{Square-root rate of parameters:}
    \begin{align*}
        \sup_{P \in \MC{P}} \big( \betahat{1:T-1} - \betastar{1:T-1}(P) \big) = \sup_{P \in \MC{P}} O_P(1/\sqrt{n}).
    \end{align*}
\end{enumerate}
We now verify that the above conditions hold:
\begin{enumerate}
    \item \bo{Bounded Moments:} For any dimension $k$,
    \begin{align*}
        &\sup_{P \in \MC{P}} \E_{P, \pi^*} \left[ \left| \dot{\ell}_{\thetastar}(\HH_{i,T})_{[k]} + Q_{P,2:T} \psi_{\betastar{1:T-1}} (\HH_{i,T-1})_{[k]} \right|^{2+\alpha} \right] \\
        &\leq 2^{1+\alpha} \sup_{P \in \MC{P}} \E_{P, \pi^*} \left[ \left| \dot{\ell}_{\thetastar}(\HH_{i,T})_{[k]} \right|^{2+\alpha} + \left| Q_{P,2:T} \psi_{\betastar{1:T-1}} (\HH_{i,T-1})_{[k]} \right|^{2+\alpha} \right] < \infty.
    \end{align*}
    The first inequality above holds because by Jensen's inequality $\big( \frac{|a+b|}{2} \big)^{2+\alpha}
    \leq \big( \frac{|a|+|b|}{2} \big)^{2+\alpha} \leq \frac{ |a|^{2+\alpha} + |b|^{2+\alpha} }{2}$. \\
    The final inequality above holds by Assumptions \print{\ref{assump:moments}}{3} and \print{\ref{assump:linearity}}{5}.
    \item \bo{Minimum Exploration:} This condition holds since our bandit algorithm has a constant exploration rate $\{ \pi_{\min}^{(n)} \}_{n \geq 1}$ where $\pi_{\min}^{(n)} = \pi_{\min}$ for a constant $\pi_{\min} > 0$.
    \item \bo{Lipschitz Policy Function:} This holds since by Assumption \print{\ref{assump:diffPolicy}}{4}, for each $x \in \MC{X}, a \in \MC{A}$, $\pi(x,a; \beta)$ is differentiable with respect to $\beta$. Also for each $\beta \in \real^{d_\beta}$, $\| \dot{\pi}(\cdotspace; \beta) \|_\infty < \infty$.
    \item \bo{Square-root rate of parameters:} By Assumption \print{\ref{assump:linearity}}{5}, $\sup_{P \in \MC{P}} \big( \betahat{1:T-1} - \betastar{1:T-1}(P) \big) = \sup_{P \in \MC{P}} O_P(1/\sqrt{n})$.
\end{enumerate}

By Slutsky's Theorem, continuous mapping theorem, \eqref{eqn:mNormalityApplyAlmost}, and \eqref{eqn:mNormalityApply} we have that
\begin{align*}
    \sqrt{n} \big( \thetahat - \thetastar \big)
    &= -\ddot{L}_{P}^{-1} \frac{1}{\sqrt{n}} \sum_{i=1}^n \frac{ \pi_{i,2:T}^* }{ \hat{\pi}_{i,2:T}^{(n)} } \left( \dot{\ell}_{\thetastar}(\HH_{i,T}) - Q_{P,2:T} \psi_{\betastar{1:T-1}} (\HH_{i,T-1}) \right) + \sup_{P \in \MC{P}} o_P (1) \\
    &\Dto \N \left( 0, \ddot{L}_{P}^{-1} \Sigma_P (\ddot{L}_{P}^{-1})^\top \right)
\end{align*}
uniformly over $P \in \MC{P}$.
Above we use
\begin{align*}
    \Sigma_P \triangleq \E_{P, \pi^*} \left[ \left( \dot{\ell}_{\thetastar}(\HH_{i,T}) - Q_{P,2:T} \psi_{\betastar{1:T-1}} (\HH_{i,T-1}) \right) \left( \dot{\ell}_{\thetastar}(\HH_{i,T}) - Q_{P,2:T} \psi_{\betastar{1:T-1}} (\HH_{i,T-1}) \right)^\top \right]
\end{align*}
which holds by definition.
\end{proof}

\clearpage
\section{Additional Simulation Details}
\label{app:simulations}

\subsection{Synthetic Simulation Details}
\label{app:syntheticEnv}

\subsubsection{Additional Simulation Environment Details.}

In the generation of outcomes $Y_{i,t}$ from \print{\eqref{eqn:Ysynthetic}}{(27)}, we use $\alpha_0 = 0$, $\alpha_1 = 1$, and $\alpha_2 = 0$. The reward noise $\epsilon_{i,t} \sim \N(0, 1)$ marginally; however, for each individual $i$ the reward noise is correlated over time. Specifically, $\TN{Corr}(\epsilon_{i,t}, \epsilon_{i,s}) = 0.5^{|t-s|/2}$.

\subsubsection{Additional Algorithm Details.}

All algorithms are initialized with default settings, run for $T=50$ decision times, and updated after every decision time.
\begin{itemize}

\item {\textbf{Boltzmann Sampling.}}
This algorithm is a modification of that described in Section \print{\ref{sec:boltzmann}}{4.4.1}. The algorithm parameters $\beta_t^{(n)} \in \mathbb{R}^4$ are initialized to $\bf{0}$ then updated to  after each decision time $t \in [1:T-1]$ according to the ridge regression criterion in display \print{\eqref{eqn:boltzmannRegression}}{(15)}, where $\lambda=1$ and $\phi_{\TN{alg}}(X_{i,t}) = [1, R(Y_{i,t-1})]$ for $t \in [2:T]$. The probability of taking action $1$ at time $t$ is computed similarly to \print{\eqref{eqn:boltzmannProb1}}{(16)}, but the action selection probabilities asymptote to $\pi_{\min} = 0.1$ and $1 - \pi_{\min} = 0.9$. This is enacted using the following generalized logistic function:
\begin{align*}
    \PP \big( A_{i,t} = 1 \mid \betahat{t-1}, X_{i,t} \big) 
    = \pi_{\min} + \frac{1 - 2\pi_{\min} }{1 + \exp(- s \cdot \phi_{\TN{alg}}(X_{i,t})^\top \betahat{t-1,1} )}, 
\end{align*}
with steepness $s=5$. See Section \print{\ref{sec:consistency}}{5.1} for a discussion of this type of ``minimum exploration'' constraint. \\

\item {\textbf{Smoothed Thompson Sampling.}}
This algorithm is as described in Section \print{\ref{sec:smoothed_thompson_sampling}}{4.4.2}, where $\phi_{\TN{alg}}(X_{i,t}) \triangleq [1, R(Y_{i,t-1})]$ and the allocation function is given by 
\begin{align*}
    \rho(x)
    = \pi_{\min} + \frac{1 - 2\pi_{\min} }{1 + \exp(- s \cdot x )}
\end{align*}
with $\pi_{\min} = 0.1$ and $s=5$. This is essentially the same generalized logistic function used for Boltzmann sampling.
The algorithm parameters $\beta_t^{(n)} \in \mathbb{R}^{14}$ encode the current multivariate normal distribution of the regression parameter vector $\alpha$ at each decision time, concatenating the mean vector and the upper triangle of the covariance matrix for this distribution into a single vector.  A standard normal prior distribution is used: $\alpha \sim N(\textbf{0}, I)$.  The scale of the prior variance is chosen to provide the same level of regularization as $\lambda$ does in the ridge regression used in Boltzmann sampling above. % see the discussion of the correspondence between Bayesian linear regression and ridge regression in Section 3.4.1 of \print{\citep{hastie2009elements}}{[32]}.  
Finally, the noise variance $\sigma^2$ in display \print{\eqref{smooth_thompson_sampling_blr_reward_model}}{(17)} is taken to be 1. \\

\item {\textbf{$\epsilon$-Greedy.}}
This algorithm is as described in Section \print{\ref{sec:non_repl_e_greedy}}{3.2.2}, with $\epsilon = 0.1$ and $\phi_{\TN{alg}}(X_{i,t}) \triangleq [1, R(Y_{i,t-1})]$.
\end{itemize}

\subsection{Oralytics Simulation Details}
\label{app:oralytics}

\subsubsection{Additional Simulation Environment Details.}
\label{sec_sim_env}
To create $n$ simulated environments (one per user), we first draw with replacement from $N_{\text{pilot}} = 9$ individuals from the Oralytics pilot study.\footnote{Oralytics pilot data is publicly accessible in GitHub \href{https://github.com/StatisticalReinforcementLearningLab/oralytics_pilot_data}{here}.} Each individual's data is used to create an environment model can further be represented by the following components:

\begin{itemize}
    \item Outcome Generating Process (i.e., brushing quality given covariates and action)
    \item Responsivity To Actions (i.e., delayed effect of current action on individual responsivity to future actions)
    \item App Engagement Behavior (i.e., probability of individual opening their app on any given day)
\end{itemize}

\bo{Environment State Features.} We first introduce the baseline and advantage feature spaces for each environment. Notice that these environment time-varying covariate features are a superset of the algorithm covariate features (Appendix~\ref{app_alg_state}). $X_{i,t} \in \mathbb{R}^d$ represents the $i$th individual's covariates at decision point $t$, where $d$ is the number of variables describing the individual's covariates. $g(X_{i,t}) \in \mathbb{R}^7$ denotes the features used in the environment model for both the baseline outcome and the advantage.

The features are:
\begin{enumerate}
\label{env_state_features}
    \item Intercept Term $\in \mathbb{R}$
    \item Time of Day (Morning/Evening) $\in \{0, 1\}$
    \item \label{alg_state:brushing2} $\Bar{B}$: Exponential Average of Brushing Durations Over Past 7 Days (Normalized) $\in [-1, 1]$
    \item \label{alg_state:actions2} $\Bar{A}$: Exponential Average of Engagement Prompts Sent Over Past 7 Days (Normalized)  $\in [-1, 1]$
    \item \label{alg_state:app2} Prior Day App Engagement $\in \{0, 1\}$ (if the individual has the app open and in focus, i.e., not in the background)
    \item Day of Week (Weekend / Weekday) $\in \{0, 1\}$
    \item Days Since Beginning of Oralytics Study (Normalized) $\in [-1, 1]$
\end{enumerate}

Features \print{\ref{alg_state:brushing2}}{3} and \print{\ref{alg_state:actions2}}{4} are 
\begin{align}
    \bar{B}_{i,t} = c_{\gamma}\sum_{j = 1}^{14} \gamma^{j-1} \Y_{i, t - j}
    \quad \TN{and} \quad
    \bar{A}_{i,t} = c_{\gamma}\sum_{j = 1}^{14} \gamma^{j-1} A_{i, t - j}
    \label{eqn:BbarAbar}
\end{align}
respectively, where $\gamma=13/14$. Recall that $\Y_{i, t}$ is the brushing quality defined in \print{\eqref{eqn:qualityDef}}{(28)} and $A_{i,t}$ is the treatment indicator. \\ %This is the same $\bar{B}_t, \bar{A}_t$ used in the cost term of the reward as described in Section~\ref{sec_app_reward}. \\

\bo{Outcome Generating Process.}
We use a zero-inflated Poisson to model each individual's outcome generating process because of the zero-inflated nature of brushing quality found in the Oralytics pilot data set. Each individual's outcome generating process is as follows:
$$
Z \sim \text{Bernoulli} \bigg(1 - \mathrm{sigmoid} \big( g(X_{i, t})^\top w_{i,b} - A_{i, t} \cdot \max \big[ \Delta_{i,B}^\top g(X_{i, t}), 0 \big] \big) \bigg)
$$
$$
S \sim \text{Poisson} \big( \exp \big( g(X_{i, t})^\top w_{i,p} + A_{i, t} \cdot \max\big[ \Delta_{i,N}^\top g(X_{i, t}), 0 \big] \big) \big)
$$
$$
\Y_{i, t} = ZS
$$

$w_{i,b}, w_{i,p}$ are individual-specific baseline weight vectors, $\Delta_{i,B}, \Delta_{i,N}$ are individual-specific advantage (or effect size) weight vectors. $g(X_{i, t})$ is the baseline feature vector and the feature vector that interacts with the effect size as described in the subsequent section, and $\mathrm{sigmoid}(x) = \frac{1}{1 + e^{-x}}$.

The individual's intent to brush is modeled by the Bernoulli outcome $Z$ and the individual's brushing quality when they intend to brush is modeled by the Poisson outcome $S$. Notice that the models for $Z, S$ means the effect size on $\Y$ must be non-negative. Otherwise, not sending an engagement prompt could yield a higher immediate brushing quality than sending an engagement prompt, which is nonsensical in this oral health setting. We ensure that $\max\big[ \Delta_{i,B}^\top h(X_{i, t}), 0 \big]$ and $\max\big[ \Delta_{i,N}^\top h(X_{i, t}), 0 \big]$ are non-negative to prevent the effect size from switching signs and having a negative effect on brushing quality.

We jointly fit weights $w_{i,b}, w_{i,p}, \Delta_{i,B}, \Delta_{i,N}$ for both the Bernoulli and the Poisson components using each individual's data. Specifically, we used MAP with prior $w_{i,b}, w_{i,p}, \Delta_{i,B}, \Delta_{i,N} \sim \mathcal{N}(0, I)$ as a form of regularization because we have sparse data for each individual. Finalized weight values were chosen by running random restarts and selecting the weights with the highest log posterior density. \\

%%%%%%%%%%%%%%%%%%%%%%%%%%%%%%%%%%%%%%%%%%%%

\bo{Responsivity to Actions.}
We model delayed effects of actions by shrinking individuals' responsiveness to actions, i.e., their initial treatment effect sizes. We shrink individuals' effect sizes by a factor $E=0.5$ when a certain criterion is met. The choice of $0.5$ was chosen in discussion with domain experts. Specifically, the criterion if either of the two scenarios holds: (a) $\mathbb{I}[\bar{B}_{i, t} > b]$ (individual brushes well) and $\mathbb{I}[\bar{A}_{i, t} > a_1]$ (individual was sent too many engagement prompts for a healthy brusher), or (b) $\mathbb{I}[\bar{A}_{i, t} > a_2]$ (the individual has been sent too many engagement prompts).  See display \eqref{eqn:BbarAbar} for definitions of $\bar{B}$ and $\bar{A}$; we define the values of hyperparamters $b, a_1, a_2$ in the next section (under the subheading ``Reward''). Notice that the same definitions and values of $\bar{B}, \bar{A}, b, a_1, a_2$ used in the environment models are the same as will be used the reward definition.

The first time an individual's criterion has been met, the individual's future effect sizes $\Delta_{i, B}, \Delta_{i, N}$ starting at time $t + 1$ will be scaled down proportionally by $E = 0.5$ (the individual is less responsive to treatment). Namely, the effect sizes will be shrunken to $E\cdot\Delta_{i,B}, E\cdot\Delta_{i,N}$. Then after a week, at time $t + 14$, we will check the criterion again. If the criterion is met again, the effect sizes will be further shrunk by a factor of $E$ down to $E^2\cdot\Delta_{i,B}, E^2\cdot\Delta_{i,N}$ starting at time $t + 15$. However, if the criterion is not fulfilled, then the individual recovers their original effect size $\Delta_{i,B}, \Delta_{i,N}$ starting at time $t + 15$. 
This procedure continues until the individual finishes the study. Notice that this means the individual can only have their effect size shrunk at most once a week. This procedure simulates how the individual may experience a reduction in responsiveness, but after a week, if the RL algorithm does not intervene too much, the individual may recover their prior responsivity. \\

\bo{App Engagement Behavior.}
We simulate individual app opening as follows. For each individual in the pilot study, we calculated $p^{\text{app}}_i$, the proportion of days that the individual opened the app during the pilot study (i.e., number of days the individual opened the app divided by 35, the number of days in the pilot study). Then at the end of each day in the simulation study, we independently sample from a Bernoulli distribution with probability $p^{\text{app}}_i$ for every individual $i$. \\

\subsubsection{Additional Algorithm Details.}
~ \\

\bo{Algorithm Features}
\label{app_alg_state}
$\phi_{\TN{alg}}(X_{i,t}) \in \mathbb{R}^5$ denotes the features used in the algorithm's model for both the baseline reward function and the advantage.

 The features are:
\begin{enumerate}
\label{alg_state_features}
     \item Intercept Term $\in \mathbb{R}$
    \item Time of Day (Morning/Evening) $\in \{0, 1\}$
    \item \label{alg_state:brushing} $\Bar{B}$: Exponential Average of Brushing Quality Over Past 7 Days (Normalized) $\in [-1, 1]$
    \item \label{alg_state:actions} $\Bar{A}$: Exponential Average of Engagement Prompts Sent Over Past 7 Days (Normalized)  $\in [-1, 1]$
    \item \label{alg_state:app} Prior Day App Engagement $\in \{0, 1\}$ (if the individual has the app open and in focus (i.e. not in the background))
\end{enumerate}

Features \print{\ref{alg_state:brushing}}{3} and \print{\ref{alg_state:actions}}{4} are $\bar{B}_{i,t} = c_{\gamma}\sum_{j = 1}^{14} \gamma^{j-1} \Y_{i, t - j}$ and $\bar{A}_{i,t} = c_{\gamma}\sum_{j = 1}^{14} \gamma^{j-1} A_{i, t - j}$ respectively, where $\gamma=13/14$. Recall that $\Y_{i, t}$ is brushing quality as defined in \print{\eqref{eqn:qualityDef}}{(28)} and $A_{i,t}$ is the treatment indicator. This is the same $\bar{B}_t, \bar{A}_t$ as defined in \print{\eqref{eqn:qualityDef}}{(28)}. \\

\bo{Treatment Action.}
\label{app:action_space}
The algorithm decided between sending a prompt $A_{i, t} = 1$ or not $A_{i, t} = 0$. If $A_{i, t} = 1$, the content of the prompt is randomly selected from three categories: (1) individual winning a gift (direct reciprocity), (2) individual winning a gift for their favorite charity (reciprocity by proxy), and (3) Q\&A for the morning decision time or feedback on prior brushing for the evening decision time. \\

\bo{Reward.}
\label{sec_app_reward}
The reward is a function of brushing quality $\Y_{i, t}$. $R(Y_{i,t}) := \Y_{i, t} - C_{i, t}$ where $C_{i , t}$ is a cost term that represents a crude approximation approximation of the delayed effects of actions. We now discuss in detail about the design of the cost term $C_{i, t}$. The inclusion of cost term $C_{i, t}$ allows the bandit algorithm to optimize for immediate healthy brushing behavior, while simultaneously considering the delayed effect of the current action on the effectiveness of future actions. The cost term can be interpreted as a function (with parameters $\xi_1, \xi_2$) which takes in current covariates $X_{i, t}$ and action $A_{i, t}$ and returns the delayed negative effect of currently sending an engagement prompt.

Recall in the time-varying covariate features, we defined $\bar{B}_{i,t} := c_{\gamma}\sum_{j = 1}^{14} \gamma^{j-1} \Y_{i, t - j}$ and $\bar{A}_{i,t} := c_{\gamma}\sum_{j = 1}^{14} \gamma^{j-1} A_{i, t - j}$. Note that in the algorithm time-varying covariate features, $\bar{B}, \bar{A}$ are normalized, but in the cost term below they are not. We set $\gamma = \frac{13}{14}$ to represent looking back 14 decision point points and scale each sum by constant $c_{\gamma}=\frac{1-\gamma}{1-\gamma^{14}}$ so that the weights sum to 1. Notice that our choice of $\gamma$ and the scaling constant means $0\le\bar{B}_{i,t}\le 180$ and $0\le \bar{A}_{i,t}\le 1$.
$\bar{B}_{i,t}$ captures the individual's exponentially discounted brushing quality in the past week. $\bar{A}_{i,t}$ captures the number of actions that were sent over the past week. 
Both terms are exponentially discounted because we expect that actions sent and individual brushing in the near past will be more predictive of the delayed impact of the actions (i.e., affecting a individual's responsivity to future actions) than those in the further past.
% Billie Note for Susan: I know you are interested in habit formation and I think there is a connection here. If the individual is a high performer during a week in which they received many prompts, it would make sense to penalize action delivery because its possible that the person has formed a habit and it is time to tone down action delivery to facilitate more independence; here, the disadvantage of sending a prompt is not related to habituation or burden, but to a missed opportunity to support independence and autonomy  

We define the cost of sending an engagement prompt (i.e. captures individual burden in sending a prompt) as:
\begin{equation}
\label{cost_term}
C_{i, t} := 
\begin{cases}
\xi_1 \mathbb{I}[\bar{B}_{i, t} > b] \mathbb{I}[\bar{A}_{i, t} > a_1] & \\
\hspace{10mm} + \xi_2 \mathbb{I}[\bar{A}_{i, t} > a_2]  & \smash{\raisebox{1.6ex}{if $A_{i, t} = 1$}} \\
0 & \hspace{-0mm} \mathrm{if~} A_{i, t} = 0
\end{cases}
\end{equation}

% explaining the terms
Notice that the algorithm only incurs a cost if the current action is to send an action, i.e., $A_{i, t} = 1$.  The first term $\xi_1 \mathbb{I}[\bar{B}_{i, t} > b] \mathbb{I}[\bar{A}_{i, t} > a_1]$ encapsulates the belief that if a high-performing individual was sent too many engagement prompts within the past week, then we want to penalize the reward. The second term $\xi_2 \mathbb{I}[\bar{A}_{i, t} > a_2]$ encapsulates the belief that regardless of individual performance, if they received too many engagement prompts within the past week, then we also want to penalize the reward. $b, a_1, a_2$ are chosen by domain experts. Notice that $a_1 < a_2$ because we believe a high-performing individual will have a lower threshold of being burdened by an engagement prompt. The scientific team decided to set the following values:

\begin{itemize}
    \item $b=111$, is set to the 50th-percentile of individual brushing durations in an earlier data set ROBAS 2.\footnote{ROBAS 2 data is publicly accessible in GitHub \href{https://github.com/ROBAS-UCLA/ROBAS.2}{here}.}
    \item $a_1 = 0.5$, represents a rough approximation of the individual getting an engagement prompt 50\% of the time (rough approximation because we are using an exponential average mean) 
    \item $a_2 = 0.8$, represents a rough approximation of the individual getting an engagement prompt 80\% of the time (rough approximation because we are using an exponential average mean) 
\end{itemize}
$\xi_1, \xi_2$ are non-negative hyperparameters that are set to $\xi_1 = 100, \xi_2 = 100$. \\

\bo{Update Times.} The Oralytics algorithm implemented a weekly update cadence (i.e., every $14$ decision times) for a total of $T=140$ decision times. During updates, the algorithm uses all the data collected for all individuals up to that time to re-fit reward model parameters. \\

\bo{Policies.} 
\begin{itemize}
\item {\textbf{Boltzmann Sampling.}}
This algorithm is a modification of that described in Section \print{\ref{sec:boltzmann}}{4.4.1}. The algorithm parameters are initialized to $\bf{0}$ and then updated to $\hat\beta_t^{(n)} \in \mathbb{R}^{10}$ after each update time $t$ according to the ridge regression criterion in \print{\eqref{eqn:boltzmannRegression}}{(15)}, where $\lambda=1.5512$ and $\phi_{\TN{alg}}(X_{i,t})$ contains the five features described at the beginning of this section. This value of $\lambda$ was chosen in order to provide approximately the same amount of regularization as the prior used in the smoothed Thompson Sampling algorithm below. \\ %; see Section 3.4.1 of \print{\citep{hastie2009elements}}{[32]} for a description of the correspondence between ridge regression and Bayesian linear regression we leverage here. \\

The probability of taking action $1$ at time $t$ is computed similarly to \print{\eqref{eqn:boltzmannProb1}}{(16)}, but with an additional constraint to be between $\pi_{\min} = 0.2$ and $1 - \pi_{\min} = 0.8$.  This is enacted via the following generalized logistic function:
\begin{align*}
    \PP \big( A_{i,t} = 1 \mid \betahat{t-1}, X_{i,t} \big) 
    = \pi_{\min} + \frac{1 - 2\pi_{\min} }{1 + \exp(- s \cdot \phi_{\TN{alg}}(X_{i,t})^\top \betahat{t-1,1} )}, 
\end{align*}
with steepness $s=0.025 \approx \frac{1}{38.83}$, where $38.83$ is the standard deviation of the reward noise. See Section \print{\ref{sec:consistency}}{5.1} for a discussion of this type of ``minimum exploration'' constraint. \\

\item {\textbf{Smoothed Thompson Sampling.}}
This algorithm is as described in Section \print{\ref{sec:smoothed_thompson_sampling}}{4.4.2}, where $\phi_{\TN{alg}}(X_{i,t})$ contains the five features described at the beginning of this section, and the allocation function is given by 
\begin{align*}
    \rho(x)
    = \pi_{\min} + \frac{1 - 2\pi_{\min} }{1 + \exp(- s \cdot x )}
\end{align*}
with $\pi_{\min} = 0.2$ and $s=0.025 \approx \frac{1}{38.83}$, where $38.83$ is the standard deviation of the reward noise. This is essentially the same generalized logistic function used for Boltzmann sampling. \\

The algorithm parameters $\hat\beta_t^{(n)} \in \mathbb{R}^{65}$ encode the current multivariate normal distribution of the regression parameter vector $\alpha$ at each decision time, concatenating the mean vector and the upper triangle of the covariance matrix for this distribution into a single vector. A conjugate prior distribution inspired by that used in \print{\citep{trella2024oralytics}}{[88]} is used: 
$$\alpha \sim N(\textbf{0}, \mathrm{diag}(5329,\ 2500,\ 9025,\ 2916,\ 6889,\ 529,\ 4225,\ 4761,\ 3136,\ 1156)).$$  
Finally, the noise variance $\sigma^2$ display in \print{\eqref{smooth_thompson_sampling_blr_reward_model}}{(17)} is set to the estimated reward variance fitted using Oralytics pilot data ($\sigma^2 = 3878$) from \print{\citep{trella2024oralytics}}{[88]}.  \\

\item {\textbf{$\epsilon$-Greedy.}}
This algorithm is as described in Section \print{\ref{sec:non_repl_e_greedy}}{3.2.2}, with $\epsilon = 0.2$ and $\phi_{\TN{alg}}(X_{i,t})$ containing the five features mentioned at the beginning of this section.
\end{itemize}

\end{document}